\documentclass{article}

\usepackage{arxiv}
\usepackage{verbatim}
\usepackage{graphicx}%
\usepackage{multirow}%
\usepackage{amsthm}%
\usepackage{mathrsfs}%
\usepackage{booktabs}
\usepackage{nicefrac}
\usepackage{epstopdf}
\usepackage[title]{appendix}%
\usepackage{xcolor}%
\usepackage{textcomp}%
\usepackage{manyfoot}%
\usepackage{booktabs}%
\usepackage{algorithm}%
\usepackage{subfigure}
\usepackage{algorithmicx}%
\usepackage{algpseudocode}%
\usepackage{listings}%
\usepackage{xurl}
\usepackage{tikz}
\usepackage{caption}
\usepackage{subcaption}
\usetikzlibrary{arrows.meta, positioning}
\usepackage[normalem]{ulem}

\RequirePackage{amsthm,amsmath,amsfonts,amssymb}
\RequirePackage[numbers]{natbib}
\RequirePackage[colorlinks,citecolor=blue,urlcolor=blue,backref=page,backref=page]{hyperref}
\RequirePackage{graphicx}
\theoremstyle{plain}

\newtheorem{theorem}{Theorem}[section]
\newtheorem{proposition}{Proposition}[section]

\theoremstyle{definition}

\theoremstyle{remark}

\title{Scalable Bayesian Spatiotemporal Tensor Modeling for Censored and Missing Areal Data}

\author{
  Jose A. Ordoñez \\
  Department of Statistics\\
  Pontificia Universidad Católica de Chile\\
  Santiago, PA 782 0436 \\
  \texttt{jose.ordonez@uc.cl} \\
   \And
 Tsung-I Lin \\
  Institute of Statistics, National Chung Hsing University\\
  Department of Public Health, China
Medical University \\
  \texttt{tilin@nchu.edu.tw} \\
  \And
  Victor H. Lachos \\
  Department of Statistics\\
  University of Connecticut\\
  \texttt{hlachos@uconn.edu} \\
  \And
  Luis M. Castro \\
  Department of Statistics\\
  Pontificia Universidad Católica de Chile\\
  Santiago, PA 782 0436 \\
  \texttt{lmcastro@uc.cl} \\
}

\begin{document}
\maketitle
\begin{abstract}
We propose a new Bayesian approach for spatiotemporal areal data with censored and missing observations, where the response is represented as a tensor indexed by spatial units and time points. The method introduces a flexible random effect that combines the spatial dependence structures of the Simultaneous Autoregressive (SAR) and Directed Acyclic Graph Autoregressive (DAGAR) models with a temporal autoregressive component. This formulation brings both spatial models into a unified spatiotemporal framework by expressing them as Gaussian Markov random fields in innovation form, providing an interpretable representation of spatial, temporal, and joint spatiotemporal dependence while preserving the multiway structure of the data. The innovation-based structure enables scalable Bayesian implementation in \texttt{Stan} for datasets of moderate size. Simulation studies show that the proposed model outperforms common ad hoc imputation strategies for censored and missing data. We further illustrate its practical value through an application to carbon monoxide (CO) concentrations, where the proposed framework captures complex spatiotemporal dependence in the presence of censoring and missingness.
\end{abstract}


\section{Introduction}\label{sec1}

In recent years, rapid technological advances have substantially improved the collection of spatially and temporally referenced data, making this information increasingly available and valuable for scientific research. However, incorporating the temporal dimension into spatial analyses introduces additional methodological and computational challenges, particularly when flexible statistical models are required. These challenges become even more pronounced in large datasets, where both spatial and temporal dimensions can substantially increase the computational burden.

Such methodologies have been widely used in environmental and pollution-related applications, where large monitoring networks, satellite products, and sensor-based measurements generate high-dimensional spatiotemporal datasets. For instance, \citet{gaedkemerzhaeuser2025accelerated} proposed an accelerated spatiotemporal Bayesian modeling framework for multivariate Gaussian processes, motivated by the need for scalable inference in complex environmental settings. Similarly, \citet{deLuna2005Predictive} introduced a family of predictive spatiotemporal models specifically designed for environmental data, focusing on carbon monoxide concentrations in Venice, which are characterized by spatial sparsity and temporal richness. Their methodology relies on a vector autoregressive specification, treating each monitoring station as a time series, and a model-building strategy that facilitates the identification of spatial dependence. In a related context, \citet{Wang2019} proposed a penalized local polynomial regression model for spatial data analysis, aiming to capture spatial heterogeneity in regression coefficients. Their methodology was applied to environmental variables, including PM2.5 concentrations and several pollutant gases, by representing spatially varying parameters through local polynomials defined at selected anchor points and estimating them using a penalized least-squares procedure. Together, these contributions illustrate the growing interest in developing flexible and computationally feasible tools for modeling complex environmental processes, particularly when spatial heterogeneity, temporal evolution, or high-dimensional observations must be considered.

Within this broader line of research, several contributions have focused more directly on the explicit modeling of spatiotemporal dependence in air pollution studies. \citet{Deb2019SpatioTemporal} introduced a spatiotemporal model for PM2.5 concentrations that accounts for space-time interactions by allowing temporal pollution trends to be more similar among sites that are close in space. Their model was applied to a ten-year dataset collected from 66 monitoring stations across Taiwan, addressing an important gap in the literature on air pollution modeling in Asian contexts. \citet{valeriano2021}, on the other hand, extended the spatiotemporal modeling framework by introducing separable correlation structures and implementing the Stochastic Approximation Expectation-Maximization algorithm for parameter inference. Their approach facilitates maximum likelihood estimation in complex settings by approximating the expectation step using stochastic simulations. Missing and censored observations were accommodated by augmenting the complete data formulation and modeling the unobserved values using truncated normal distributions. Although their empirical application did not involve carbon monoxide (CO) or particulate matter, the methodology was illustrated using ozone concentration data from New York, USA.

Our proposal introduces a Bayesian spatiotemporal model for areal data with censored and missing observations, treating these mechanisms as informative features of the data-generating process, rather than as nuisance features. In this setting, the response is naturally viewed as a tensor indexed by areal units and time points, which preserves the multiway structure of the data and provides a convenient representation for modeling spatial, temporal, and joint spatiotemporal dependencies. Since one mode of this tensor corresponds to areal units, we build on the directed acyclic graph autoregressive model \citep[DAGAR; see][]{datta2019}, which represents spatial dependence through a directed acyclic graph. This structure provides a natural way to encode neighborhood relationships among zones while inducing sparse and positively definite precision-based representations. Such sparsity is particularly important for scalable computation because it avoids the direct manipulation of dense spatial covariance matrices and offers an efficient alternative to classical areal models such as the conditional autoregressive \citep[CAR;][]{Besag1974} and simultaneous autoregressive \citep[SAR;][]{Whittle1954} models.

A key contribution of our paper is the construction of a flexible spatiotemporal random effect defined over this tensor structure, combining areal spatial dependence with a temporal autoregressive component. In this formulation, the spatial component captures dependence among areal units via either the DAGAR or SAR specification, whereas the temporal component describes persistence through an innovation-based autoregressive mechanism. By expressing the resulting process as a Gaussian Markov random field in the innovation form, the proposed framework exploits sparsity and conditional independence across both space and time, leading to a scalable representation of spatiotemporal dependence. This formulation reduces the computational burden associated with high-dimensional covariance matrices, facilitates Bayesian implementation in \texttt{Stan}, and provides interpretable parameters for spatial, temporal, and joint spatiotemporal dependencies. Therefore, both DAGAR- and SAR-based spatiotemporal models can be studied using a unified tensor-based inferential strategy. Simulation studies and an application to air quality data on CO concentrations further demonstrated the practical utility of the proposed framework, showing its ability to handle censored and missing observations while providing accurate predictions and an interpretable characterization of the underlying spatiotemporal dependence.

The remainder of this paper is organized as follows. Section \ref{modeldef} introduces the model and defines the proposed spatiotemporal random effect based on the SAR and DAGAR specifications. Section \ref{section5} presents the Bayesian inference framework under the specified priors, treating censored and missing values as latent random variables. Section \ref{section6} reports a simulation study that evaluates the estimation and prediction performance. Section \ref{section7} illustrates the proposed methodology using a multi-station air quality dataset. Finally, Section \ref{section8} closes the paper with some remarks.

\section{The spatiotemporal censored tensor model}\label{modeldef}

Let $\mathcal{Y}\in\mathbb{R}^{n\times T}$ be a second-order response tensor,
where $n$ denotes the number of spatial units and $T$ denotes the number of
time points. We consider the following spatiotemporal tensor model:
\begin{equation}\label{tensor_model}
\mathcal{Y}
=
\left\langle \mathcal{B},\mathcal{X}\right\rangle_3
+
\mathcal{W}
+
\mathcal{E},
\end{equation}

where $\mathcal{B}$ is an order-3 tensor of dimension $(n,T,p+1)$ containing
the unknown regression coefficients associated with the fixed effects, and
$\mathcal{X}$ is an order-3 tensor of the same dimension collecting the
covariates over space and time. Note that the term
$\langle \mathcal{B},\mathcal{X}\rangle_3$ denotes the contraction of
$\mathcal{B}$ and $\mathcal{X}$ along their third mode, which defines the mean
structure of the model. Specifically, if $x_{ijk}=x_k(s_i,t_j)$ denotes the
value of the $k$-th covariate at spatial location $s_i$ and time point $t_j$,
then the $(i,j)$-th entry of the mean tensor is given by,
$$
\left(\langle \mathcal{B},\mathcal{X}\rangle_3\right)_{ij}
=
\sum_{k=0}^{p} \mathcal{B}_{ijk}x_{ijk}.
$$
This formulation allows the effect $\mathcal{B}_{ijk}$ of the $k$-th covariate to
vary over both space and time. The term $\mathcal{W}\in\mathbb{R}^{n\times T}$ represents the structured spatiotemporal random effect
and is assumed to follow a tensor normal distribution, i.e.,
$
\mathcal{W}\sim TN_{n\times T}(\mathscr{O},\sigma^2\Phi,\Gamma)
$, where $\mathscr{O}\in\mathbb{R}^{n \times T}$ is the zero-location tensor of order $n\times T$, $\sigma^2$ controls the overall magnitude of the structured spatiotemporal variability, $\Phi$ is the $T\times T$ correlation matrix describing
temporal association and $\Gamma$ is the $n\times n$ covariance matrix describing spatial
association. In addition, $\mathcal{E}$ denotes an independent
unstructured error term, also assumed to follow a tensor normal distribution, i.e.,
$
\mathcal{E}\sim TN_{n\times T}(\mathscr{O},\mathbf{I}_T,\tau^2 \mathbf{I}_n),
$
where $\tau^2$ controls the amount of independent noise, and $\mathbf{I}_q$ denotes a $q\times q$ identity matrix. 

For notational convenience, let $N=nT$ and define
$
\operatorname{vec}(\mathcal{Y})
=
\mathbf{y}
=
(y_{11},\ldots,y_{1T},\ldots,y_{n1},\ldots,y_{nT})^\top
$. We adopt the same vectorization convention for
$\mathcal{W}$ and $\mathcal{E}$, and write $\mathbf{\mu}=\operatorname{vec}\left(\langle \mathcal{B},\mathcal{X}\rangle_3\right)$, 
$\boldsymbol{\omega}=\operatorname{vec}(\mathcal{W})$, and
$
\boldsymbol{\epsilon}=\operatorname{vec}(\mathcal{E})
$. It follows from the model assumptions that
$
\boldsymbol{\omega}
\sim
N_N(\mathbf{0},\Gamma\otimes\sigma^2\Phi)
$, $
\boldsymbol{\epsilon}
\sim
N_N(\mathbf{0},\tau^2 \mathbf{I}_N),
$
and therefore
$
\mathbf{y}
\sim
N_N
\left(
\boldsymbol{\mu},
\Gamma\otimes\sigma^2\Phi+\tau^2 \mathbf{I}_N
\right),
$ where $\otimes$ denotes the Kronecker product. 

In some situations, and depending on the measurement instrument, observations may be subject to lower and/or upper detection limits, beyond which they cannot be accurately quantified. In addition, missing data may occur for several reasons. For example, in the case of CO measurements, \citet{KIM2024124165} reports that missingness can arise from instrument calibration and verification procedures, equipment malfunctions, repairs or replacements, and outlier removal during data preprocessing.

In this context, the observed information associated with the response $y_{ij}$ can be represented as $(Z_{ij}, C_{ij})$. Here, $Z_{ij}$ denotes either the observed value $Z_{ij0}=y_{ij}$, when the response is fully observed, or an interval $[Z_{ij1}, Z_{ij2}]$ containing the latent true value when the response is censored or missing. The variable $C_{ij}$ is the censoring indicator, defined as
\begin{equation}\label{censoring}
 C_{ij}=\left\{\begin{array}{lcl}
1 & \text { if } & Z_{ij1} \leq y_{ij} \leq Z_{ij2}, \\
0 & \text { if } & y_{ij}=Z_{ij0}.
\end{array}\right.    
\end{equation}
As particular cases of \eqref{censoring}, we have left censoring when $C_{ij} = 1$ and $Z_{ij} = [-\infty, Z_{ij2}]$, right censoring when $C_{ij} = 1$ and $Z_{ij} = [Z_{ij1}, \infty]$, and missingness when $C_{ij} = 1$ and $Z_{ij} = [-\infty, \infty]$. The model defined by \eqref{tensor_model} and \eqref{censoring} will be referred to as the normal spatio-temporal censored linear model over graphs (NST-CLG).

\subsection{Spatial correlation matrix $\Gamma$}\label{gammaspasec}

Because our model is based on areal data, it is necessary to construct a correlation structure from a proximity matrix. In this context, the proximity matrix is defined through a graph representation, where the neighboring relationships between regions are encoded as edges connecting the corresponding vertices. In this setting, let ${\cal G} = \{{\cal S},{\cal E}\}$ denote a graph with vertex set (regions) ${\cal S} = \{ s_1, \ldots, s_n \}$ and edge set ${\cal E} = \{ (s_i,s_j): s_i, s_j \in {\cal S} \}$, with cardinality $|{\cal E}| = n-1$. We write $s_i \sim s_k$ to indicate an edge connecting $s_i$ and $s_k$. Therefore, we consider 
\begin{equation}\label{covdagar}
  \Gamma = 
\left[(\mathbf{I}_n-\mathbf{B}_{n})^\top \mathbf{F}_{n}(\mathbf{I}_n-\mathbf{B}_{n})\right]^{-1},  
\end{equation}
under two cases. 

\begin{enumerate}
\item The first structure is based on the SAR model. In this setup, $\mathbf{F}_n = \kappa\mathbf{I}_n$, with $\kappa >0$ is an unknown parameter which is fixed equal to $1$ to avoid idenfiability issues, and $\mathbf{B}_{n} = \rho \mathbf{A}_\mathcal{S}$, with $\rho$ a spatial autoregression parameter, and $\mathbf{A}_\mathcal{S}$ a $n\times n$ proximity matrix, with elements $a^{s}_{ik}=1$ if $s_i \sim s_k$ and $0$ elsewhere, for $i,k=1,\ldots,n$. As noted in \citet{banerjeebook2025}, it is necessary to ensure the non-singularity of $(\mathbf{I} - \mathbf{B}_{n})$ in order for $\Gamma$ to define a valid (positive definite) covariance matrix. This can be achieved by setting $\rho \in \left(\nicefrac{1}{\lambda_{(1)}}, \nicefrac{1}{\lambda_{(n)}}\right)$, where  $\lambda_{(1)} < \lambda_{(2)} < \ldots \lambda_{(n)}$ are the ordered eigenvalues of $\mathbf{A}_\mathcal{S}$. A convenient alternative is to adopt the symmetric normalization 
$\tilde{\mathbf{A}}_{\mathcal{S}} = \mathbf{D}^{-1/2} 
\mathbf{A}_{\mathcal{S}} \mathbf{D}^{-1/2} = (\tilde{a}^{s}_{ik})$, where 
$\mathbf{D} = \mathrm{diag}(a^{s}_{1+}, \ldots, a^{s}_{n+})$ and 
$a^{s}_{i+} = \sum_{k=1}^{n} a^{s}_{ik}$ denotes the number of 
neighbors of region $s_i$. This specification preserves symmetry 
and ensures that the spatial operator has a real-valued spectrum, 
a feature that simplifies theoretical analysis and improves numerical 
stability. In the case of undirected 
graphs, this normalization implies that the admissible range for 
the spatial dependence parameter reduces to $\rho \in (-1,1)$ 
\citep{Chung1997}. Overall, this symmetric formulation provides a stable 
and conceptually clean basis for defining valid covariance or precision 
structures.

\item The second structure is based on the DAGAR model. Consider $N_\mathcal{S}(s_i) = \lbrace s_k: s_k \sim s_i, \textrm{with} \,\, k< i \rbrace$ as the set of neighbors of $s_i$. Define, 

\begin{eqnarray*}
b_{i k}&=&\left\{\begin{array}{cl}
0 & \text { for } k \text{ such that } s_k \notin {N}_\mathcal{S} \left(s_i\right) ; \\
 & \\
\frac{\rho}{1+\left(n_{s_i}-1\right) \rho^2} & \text { for } i=2, \ldots, n \text{ and } k \text{ such that } s_k\in {N}_{\mathcal{S}}\left(s_i\right)
\end{array}\right.\\
f_{i i}&=&\frac{1+\left(n_{s_i}-1\right) \rho^2}{1-\rho^2} ,
\end{eqnarray*}
with a spatial correlation parameter $\rho >0$, and $n_{s_i}$, the number of elements in ${N}_{\mathcal{S}}(s_i)$. Here, the elements of the matrix $\mathbf{B}_{n}$ are denoted by $b_{ik}$, and the matrix $\mathbf{F}_{n}$ is diagonal with entries $f_{ii}$.

\end{enumerate}

Note that we include only areal models whose covariance matrices can be expressed as in \eqref{covdagar}. The distributions of these models are induced through a noise term, analogous to the formulation of an autoregressive process of order $p$, AR($p$). In contrast, CAR models specify the distribution directly on the variable of interest, leading to a covariance structure that differs from that of the SAR and DAGAR models. As our analysis focuses exclusively on models of the form \eqref{covdagar}, the CAR specification is not considered further in this study.

\subsection{Temporal correlation matrix $\Phi$}

In this setup, we assume that $T_i = T$ for all $i = 1, \ldots, n$. For the temporal component $\Phi$, we adopt a correlation structure induced by an autoregressive process of order $p$, denoted $\text{AR}(p)$. In this case, $\Phi$ takes the Toeplitz form
\begin{equation}
\label{arpcorr} 
\Phi = \begin{bmatrix} 
\phi_0 & \phi_1 & \cdots & \phi_{T-1} \\ 
\phi_1 & \phi_0 & \cdots & \phi_{T-2} \\ 
\vdots & \vdots & \ddots & \vdots \\ 
\phi_{T-1} & \phi_{T-2} & \cdots & \phi_0 
\end{bmatrix},
\end{equation}
where $\phi_k$ denotes the correlation at lag $k$. These correlations are determined by the Yule-Walker equations \citep{bandorff1973},
$\phi_k = \gamma_1\phi_{k-1} + \ldots + \gamma_p\phi_{k-p}$,
with $\gamma_1,\ldots,\gamma_p$ representing the autoregressive parameters. For this process, the stationarity conditions depend on the roots of its characteristic polynomial $A(z) = 1 - \sum_{i=1}^{p}\gamma_i z^i$. The process is stationary if and only if all the roots of $A(z) = 0$ lie outside the unit circle in the complex plane. This condition guarantees that the process has a finite variance and an autocovariance function that depends only on the temporal lag rather than on the specific time index.

\subsection{Innovation-form Gaussian Markov random field representation of $\mathcal{W}$ }

First, we introduce some notation. For a fixed spatial location $s_j$, let
$\boldsymbol{\omega}_{s_j}$ denote the vector collecting the temporal components of
$\boldsymbol{\omega}$ associated with that location, that is,
$\boldsymbol{\omega}_{s_j}=(\omega(s_j,t_1),\ldots,\omega(s_j,t_T))^\top$.
Similarly, for a fixed time point $t_j$, with $t_1<t_2<\cdots<t_T$, define
$\boldsymbol{\omega}_{t_j}$ as the vector of spatial components observed at that
time,
$\boldsymbol{\omega}_{t_j}=(\omega(s_1,t_j),\ldots,\omega(s_n,t_j))^\top$. This notation allows the spatial process at each time point to be written in
Gaussian Markov random field in innovation (GMRFI) form as
\begin{equation}\label{GMRFinnov}
    \boldsymbol{\omega}_{t_j}
    =
    \mathbf{B}_n\boldsymbol{\omega}_{t_j}
    +
    \boldsymbol{\epsilon}_{t_j},
\end{equation}
where
$\boldsymbol{\epsilon}_{t_j}
=
(\epsilon(s_1,t_j),\ldots,\epsilon(s_n,t_j))^\top
\sim
N_n(\boldsymbol{0},\mathbf{F}_n)$. Note that, analogously, we use the same vectorization convention for the error tensor
$\mathcal{E}$, so that its spatial and temporal components can be defined in the
same way as those of $\mathcal{W}$.

The following results demonstrate that the Kronecker structure allows us to represent $\boldsymbol{\omega}$ as a GMRFI form, which, through appropriate adjacency relations, captures both temporal and spatial dependencies. We begin by showing that when $\boldsymbol{\omega}_{s_j}$ follows an AR($p$) structure, it admits such a representation. In particular, it can be expressed as a DAGAR process, where each coefficient $b_{jl}$ is defined in terms of the coeficients obtained through the Durbin Levinson recursion.

\begin{proposition}\label{prop1}
Assuming that $\boldsymbol{\omega}_{s_j}$ follows and AR($p$) structure, let $\kappa_1, \ldots, \kappa_T \in (-1,1)$ the partial autocorrelations associated to the process where $\kappa_\ell = 0$ for $\ell >p$. Let $\gamma^{(m)}_{k}$ and $\nu_m$ denote the coefficients and factor variances defined by the Durbin - Levinson recursion, 
\begin{align*}
\gamma^{(1)}_1 &= \kappa_1, \\
\gamma^{(m)}_m &= \kappa_m, \qquad m=2, 3, \dots, T \\
\gamma^{(m)}_k &= \gamma^{(m-1)}_k - \kappa_m\,\gamma^{(m-1)}_{m-k},
\qquad k=1,\dots,m-1.\\
\nu_1 &= 1,\,\, \nu_{m +1} = \nu_m(1- \kappa^2_m).
\end{align*}

 Then, $\boldsymbol{\omega}_{s_j}$ follows a $T$-variate normal distribution with zero mean and covariance matrix given by
\begin{eqnarray*}
[(\mathbf{I}_T-\mathbf{B}_T)^\top \mathbf{F}_T(\mathbf{I}_T-\mathbf{B}_T)]^{-1},
\end{eqnarray*}
where $\mathbf{B}_T$ has elements $b_{ij}$, with $i,j\in \{1,\ldots,T\}$, given by
\begin{eqnarray*}
b_{jl}=\left\{\begin{array}{ccl}
0 & \text { for } & t_j \notin {{N}}_\mathcal{T} \left(    t_i\right) ; \\
\gamma^{(j-1)}_{j-l} & \text { for } & j=2, \ldots, T \text{ and } t_l \in {N}_{\mathcal{T}}\left(t_j\right),
\end{array}\right. 
\end{eqnarray*}
${N}_{\mathcal{T}}(t_j) = \lbrace t_{j-h}: h = 1,2, \ldots,\text{min}(p,j-1)\rbrace$ and $\mathbf{F}_T = \sigma^{-2} \mathbf{D}$, with $\mathbf{D} = diag(\nu_1,\ldots,\nu_T)$.
\end{proposition}





\noindent{\bf Proof}: The proof of this proposition is provided in the Supplementary Material. 

Note that, for a fixed region $s_i$, the Durbin--Levinson recursion provides a coherent initialization of the first values
$\omega(s_i,t_1),\ldots,\omega(s_i,t_p)$ such that the process $\boldsymbol{\omega}_{s_i}$ satisfies the weak stationarity conditions.
Moreover, once the effective autoregressive order reaches $p$, the recursion stabilizes and, for all subsequent times,
the coefficients $\gamma_k^{(m)}$, $k=1,\ldots,p$, coincide exactly with the coefficients of the autoregressive polynomial,
that is,
\begin{eqnarray*}
\gamma_k^{(m)} = \gamma_k, \qquad k=1,\ldots,p,\ \ m \ge p.
\end{eqnarray*}

It is also important to stress that the random effect $\boldsymbol{\omega}_{s_j}$ admits a DAG representation, where the adjacency matrix is given by
$\mathbf{A}_T = \sum_{k=1}^p \mathbf{J}^k$,
with $\mathbf{J}^k$ the $k$-th backward shift matrix, whose elements are $(\mathbf{J}^k)_{jl} = 1$ if $l = j - k$, and $0$ otherwise. This matrix encodes a neighborhood structure by connecting each node $j$ with its $p$ predecessors $(j - k)$, allowing $\boldsymbol{\omega}_{s_j}$ to be defined as an AR($p$) process. A graphical representation of this relationship is shown in Figure~\ref{ardag} for the cases $p = 1, 2$.

Next, we show that the random effect $\boldsymbol{\omega}$, under the specifications given above, can be expressed as a GMRFI form, thus extending the SAR and DAGAR models to the spatiotemporal framework.

 \begin{theorem}\label{thm1}
The process $\boldsymbol{\omega} \sim N_{N} (\mathbf{0},  \boldsymbol{\Gamma}\otimes\boldsymbol{\Phi})$ can be written as a GMRFI process. Specifically,
\begin{enumerate}
\item Let $\mathbf{A} = \mathbf{I}_n \otimes \mathbf{A}_T + \mathbf{A}_S \otimes \mathbf{I}_T + \mathbf{A}_S \otimes \mathbf{A}_T$ denote the adjacency matrix, with elements $a_{(ik,jl)} = \delta^{S}_{ik} a^{T}_{jl} + a^{S}_{ik} \delta^{T}_{jl} + a^{S}_{ik} a^{T}_{jl}$, where $i,k \in \{1,\ldots,n\}$, $j,l \in \{1,\ldots,T\}$, and $\delta^{S}_{ik}$ and $\delta^{T}_{jl}$ are Kronecker delta functions representing the components of $\mathbf{I}_n$ and $\mathbf{I}_T$, respectively. We define the following relationship:
$(s_i,t_j) \sim (s_k,t_l)$ if $a_{(ik,jl)} = 1$.
Note that the relation $\sim$ defines a pair of neighbors in the spatiotemporal context. If $\Gamma$ is chosen as the covariance matrix of a SAR process, then $\boldsymbol{\omega}$ is a GMRFI with,
\begin{equation}\label{SARST}
    \omega(s_i,t_j) = \sum_{\{(s_k,t_l): (s_i,t_j) \sim(s_k,t_l)\}} b_{(ik,jl)} \omega{(s_k,t_l)} + \epsilon(s_i,t_j)
\end{equation}
where $\epsilon(s_i,t_j) \sim N(0,\sigma^2\nu_j)$ and
\begin{equation*}
 b_{(ik,jl)}=\left\{\begin{array}{ccl}
0 & \text { for } & (s_i,t_j) \nsim (s_k,t_l) ; \\
\gamma^{(j-1)}_{j-l}  & \text { for } & s_i = s_k\,\,\text{and}\,\,t_l \in { N}_{\mathcal{T}}\left(t_j\right),\\
\tilde{a}^{s}_{ik}\rho & \text { for } & t_j = t_l\,\,\text{and}\,\, s_i \sim s_k, \\
-\tilde{a}^{s}_{ik}\rho \gamma^{(j-1)}_{j-l} & \text { for }  & s_i \sim s_k,\,t_l \in { N}_{\mathcal{T}}\left(t_j\right),\, i \neq k,\, j \neq l
\end{array} \right.
\end{equation*}

with $\tilde{a}^{s}_{ik}$ being the element $(i,k)$ of the matrix $\tilde{\mathbf{A}}_S$ as defined in Section \ref{gammaspasec}. 

\item If $\Gamma$ is chosen as the covariance structure of a DAGAR process, then $\boldsymbol{\omega}$ is also a DAGAR process, i.e.,
\begin{eqnarray}
 \omega(s_1,t_1) &=& \epsilon(s_1,t_1)\label{DAGARST.1}\\
 \omega(s_i,t_j) &=& \sum_{\mathcal{C}} b_{(ik, jl)}\omega(s_k,t_l)+ \epsilon(s_i,t_j)\label{DAGARST.2} 
\end{eqnarray} 
where $\mathcal{C}=\{(s_k,t_l): (s_k,t_l) \,\in\,{N}_{(\mathcal{S},\mathcal{T})}\}$, 
$\epsilon(s_i,t_j) \sim N(0, f_{ii}^{-1} \sigma^2\nu_j)$, ${N}_{(\mathcal{S},\mathcal{T})} (s_i,t_j) = {N}_{\mathcal{S}}(s_i) \times {N}_{{\cal T}}(t_j)$ and
\begin{equation*}
b_{(ik, jl)}=\left\{\begin{array}{ccl}
0 & \text { for } & (s_k, t_l) \notin {N}_{\mathcal{S},\mathcal{T}} (s_i, t_j) ; \\
\gamma^{(j-1)}_{j-l}  & \text { for } & s_i = s_k\,\,\text{and}\,\,t_l \in { N}_{\mathcal{T}}\left(t_j\right),\\
\frac{\rho}{1+\left(n_{<i}-1\right) \rho^2} & \text { for } & t_i = t_j\,\,\text{and}\,\, s_k \in {N}_{\mathcal{S}}\left(s_i\right), \\
-\frac{\rho}{1+\left(n_{<i}-1\right) \rho^2} \gamma^{(j-1)}_{j-l} & \text { for }  & s_k \in {N}_{\mathcal{S}}\left(s_i\right),\,t_l \in {N}_{\mathcal{T}}\left(t_j\right),\, k \neq i,\, l \neq j 
\end{array} \right.
\end{equation*}
  \\
\end{enumerate}
\end{theorem}

\noindent{{\bf Proof}}: The proof is provided in the Supplementary Material.

\begin{figure}
\centering
\subfigure[AR(1): $\omega(s_i,t_j)$ depends on $\omega(s_i,t_{j-1})$.]{
\begin{minipage}{0.9\textwidth}  
\centering
\begin{tikzpicture}[->,>=latex, node distance=2cm, thick, scale=1, transform shape]
  \node[circle,draw](xm2){$j-2$};
  \node[circle,draw,right of=xm2](xm1){$j-1$};
  \node[circle,draw,right of=xm1](x){$\,\,\,\,j\,\,\,\,$};
  \node[circle,draw,right of=x](xp1){$j+1$};
  \node[circle,draw,right of=xp1](xp2){$j+2$};

  \path (xm2) edge (xm1);
  \path (xm1) edge (x);
  \path (x) edge (xp1);
  \path (xp1) edge (xp2);
\end{tikzpicture}
\end{minipage}
}

\vspace{1cm}  

\subfigure[AR(2): $\omega(s_i,t_j)$ depends on $\omega(s_i,t_{j-1}),\,\omega(s_i,t_{j-2})$.]{
\begin{minipage}{0.9\textwidth}
\centering
\begin{tikzpicture}[->,>=latex, node distance=2cm, thick, scale=1, transform shape]
  \node[circle,draw](xm2){$t-2$};
  \node[circle,draw,right of=xm2](xm1){$t-1$};
  \node[circle,draw,right of=xm1](x){$\,\,\,\,t\,\,\,\,$};
  \node[circle,draw,right of=x](xp1){$t+1$};
  \node[circle,draw,right of=xp1](xp2){$t+2$};
  \path (xm2) edge (xm1);
  \path (xm1) edge (x);
  \path (xm2) edge[bend left=20] (x);
  \path (x) edge (xp1);
  \path (xm1) edge[bend left=20] (xp1);
  \path (xp1) edge (xp2);
  \path (x) edge[bend left=20] (xp2);
\end{tikzpicture}
\end{minipage}
}
\caption{Directed acyclic graph representations of AR(1) and AR(2) processes.}\label{ardag}
\end{figure}

In the first statement of Theorem \ref{thm1}, the choice of $\tilde{a}^{s}_{ik}$ is made purely for convenience when verifying the nonsingularity of the matrix $(\mathbf{I}_n - \mathbf{B}_n)$. The statements continue to hold if $a^{s}_{ik}$, the $(i,k)$-th element of the matrix $\mathbf{A}_S$, is used instead of $\tilde{a}^{s}_{ik}$. Using $\Gamma$ as the SAR covariance structure, \eqref{SARST} allows us to rewrite the process for $t>p$ as
\begin{eqnarray}\label{sarinterpret}
     \omega(s_i, t_j) &=&  \underbrace{\sum_{q = 1}^{p}  
     \gamma_{q}\omega(s_i, t_{j-q})}_{\mathbb{A}} + \underbrace{\sum_{ k\neq i} \tilde{a}^{s}_{ik}\rho \omega(s_k, t_j)}_{\mathbb{B}} \nonumber \\
     &- & \underbrace{\sum_{k \neq i}\sum_{q = 1}^{p}\tilde{a}^{s}_{ik}\gamma_{q}\rho\omega(s_k, t_{j-q})}_{\mathbb{C}} 
     \,\, +\,\, \epsilon(s_i, t_j).
\end{eqnarray}

On the other hand, using $\Gamma$ as the DAGAR covariance structure, equations \eqref{DAGARST.1}-\eqref{DAGARST.2} allow us to rewrite the process for $t>p$ as
\begin{eqnarray}\label{dagarinterpret}
     \omega(s_i, t_j) &=&  \underbrace{\sum_{q = 1}^{p}  
     \gamma_{q}\omega(s_i, t_{j-q})}_{\mathbb{A}} + \underbrace{\sum_{ s_k\,\in\,{N}_{\mathcal{S}} (s_i)} \frac{\rho}{1+\left(n_{s_i}-1\right) \rho^2}\omega(s_k, t_j)}_{\mathbb{B}} \nonumber \\
     &-&  \underbrace{\sum_{s_k \in {N}_{s}(s_i)}\sum_{q = 1}^{p}\frac{\gamma_{q}\rho}{1+\left(n_{s_i}-1\right) \rho^2}\omega(s_k, t_{j-q})}_{\mathbb{C}} 
     \,\, +\,\, \epsilon(s_i, t_j)
\end{eqnarray}

Regarding equation \eqref{sarinterpret}, $\gamma_{q}$ in component $\mathbb{A}$ denotes the autoregressive parameter for the fixed location $s_i$. Moreover, in component $\mathbb{B}$, the parameter $\rho$, which is associated with spatial correlation, poses interpretational challenges similar to those documented for SAR models at a fixed time point \citep[see][]{banerjeebook2025,wall2004}. Note that component $\mathbb{C}$ incorporates both spatial and temporal parameters.
However, the process form in \eqref{dagarinterpret} provides a more direct interpretation of the spatial and temporal parameters and their interactions, with the advantage that both parameters can be interpreted analogously to those in AR($p$) and DAGAR models. The component $\mathbb{A}$ captures temporal dependence. Marginally, the coefficients $\gamma_q$ can be interpreted as the conditional linear effect of $\omega(s_i, t_{j-q})$ on the current value $\omega(s_i, t_j)$, holding the remaining lags constant. The component $\mathbb{B}$ captures spatial dependence, where $\rho$ can be interpreted marginally as the correlation between $\omega(s_i, t_j)$ and its spatial neighbors. Finally, the term $\mathbb{C}$ accounts for the spatiotemporal cross-dependence, and the product $\gamma_q \rho$ can be interpreted as a measure of dependence between $\omega(s_i, t_j)$ and the past values of its spatial neighbors. 

It is important to emphasize that the innovation-based formulation yields a computationally efficient representation, rendering inference feasible for datasets of moderate size and enabling the implementation of this class of models in standard Bayesian software such as \texttt{Stan}.

It is worth noting that, as in the purely spatial DAGAR model, the assumptions regarding ordering rely solely on the spatial dimension. Following \citet{datta2019}, let $\pi = {\pi(1), \ldots, \pi(n)}$ denote any predetermined ordering of the spatial locations, and let $\pi^{-1}$ be its corresponding inverse permutation. Under this ordering, for $i \neq \pi(1)$, define the set of neighboring observations $\boldsymbol{\omega}_{{{N}}^{\pi}{\mathcal{S}}(s_i)}$ as the collection $\{\boldsymbol{\omega}(s_k, t_l) : \pi^{-1}(k) < \pi^{-1}(i), t_l \in {{N}}_{\mathcal{T}}(t_j)\}$, where ${{N}}^{\pi}_{\mathcal{S}}(s_i) = \{s_k : s_i \sim s_k, \pi^{-1}(k) < \pi^{-1}(i)\}$. Denote by ${\mathcal{G}}_{\pi}({\mathcal{S}}, {\mathcal{E}}_{\pi})$ the acyclic graph generated by these configurations, with ${\mathcal{E}}_{\pi}$ representing the collection of directed edges from all members of ${{N}}^{\pi}_{\mathcal{S}}(s_i)$ to $i$, for every $i \neq \pi(1)$. Then, 
\begin{eqnarray}\label{condrep}
   \omega(s_i, t_j) \mid \boldsymbol{\omega}_{{N}^{\pi}_{(\mathcal{S},\mathcal{T})}} \sim N\left( \sum_{ (s_k,t_l)\,\in\,{N}^{\pi}_{(\mathcal{S},\mathcal{T})} (s_i,t_j)} b_{(ik,jl)}\omega(s_k, t_l),\,\, f_{ii}^{-1}\sigma^2\nu_j \right),
\end{eqnarray}
with ${N}^{\pi}_{(\mathcal{S},\mathcal{T})} (s_i,t_j) = {N}^{\pi}_{\mathcal{S}}(s_i) \times {N}_{{\cal T}}(t_j)$ and $ b_{(ik,jl)}$ as defined in Theorem \ref{thm1}, but replacing $n_{s_i}$ by $n_{\pi(i)}$, the number of elements in $N^{\pi}_{\mathcal{S}}(s_i)$. Note that \citet{datta2019} reported that the ordering of regions has a negligible influence on the results. Specifically, in their simulation studies, the mean squared errors, as well as the estimates and credible intervals for $\rho$, remained nearly identical across different orders.

\section{Bayesian inference}\label{section5}

In this section, we introduce the key components required to carry out Bayesian inference for the model specified in \eqref{tensor_model}–\eqref{censoring}. We begin by detailing the prior specifications as well as the posterior analysis for the parameters of interest. Suubsequently, we describe the Bayesian prediction procedure.

 \subsection{Specification of $\pi(\boldsymbol{\theta})$ and posterior analysis}

Let $\boldsymbol{\theta} = (\boldsymbol{\beta}, \sigma^2, \rho, \gamma, \tau^2)$, where $\boldsymbol{\beta}$ denotes the linear predictor coefficients in $\mathcal{B}$, $\sigma^2$ controls the marginal variance of the structured spatio-temporal process, $\rho$ and $\gamma$ govern the spatial and temporal dependence, respectively, and $\tau^2$ represents the variance of an independent noise component accounting for unstructured variability and measurement error.
The first step is to define $\pi(\boldsymbol{\theta},\boldsymbol{\omega})$, the joint prior of the parameters and the latent process. We assume an independent joint density of the form
\begin{equation}
\pi(\boldsymbol{\theta}, \boldsymbol{\omega}) \propto \pi(\boldsymbol{\beta})\pi(\sigma^2)\pi(\rho)\pi(\boldsymbol{\gamma})\pi(\tau^2)\pi(\boldsymbol{\omega}).
\end{equation}
Under this specification, standard prior distributions are assigned to each component of $\boldsymbol{\theta}$. For the regression coefficients, we assume 
$\boldsymbol{\beta} \sim N(\boldsymbol{\mu}_{\boldsymbol{\beta}}, \boldsymbol{\Sigma}_{\boldsymbol{\beta}})$, 
a choice commonly adopted in the literature, including \citet{datta2019} and \citet{ordonez2024}, the latter considering non-Gaussian response distributions. 
For the variance components $\sigma^2$ and $\tau^2$, widely used prior specifications include the inverse-gamma distribution as well as the improper prior 
$\pi(\sigma^2)\propto \sigma^{-2}$. 
Regarding the spatial and temporal dependence parameters, we consider priors from the beta family, namely 
$\rho \sim \text{Beta}(a_\rho,b_\rho)$ and $\gamma \sim \text{Beta}(a_\gamma,b_\gamma)$. 
The beta distribution is a convenient choice due to its flexibility and computational tractability. 
Its bounded support ensures coherence with the parameter space, while the shape parameters $(a,b)$ allow a wide range of prior specifications, ranging from diffuse to highly concentrated distributions.

Concerning the latent vector $\boldsymbol{\omega}$, we consider the priors induced by the spatio-temporal extensions of the DAGAR and SAR processes, described in Theorem \ref{thm1}. For the DAGAR, the conditional representations in (\ref{dagarinterpret})-(\ref{condrep}) lead to a scalable formulation, allowing efficient Bayesian inference for responses of moderate to large dimension.

In contrast, since the SAR structure does not induce a directed acyclic graph over the spatial domain, we adopt a block-based innovation representation. Specifically, the temporal dynamics of the latent process are given by,
\begin{eqnarray*}
 \boldsymbol{\omega}_{t_1} & =  & \boldsymbol{e}_{t_1} \\
 \boldsymbol{\omega}_{t_2} &=& \gamma^{(1)}_1 \boldsymbol{\omega}_{t_{1}}  + \boldsymbol{e}_{t_2}\\
\boldsymbol{\omega}_{t_3} &=& \gamma^{(2)}_1 \boldsymbol{\omega}_{t_{2}}  + \gamma^{(2)}_2 \boldsymbol{\omega}_{t_{1}} +
\boldsymbol{e}_{t_3}\\ 
 &\vdots&\\
 \boldsymbol{\omega}_{t_p} &=& \gamma^{(p-1)}_1 \boldsymbol{\omega}_{t_{p-1}}  + \gamma^{(p-1)}_2 \boldsymbol{\omega}_{t_{p-2}} + \ldots +
 \gamma^{(p-1)}_p \boldsymbol{\omega}_{t_{1}} +
\boldsymbol{e}_{t_p}\\
\boldsymbol{\omega}_{t_j} &=& \gamma_1 \boldsymbol{\omega}_{t_{j-1}}  + \gamma_2 \boldsymbol{\omega}_{t_{j-2}} + \ldots +
 \gamma_p \boldsymbol{\omega}_{t_{j-p}} +
\boldsymbol{e}_{t_j},\qquad j > p,
\end{eqnarray*}
where $\boldsymbol{ e}_{t_j} \sim N(\mathbf 0,\, \sigma^2 \nu_j \boldsymbol{\Gamma})$, 
$\gamma_k^{(m)}$ denote the Durbin--Levinson prediction coefficients at lag $k$ and order $m$, 
$\nu_j$ are the associated innovation variance factors, and $\boldsymbol{\Gamma}$ represents 
the spatial covariance structure induced by the SAR process, as defined at the beginning of 
Subsection~\ref{gammaspasec}. Note that the Gaussian assumption on the innovation process 
 implies Gaussianity of the latent field $\boldsymbol{\omega}_{t_j}$. Moreover, the innovation representation yields the following conditional distributions:
\begin{eqnarray}\label{sarcond}
\boldsymbol{\omega}_{t_j}\mid \{\boldsymbol{\omega}_{t_\ell} : t_\ell \in N_T(t_j)\}
&\sim&
N\!\left(
\sum_{k=1}^{\min(p,j-1)} \gamma_k^{(j-1)}\, \boldsymbol{\omega}_{t_{j-k}},
\; \sigma^2 v_j\, \boldsymbol{\Gamma}
\right),
\qquad j = 2,\ldots,p,
\nonumber \\
\boldsymbol{\omega}_{t_j}\mid \{\boldsymbol{\omega}_{t_\ell} : t_\ell \in N_T(t_j)\}
&\sim&
N\!\left(
\sum_{k=1}^{p} \gamma_k\, \boldsymbol{\omega}_{t_{j-k}},
\; \sigma^2 v_j\, \boldsymbol{\Gamma}
\right),
\qquad j > p.
\end{eqnarray}

It is important to stress that this formulation avoids the explicit construction and factorization of the full spatio-temporal precision matrix of dimension $nT \times nT$, thereby reducing the computational complexity from $\mathcal{O}((nT)^3)$ to operations involving only spatial matrices of dimension $n \times n$ evaluated sequentially over time.

\subsubsection{Posterior analysis}

Conditionally on the latent vector $\boldsymbol\omega$ and the regression parameters, the likelihood is given by
\begin{equation}\label{likelihood}
L(\mathbf y \mid \boldsymbol\theta,\boldsymbol{\omega})
=
\prod_{i=1}^{N}
\prod_{j=1}^{T}
\left[
\phi\!\left(
\frac{Z_{ij0}-\mu_{ij}}{\tau}
\right)
\right]^{1-C_{ij}}
\left[
\Phi\!\left(
\frac{Z_{ij2}-\mu_{ij}}{\tau}
\right)
-
\Phi\!\left(
\frac{Z_{ij1}-\mu_{ij}}{\tau}
\right)
\right]^{C_{ij}},
\end{equation}
where $\boldsymbol{\mu} = (\mu_{11}, \ldots,\mu_{nT})^{\top} = \text{vec} (\langle \mathcal{B},\mathcal{X}\rangle_3) + \boldsymbol{\omega} $, $\mu_{ij} = \mu(s_i,t_j)$ and $\phi(\cdot)$ and $\Phi(\cdot)$ denoting the standard normal density and distribution functions, respectively. Using (\ref{likelihood}), we can deduce  the posterior distribution of the latent vector $\boldsymbol{\omega}$ and $\boldsymbol{\theta}$ as, 
\begin{equation}\label{jointpost}
\pi(\boldsymbol{\omega},\boldsymbol{\theta} \mid \mathbf{y}) \propto \pi(\boldsymbol\omega,\boldsymbol\theta)L(\mathbf{y} \mid \boldsymbol\theta, \boldsymbol{\omega}),
\end{equation}
with $\pi(\boldsymbol{\omega},\boldsymbol{\theta})$ as defined at the beginning of the section. Unfortunately, the joint posterior in (\ref{jointpost}) do not admit a closed-form solution, which makes it difficult to design an MCMC scheme for Bayesian inference. To address this, we rely on the No-U-Turn Sampler \citep{hoffman2014}, which is already implemented in the \texttt{Stan} software and available in \texttt{R} \cite{rcore} through the \texttt{rstan} package \cite{rstan}. 

\subsection{Bayesian prediction}\label{bay_pred}

Let $\widetilde{\mathcal{Y}}\in\mathbb{R}^{n\times h}$ denote the order-2 tensor of predicted responses, where $h$ is the forecasting horizon associated with the future time points $t_{T+1},\ldots,t_{T+h}$. Under the adopted vectorization convention, we write
$\textnormal{vec}(\widetilde{\mathcal{Y}})=\widetilde{\mathbf{y}}=(y(s_1,t_{T+1}),\ldots,y(s_n,t_{T+1}),\ldots,y(s_1,t_{T+h}),\ldots,y(s_n,t_{T+h}))^\top$,
which represents the vector of future responses to be predicted over the spatial locations $s_1,\ldots,s_n$ and time points $t_{T+1},\ldots,t_{T+h}$. We also consider the same notation for the tensor $\widetilde{\mathcal{W}}$, so that its vectorized form is defined consistently with that of $\widetilde{\mathcal{Y}}$. Then, prediction is carried out within a fully Bayesian framework and proceeds in two stages.

First, conditional on the model parameters $\boldsymbol{\theta}$ and on the latent field 
$\boldsymbol{\widetilde{\omega}}$—obtained recursively via (\ref{dagarinterpret})–(\ref{condrep}) for the DAGAR process 
and (\ref{sarcond}) for the SAR process—the predictive distribution of the future response vector $
\widetilde{\mathbf{y}}
$
is given by
\[
\pi \!\left(
\mathbf{\widetilde{y}}
\mid
\boldsymbol{\widetilde{\omega}}, \boldsymbol{\theta}
\right).
\]

Second, the posterior predictive density is obtained by averaging this conditional distribution with respect to the joint posterior distribution of $(\boldsymbol{\widetilde{\omega}}, \boldsymbol{\theta})$, namely
\[
\pi \!\left(
\widetilde{\mathbf{y}}
\mid
\mathbf{y}
\right)
=
\int
\pi \!\left(
\widetilde{\mathbf{y}}
\mid
\widetilde{\boldsymbol{\omega}}, \boldsymbol{\theta}
\right)
\,
\pi \!\left(
\boldsymbol{\widetilde{\omega}}, \boldsymbol{\theta}
\mid
\mathbf{y}
\right)
\, d\boldsymbol{\widetilde{\omega}}\, d\boldsymbol{\theta}.
\]
This integral formulation naturally propagates both parameter and latent process uncertainty into the predictive distribution. 
As in the estimation procedure, prediction is implemented via posterior simulation in \texttt{Stan}. 
For each posterior draw, the sampled parameters and latent states are propagated forward through the AR($p$) recursion within the \texttt{generated quantities}  block, where future latent states and corresponding observations are generated. 
By repeating this procedure across posterior samples, the predictive distribution fully reflects the hierarchical structure of the model and provides a coherent extension of the estimation step to the forecasting setting.

\section{Simulation studies}\label{section6}

In this section, we present a simulation study to examine both fitting properties and simulation performance. Data are generated from model~\eqref{tensor_model} using the covariate vector $\mathbf{x}=(1,x_1,x_2)^\top$, where $x_{1ij}\sim {N}(0,1^2)$ and $x_{2ij}\sim {N}(1,3^2)$ were independently generated. The coefficient tensor $\mathcal{B}$ is assumed to be invariant over space and time, so that $\mathcal{B}_{itk}=\beta_k$ for every $i=1,\ldots,n$, $t=1,\ldots,T$, and $k=0,1,2$, with $\boldsymbol{\beta}=(1,2,2.5)^\top$. Regarding the covariance parameters, we take $(\sigma^2,\rho,\boldsymbol{\gamma}, \tau^2) = (2, 0.8, 0.7,0.6)^\top$. 

To simulate the spatial structure, we considered grids of sizes $5$, $6$, and $7$, resulting in a total of $n_1 = 25$, $n_2 = 36$, and $n_3 = 49$ vertices (regions), respectively. With respect to the temporal dimension, we consider the number of temporal observations to be the same as the number of regions, that is, $T_1 = 25$, $T_2 = 36$, $T_3 = 49$. This gives us the total number of observations of $N_i = n_i \times T_i$ for $ i = 1,2, 3$. In total, $k = 300$ datasets were simulated for each study. 

After simulating the data, $\delta\%$ of the values were artificially left censored, as in \citet{Schelin2014}. We also generated missing values from the remaining $(100-\delta)\%$ of uncensored observations under a MCAR mechanism. The percentages of censored values were fixed at $15\%$ and $35\%$, whereas the percentage of missing values was fixed at $5\%$. For comparison purposes, we compared our proposed approach with common techniques for dealing with partial information. The first method, the LOD method, consists of substituting the censored values with the limit of detection, whereas the second method, the LOD/2 method, substitutes these values with $0.5$LOD. The missing values in these two methods were substituted by the sample mean of all the observations.

\subsection{Parameter estimation and model fitting}\label{sec:6.1}

\begin{figure}
\centering
\subfigure[]{\includegraphics[width=0.35\textwidth]{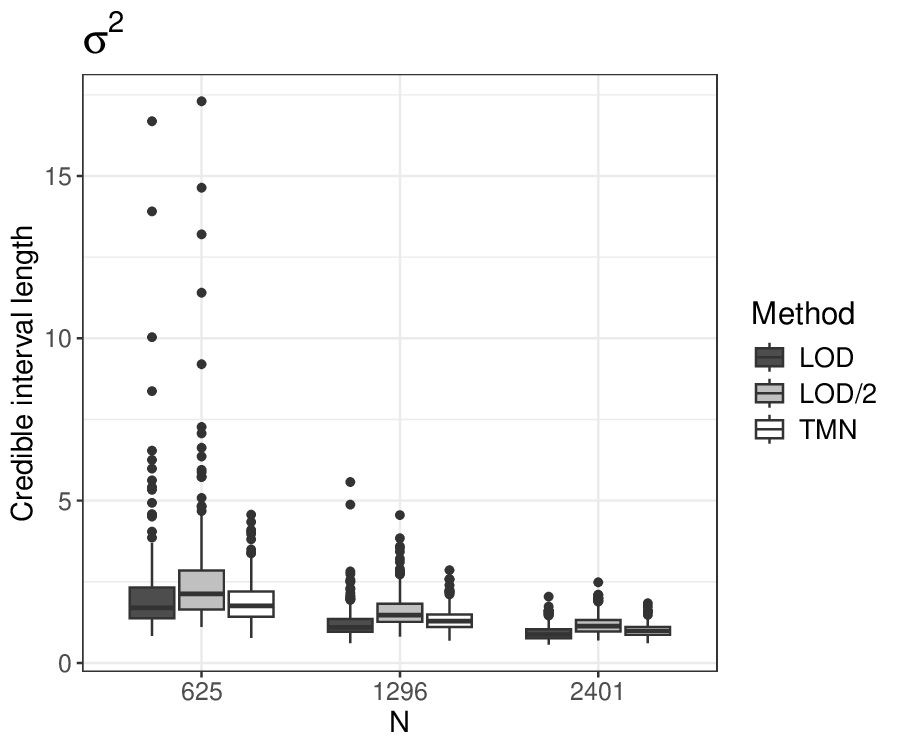}}, 
\subfigure[]{\includegraphics[width=0.35\textwidth]{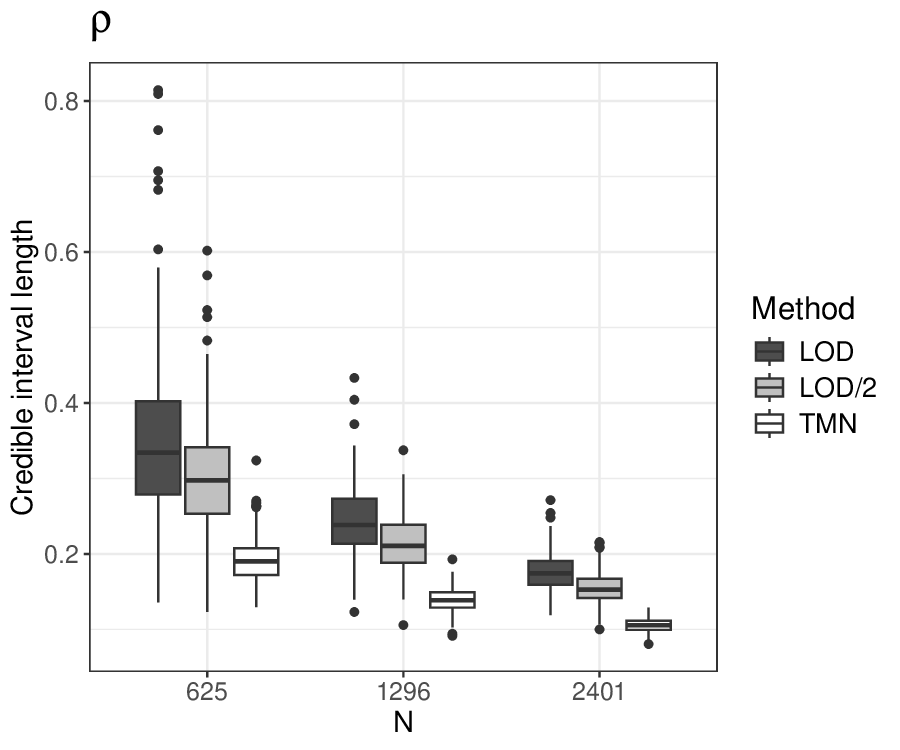}}, 
\subfigure[]{\includegraphics[width=0.35\textwidth]{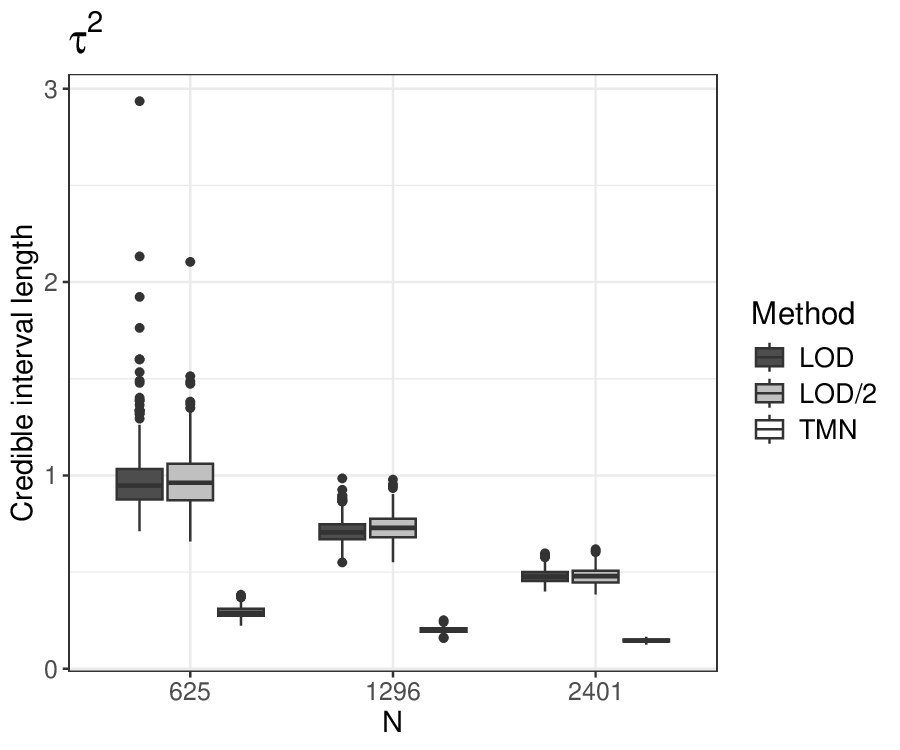}}
\subfigure[]{\includegraphics[width=0.35\textwidth]{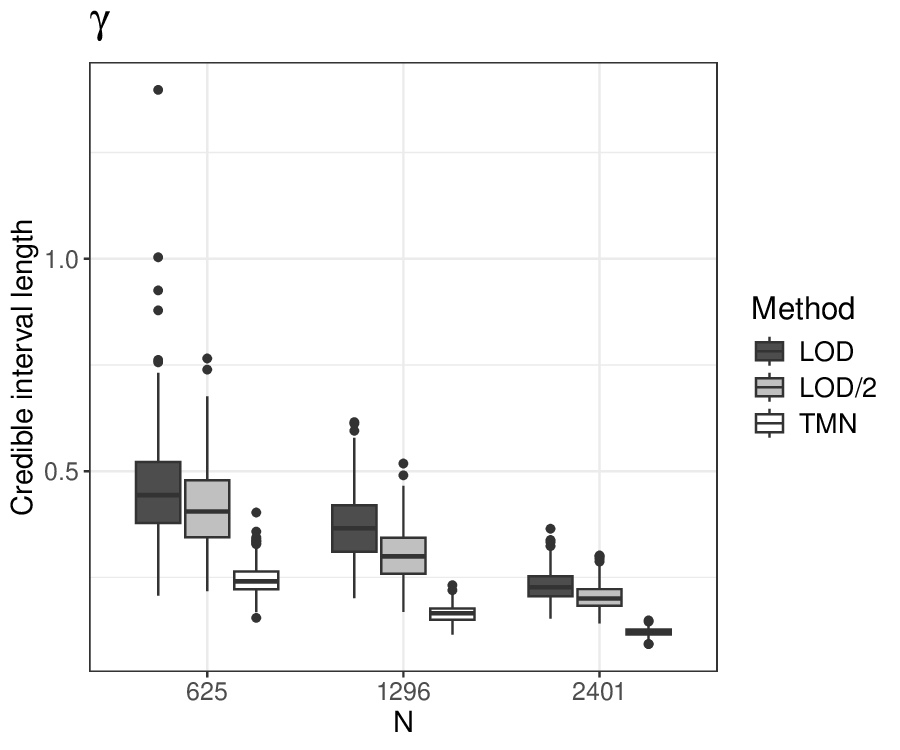}}
\caption{{\bf Simulation study I}. Credible interval lengths for the covariance structure parameters considering a censoring level of $15\%$ and a missing level of $5\%$}
\label{estcens15}
\end{figure}

\begin{figure}[!htbp]
\centering
\subfigure[]{\includegraphics[width=0.45\textwidth]{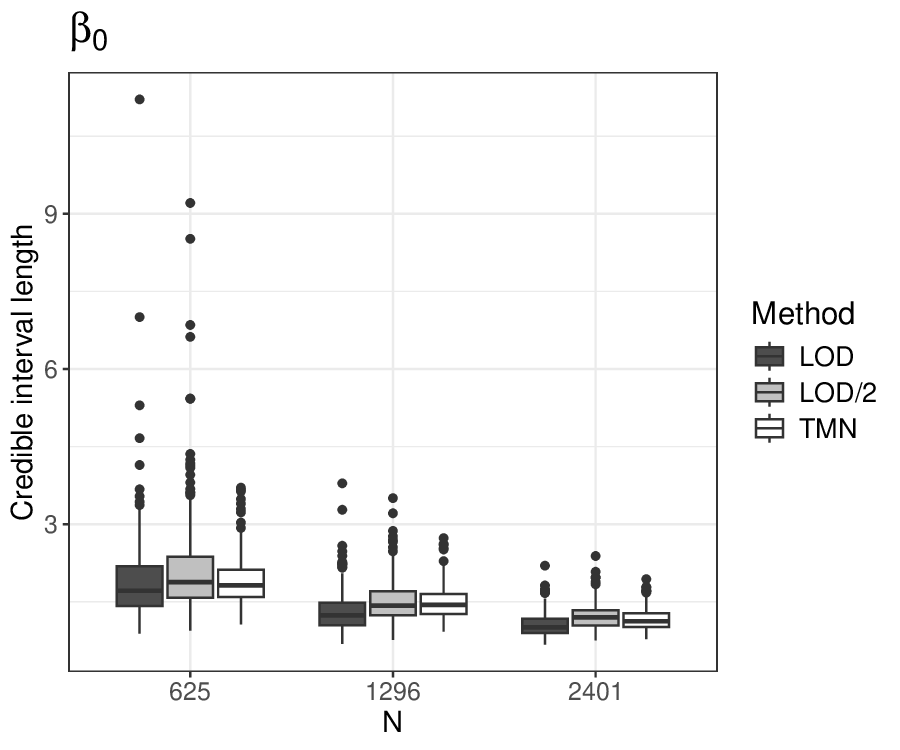}}, 
\subfigure[]{\includegraphics[width=0.45\textwidth]{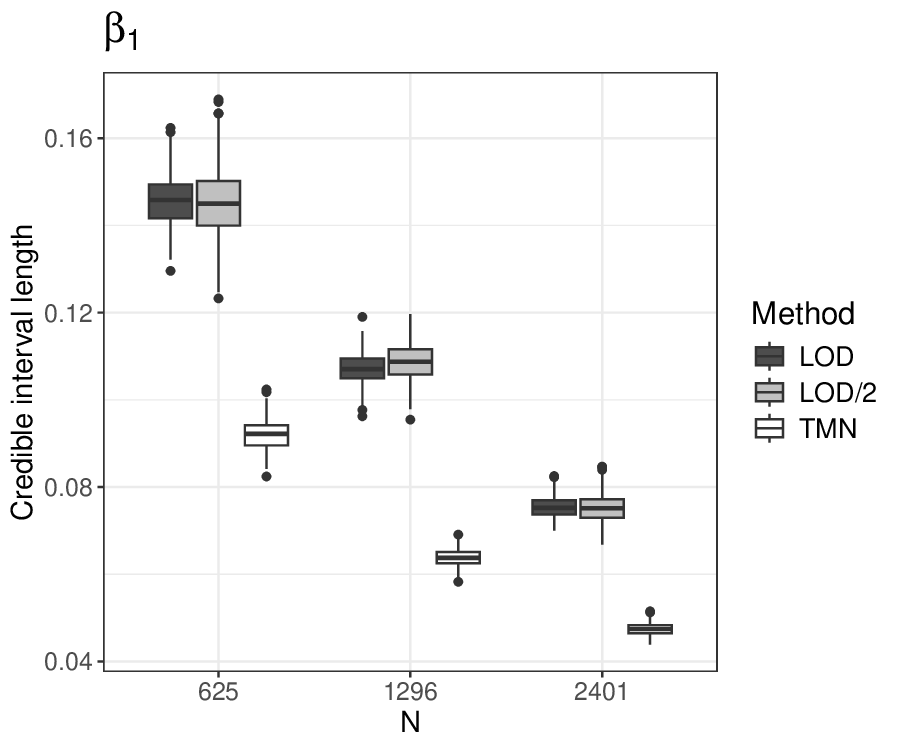}}, 
\subfigure[]{\includegraphics[width=0.45\textwidth]{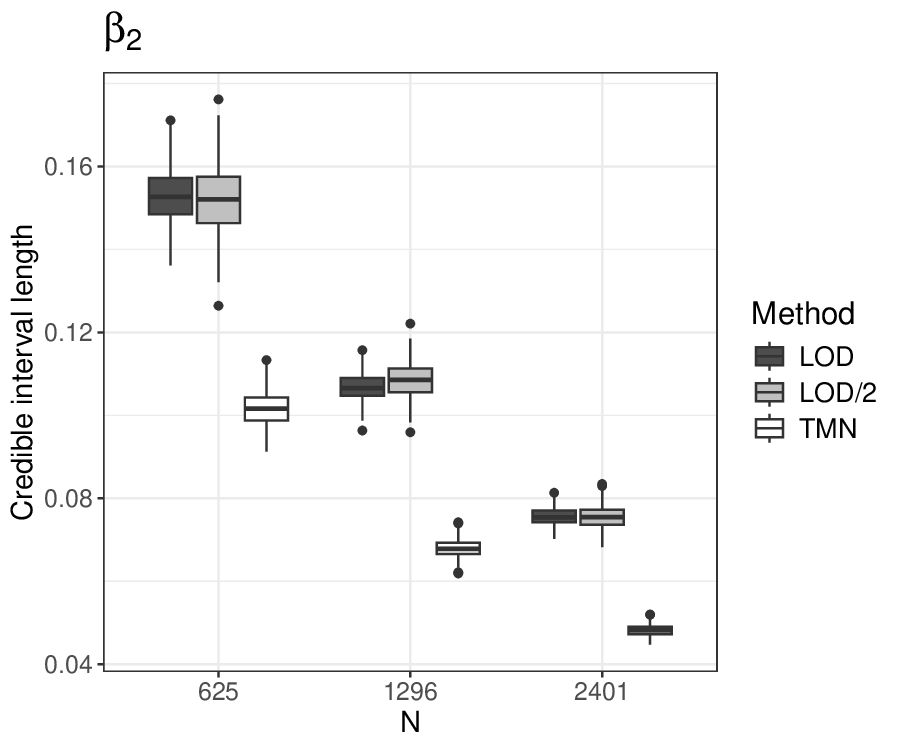}}
\caption{{\bf Simulation study I}. DAGAR model. Credible interval lenghts for the mean structure parameters considering a censoring level of $15\%$ and missing level of $5\%$}
\label{supp:est15meanpar}
\end{figure}

\begin{figure}[!htbp]
\centering
\subfigure[]{\includegraphics[width=0.45\textwidth]{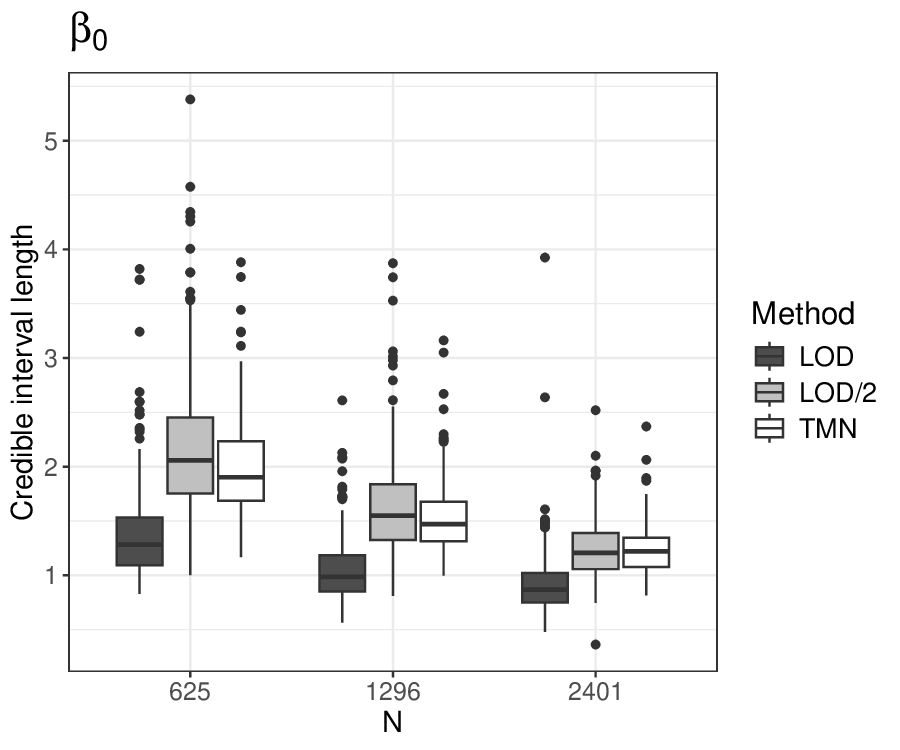}}, 
\subfigure[]{\includegraphics[width=0.45\textwidth]{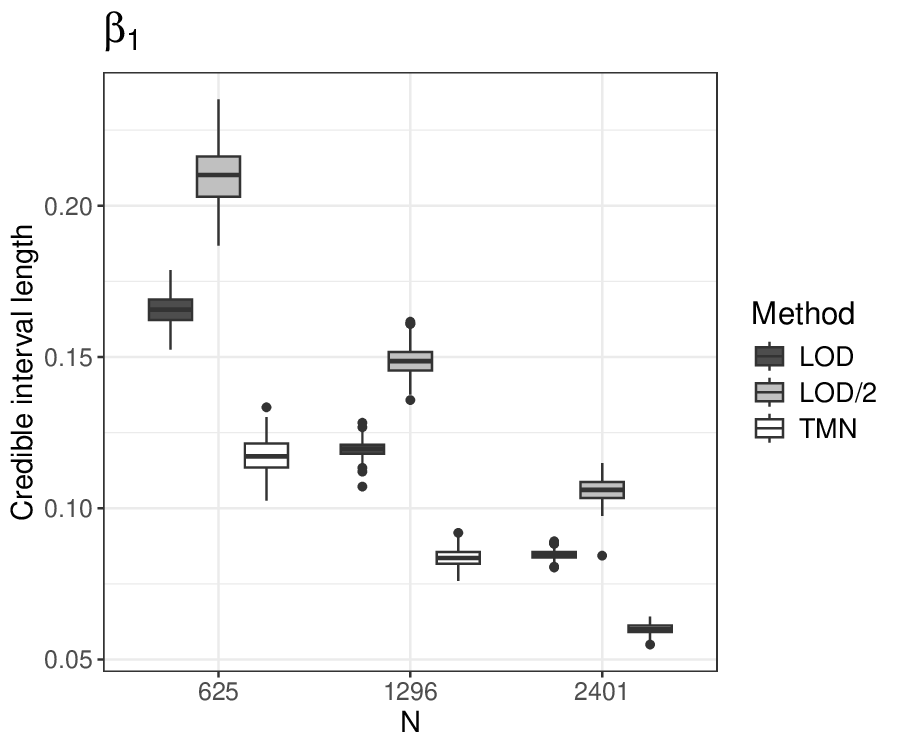}}, 
\subfigure[]{\includegraphics[width=0.45\textwidth]{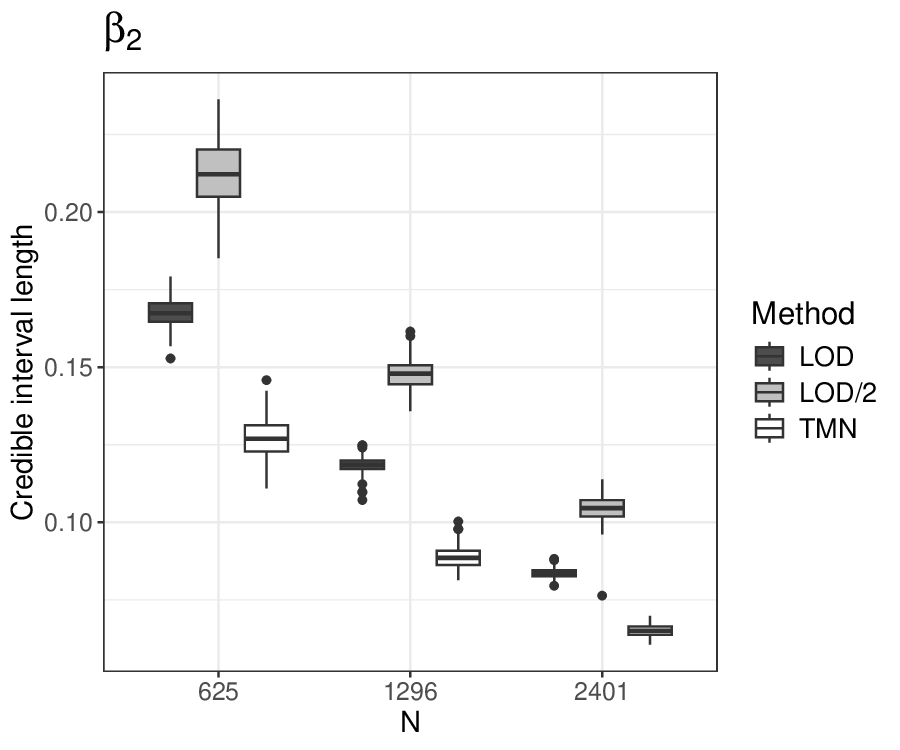}}
\caption{{\bf Simulation study I}. DAGAR model. Credible interval lengths for the mean structure parameters considering a censoring level of $35\%$ and missing level of $5\%$}
\label{supp:est35meanpar}
\end{figure}

\begin{figure}[!htbp]
\centering
\subfigure[]{\includegraphics[width=0.45\textwidth]{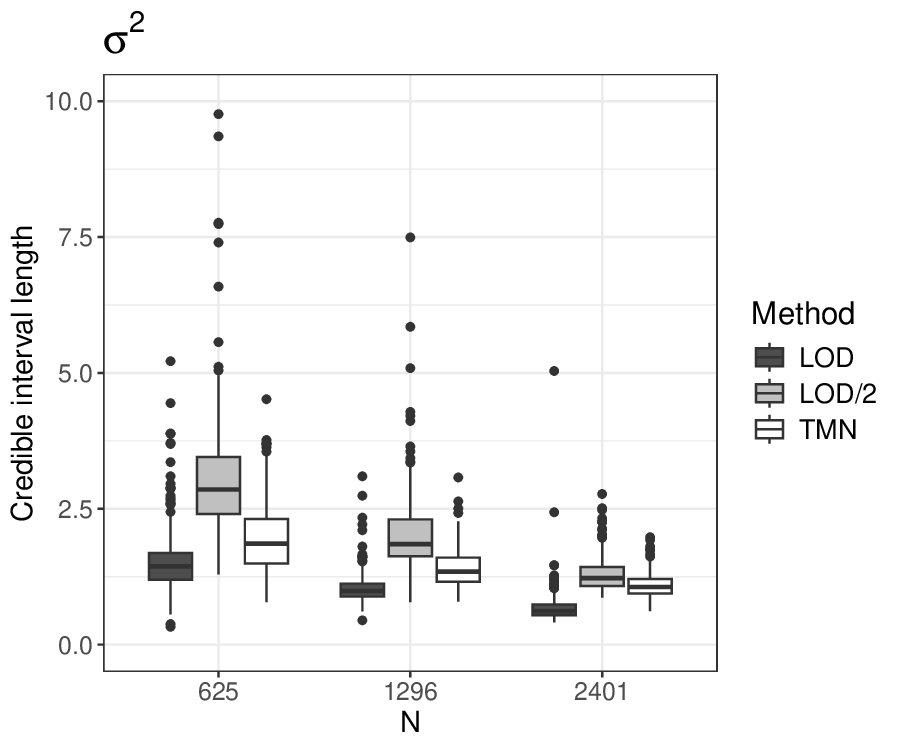}}, 
\subfigure[]{\includegraphics[width=0.45\textwidth]{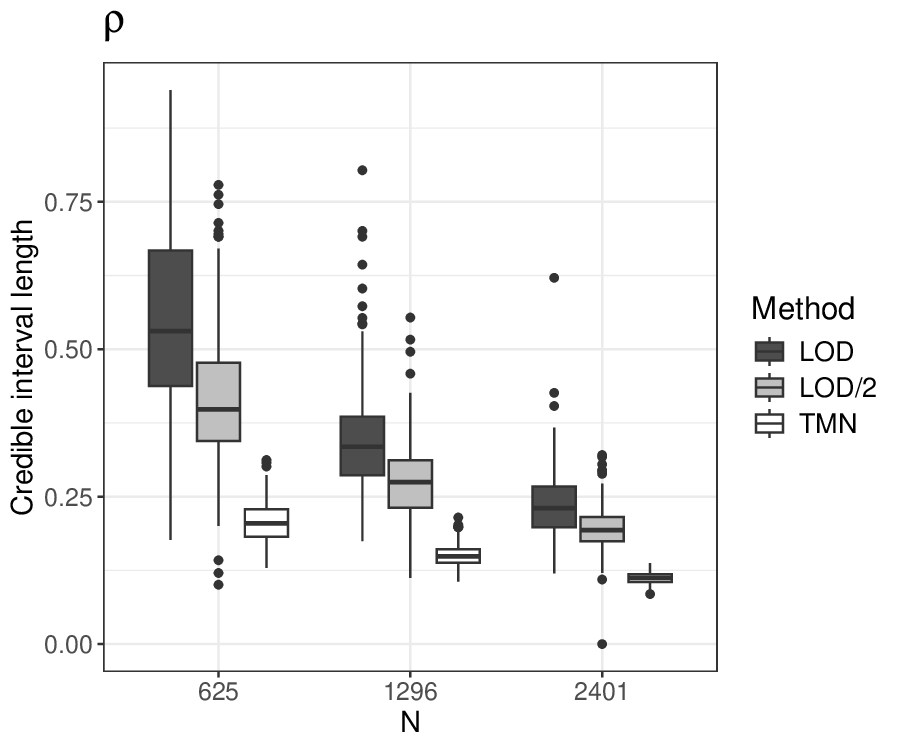}}, 
\subfigure[]{\includegraphics[width=0.45\textwidth]{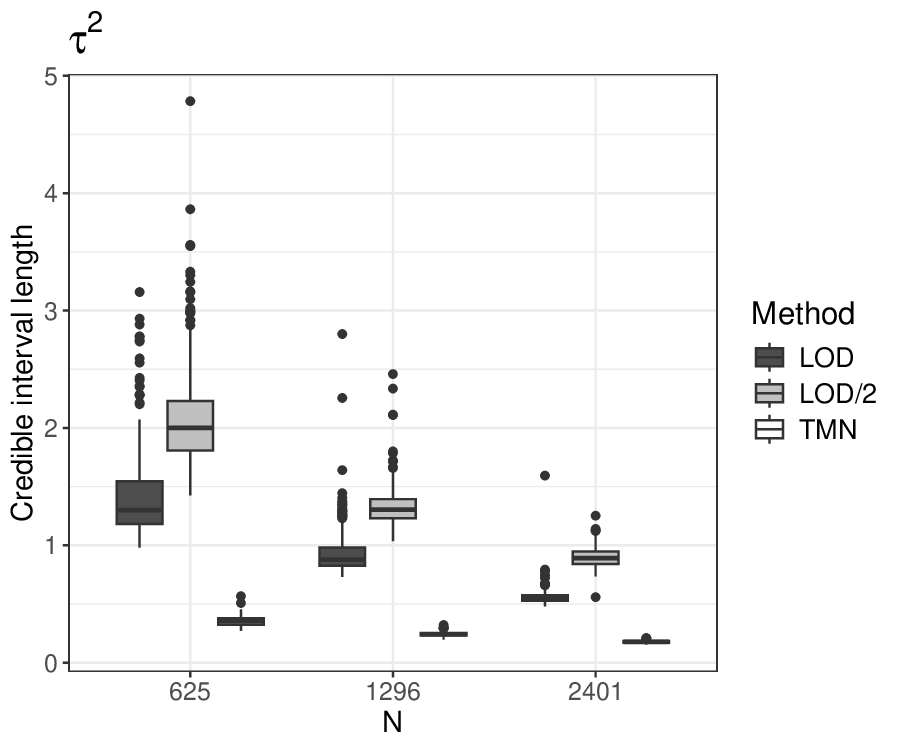}}
\subfigure[]{\includegraphics[width=0.45\textwidth]{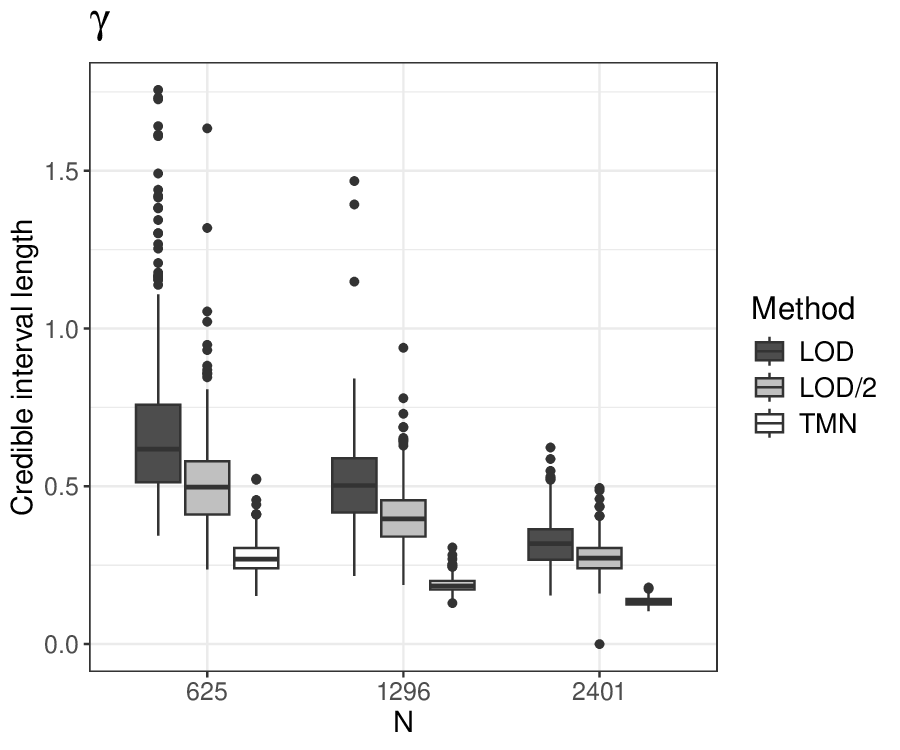}}
\caption{{\bf Simulation study I}. DAGAR model. Credible interval lengths for the covariance structure parameters considering a censoring level of $35\%$ and missing level of $5\%$}
\label{supp:est35covpar}
\end{figure}

\begin{figure}[!htbp]
\centering
\subfigure[]{\includegraphics[width=0.45\textwidth]{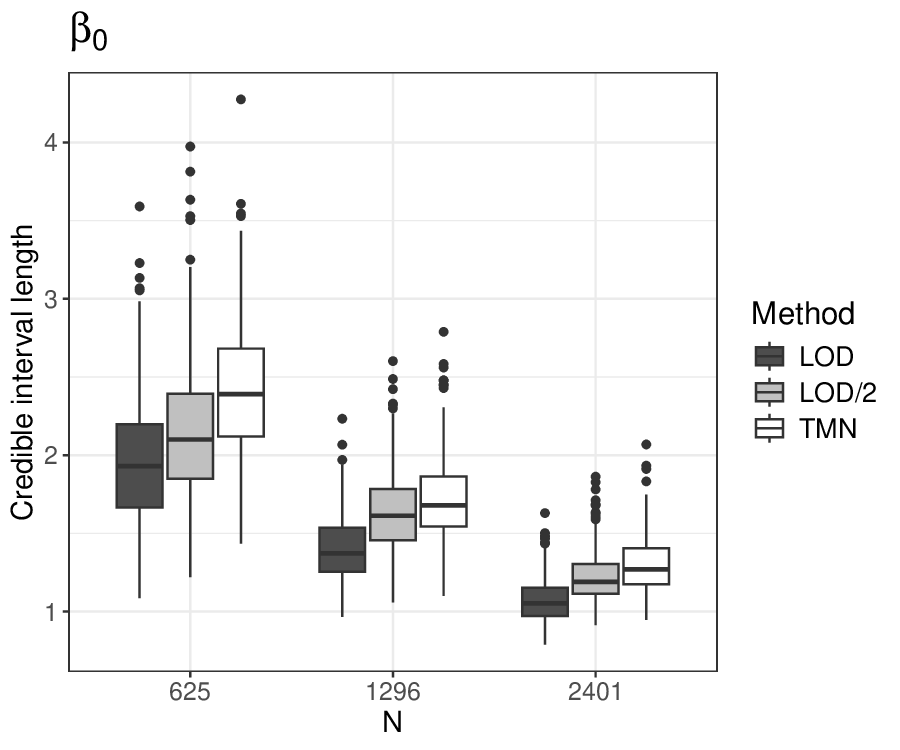}}, 
\subfigure[]{\includegraphics[width=0.45\textwidth]{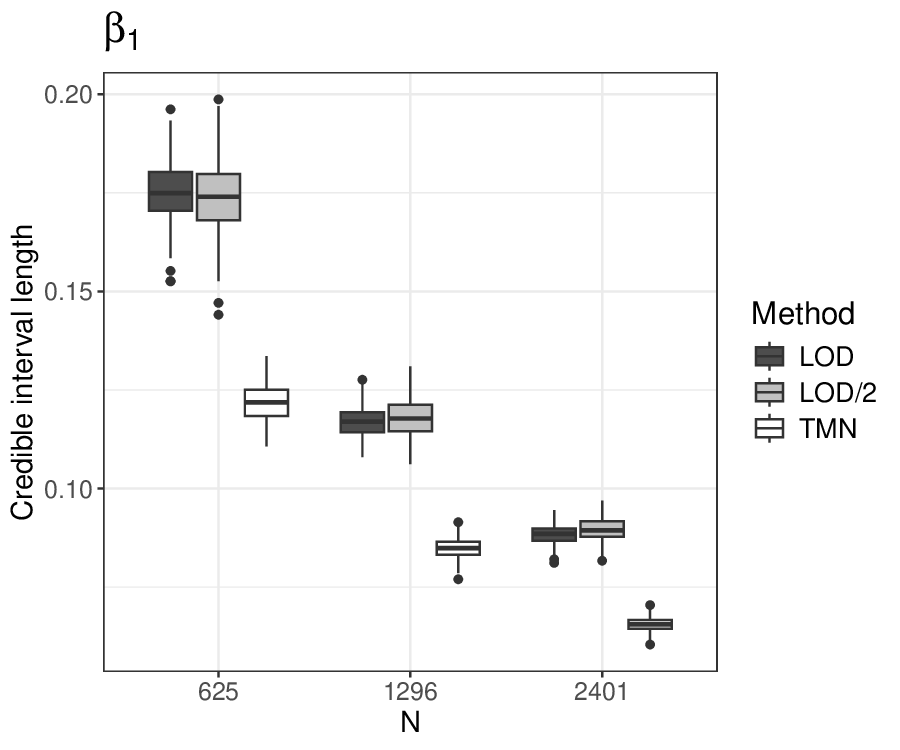}}, 
\subfigure[]{\includegraphics[width=0.45\textwidth]{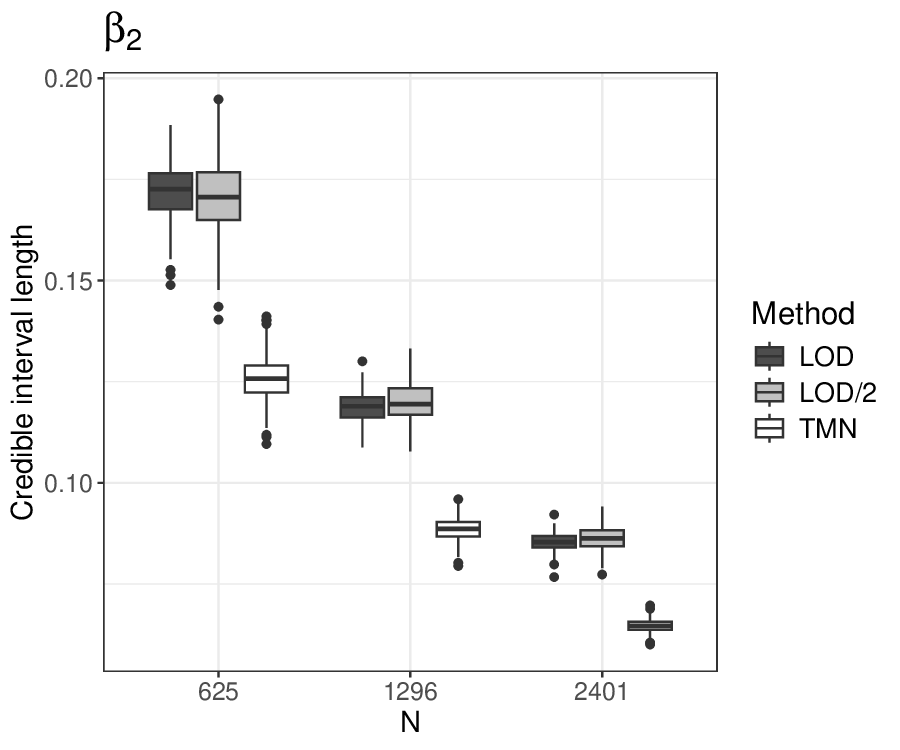}}
\caption{{\bf Simulation study}. SAR model. Credible interval lenghts for the mean structure parameters considering a censoring level of $15\%$ and missing level of $5\%$}
\label{supp:est15meanparsar}
\end{figure}

\begin{figure}[!htbp]
\centering
\subfigure[]{\includegraphics[width=0.45\textwidth]{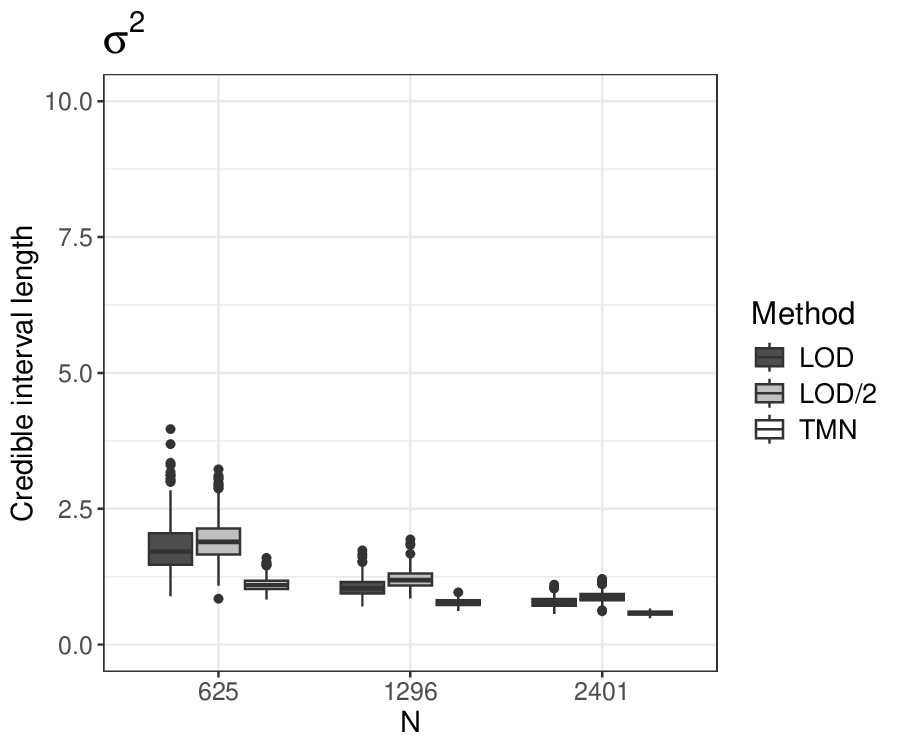}}, 
\subfigure[]{\includegraphics[width=0.45\textwidth]{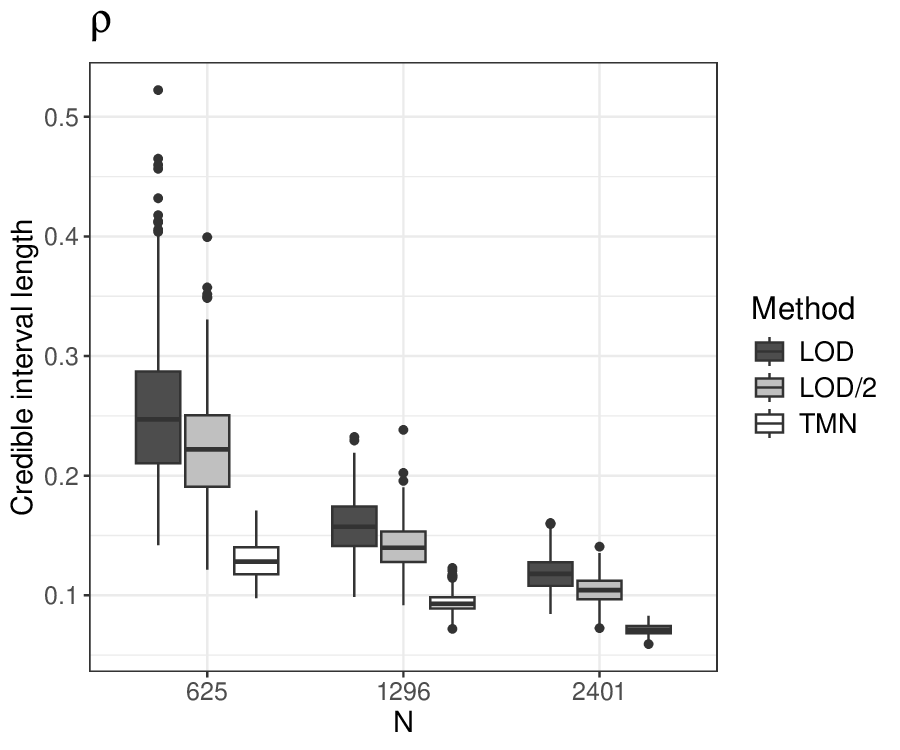}}, 
\subfigure[]{\includegraphics[width=0.45\textwidth]{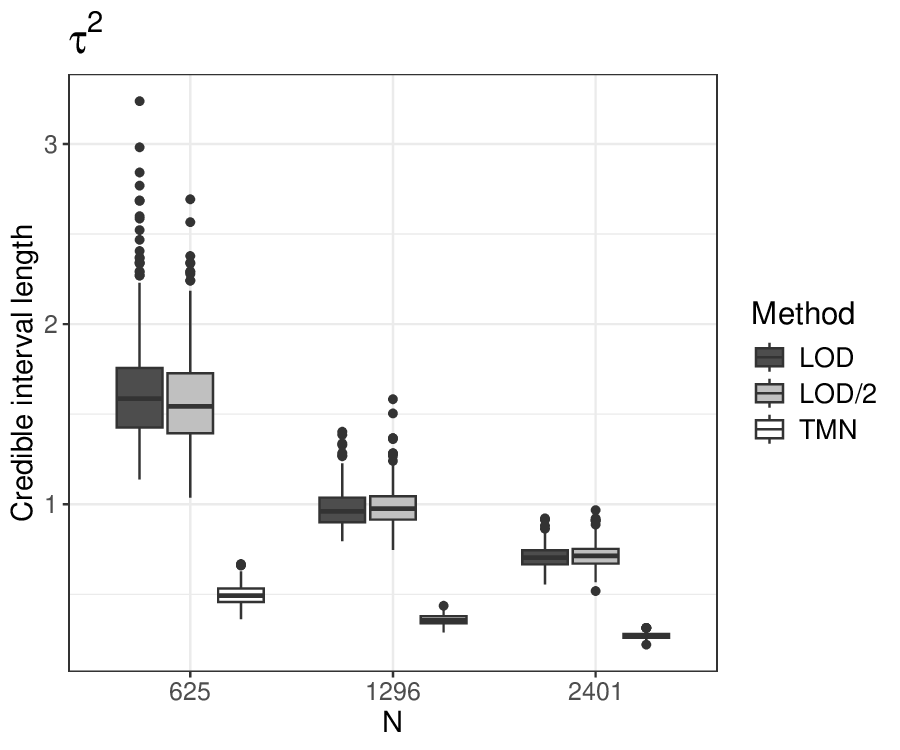}}
\subfigure[]{\includegraphics[width=0.45\textwidth]{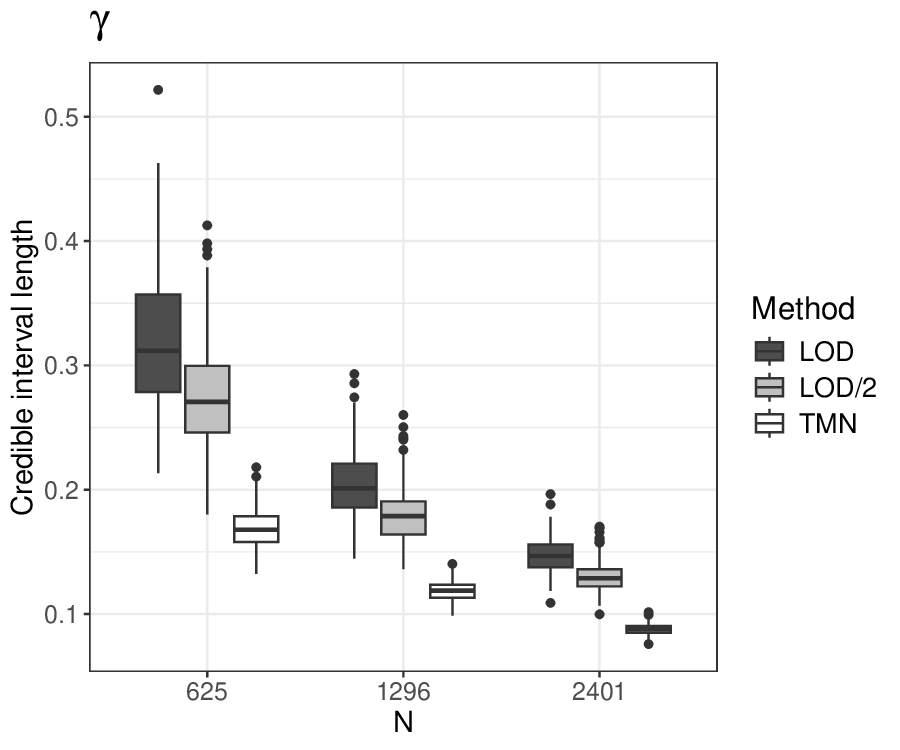}}
\caption{{\bf Simulation study I}. SAR model. Credible interval lengths for the covariance structure parameters considering a censoring level of $15\%$ and missing level of $5\%$}
\label{supp:est15covparsar}
\end{figure}

\begin{figure}[!htbp]
\centering
\subfigure[]{\includegraphics[width=0.45\textwidth]{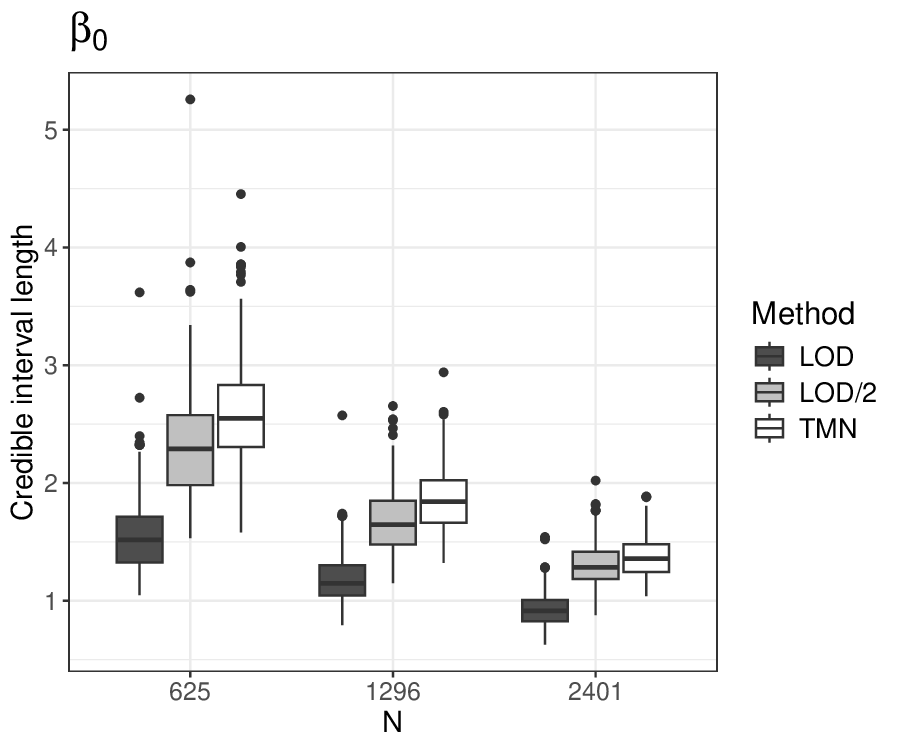}}, 
\subfigure[]{\includegraphics[width=0.45\textwidth]{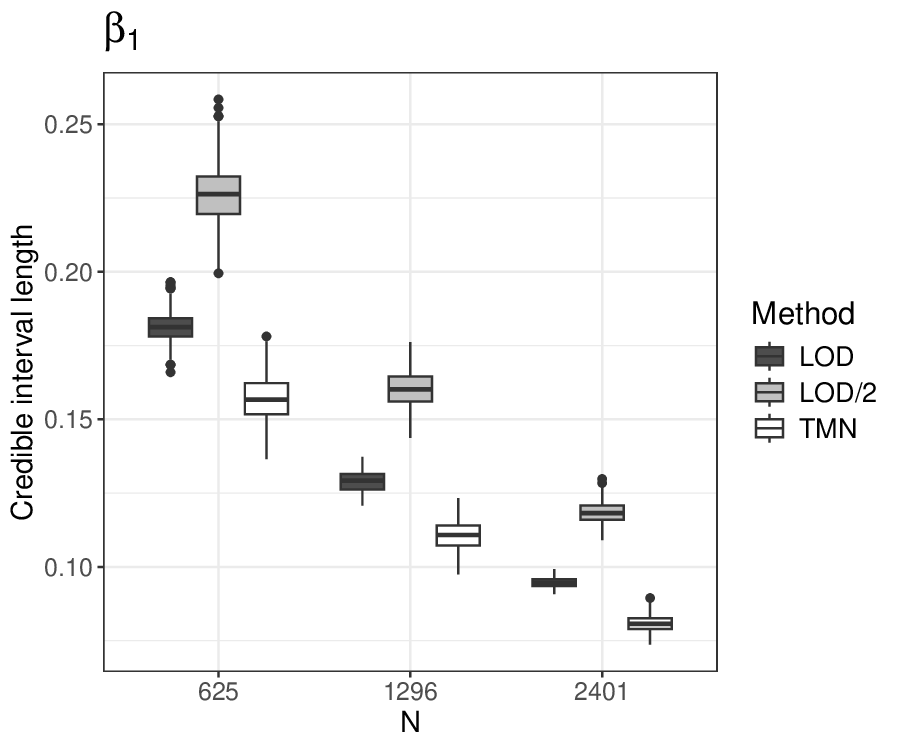}}, 
\subfigure[]{\includegraphics[width=0.45\textwidth]{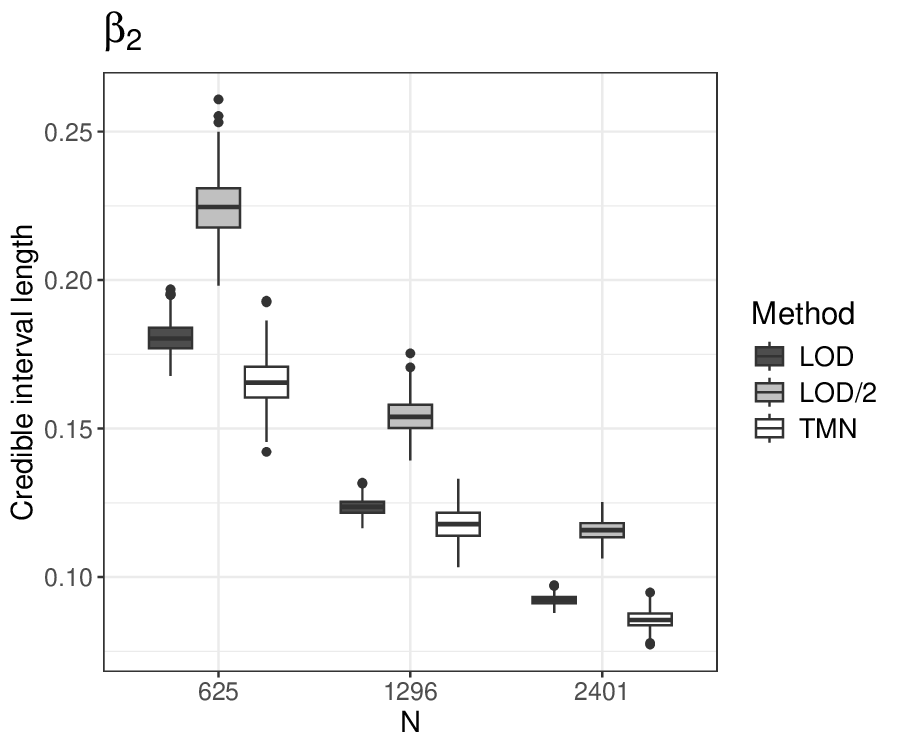}}
\caption{{\bf Simulation study I}. SAR model. Credible interval lenghts for the mean structure parameters considering a censoring level of $35\%$ and missing level of $5\%$}
\label{supp:est35meanparsar}
\end{figure}

\begin{figure}[!htbp]
\centering
\subfigure[]{\includegraphics[width=0.45\textwidth]{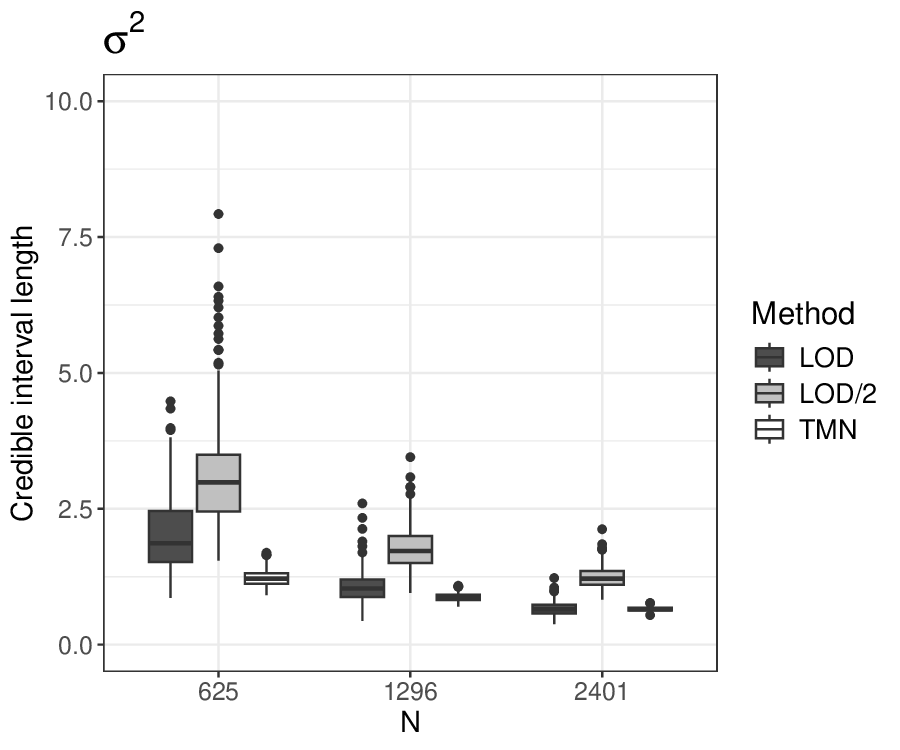}}, 
\subfigure[]{\includegraphics[width=0.45\textwidth]{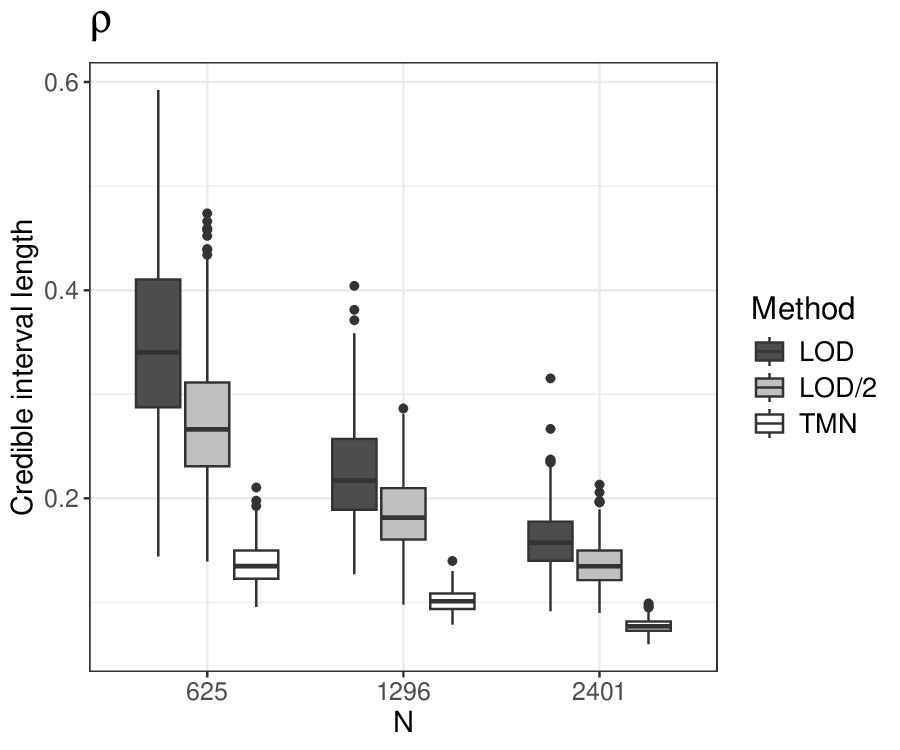}}, 
\subfigure[]{\includegraphics[width=0.45\textwidth]{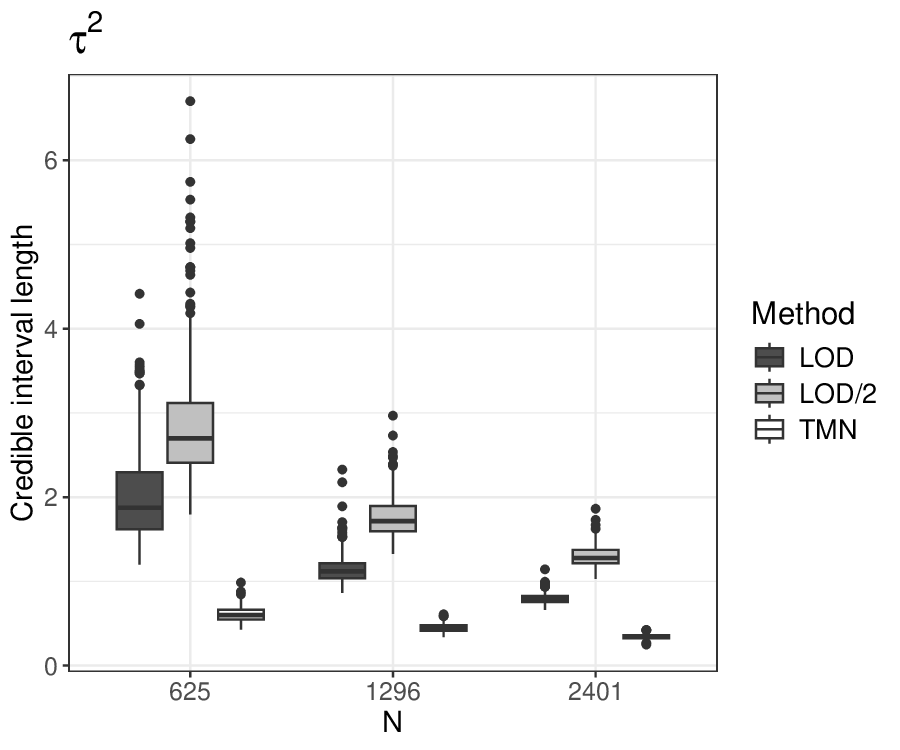}}
\subfigure[]{\includegraphics[width=0.45\textwidth]{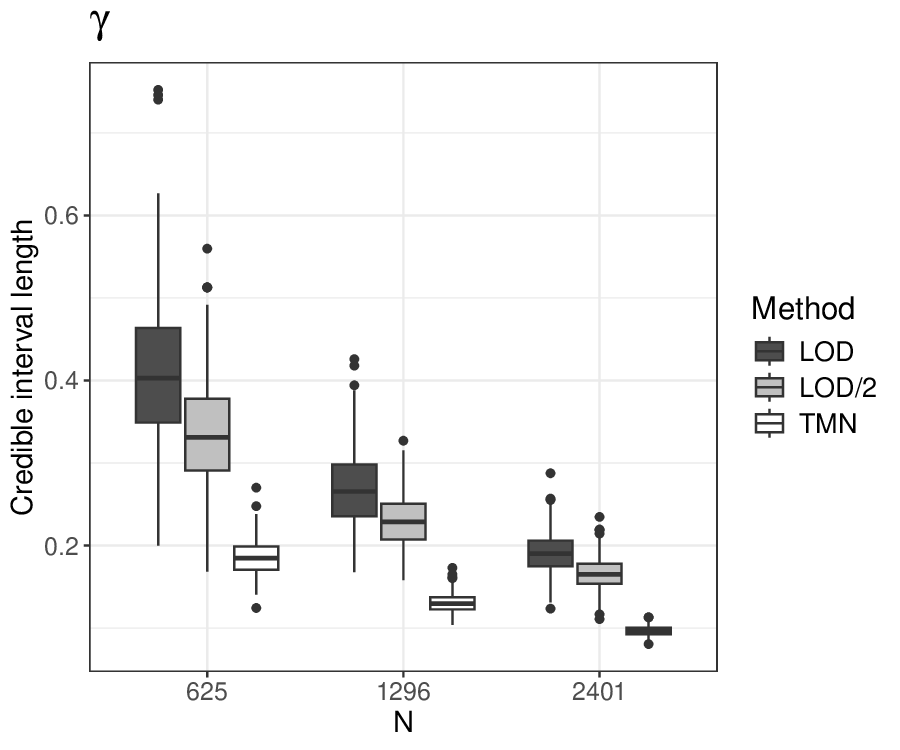}}
\caption{{\bf Simulation study I}. SAR model. Credible interval lengths for the covariance structure parameters considering a censoring level of $35\%$ and missing level of $5\%$}
\label{supp:est35covparsar}
\end{figure}

As mentioned above, our goal here is to assess model fitting in the presence of censoring and missingness. For comparison purposes, we calculated (i) the length of the $95\%$ credible interval and (ii) the coverage probability for each parameter. The first was computed as the difference between the upper and lower credible interval limits for each simulation; these limits were obtained as the $2.5\%$ and $97.5\%$ percentiles of the posterior distribution sample in each simulation. The second was computed as the proportion of simulations in which the true parameter was within the credible interval limits. Concerning the DAGAR correlation, Figure \ref{estcens15} and Table \ref{tabcens15covpar} present the distribution of the 300 interval lengths and coverage probabilities for the variance structure parameters obtained using the NST-CLG model, LOD, and LOD/2 methods under a censoring level of $15\%$ and a missingness rate of $5\%$. The corresponding distribution of the interval lengths for the mean structure parameters is shown in Figure \ref{supp:est15meanpar}. Additional results for the mean and covariance parameters under $35\%$ censoring and $5\%$ missingness are provided in Figures \ref{supp:est35meanpar} and \ref{supp:est35covpar}, as well as in Tables \ref{supp:tabest15meanpar}, \ref{supp:tabest35meanpar}, and \ref{supp:tabest35covpar}. With respect to the SAR process, Tables \ref{supp:tabest15meanparsar} and \ref{supp:tabest35meanparsar}, and Figures \ref{supp:est15meanparsar} and \ref{supp:est35meanparsar} show the distribution of lengths and coverage probabilities for the mean parameters under both censoring scenarios. On the other hand, Tables \ref{supp:tabest15covparsar} and \ref{supp:tabest35covparsar}, and Figures \ref{supp:est15covparsar} and \ref{supp:est35covparsar}, show these measures for the covariance parameters. 

For the regression coefficients $\boldsymbol{\beta}$, when censoring/missingness is $15\%/5\%$, the NST-CLG model yields the shortest posterior intervals with empirical coverage close to the nominal $95\%$ level for all sample sizes, and the same pattern holds at censoring/missing levels of $35\%/5\%$. This behavior is expected; the proposed model accounts for censored values through an appropriate truncated likelihood. In contrast, the credible intervals given by the LOD and LOD/2 substitutions narrow as $n$ increases, yet exhibit declining coverage. Constant replacements at the detection limit (or its half) and mean imputation for missingness shift the conditional mean and compress the residual spread, producing attenuation. As the sample size increases, the posterior concentrates around this biased center, so intervals become too tight and miss the target more often.

This pattern is consistent with covariance components. With $15\%$ censoring and $5\%$ missingness, the NST-CLG model produces the shortest posterior intervals and coverage closest to the nominal $95\%$ for all the parameters and sample sizes. When censoring/missingness increases to $35\%/40\%$, the intervals widen as expected, but the ranking does not change, and the NST-CLG model remains the best approach, while LOD and LOD/2 continue to under-cover. Similar results were observed for the SAR process, excepting for the intercept $\beta_0$ where the NST-CLG model presents sligthly larger interval lengths but still, better coverage probabilities.

\begin{table}[htbp]
  \centering
  \caption{{\bf Simulation study I}. Coverage probabilities of the covariance structure parameters considering a level of censoring of $15\%$}\label{tabcens15covpar}%
    \begin{tabular}{ccccc}
    \toprule
    \multicolumn{1}{l}{N} & Parameter & NST-CLG   & LOD   & LOD/2 \\
    \midrule
    \multirow{4}[2]{*}{625} & $\sigma^2$ & 0.940  & 0.703  & 0.933 \\
          & $\rho$   & 0.973  & 0.950  & 0.923 \\
          & $\tau^2$  & 0.940  & 0.000  & 0.000 \\
          & $\boldsymbol{\gamma}$   & 0.960 & 0.963  & 0.933 \\
    \midrule
    \multirow{4}[2]{*}{1296} & $\sigma^2$ & 0.947  & 0.390  & 0.907 \\
          & $\rho$    & 0.943  & 0.957  & 0.953 \\
          & $\tau^2$  & 0.950  & 0.000  & 0.000 \\
          & $\boldsymbol{\gamma}$   & 0.947  & 0.923  & 0.950 \\
    \midrule
    \multirow{4}[2]{*}{2401} & $\sigma^2$ & 0.957  & 0.347  & 0.923 \\
          & $\rho$   & 0.943  & 0.930 & 0.933 \\
          & $\tau^2$  & 0.963 & 0.000  & 0.000 \\
          & $\boldsymbol{\gamma}$   & 0.947  & 0.933  & 0.940 \\
    \bottomrule
    \end{tabular}%
\end{table}%

\subsection{Predictive perfomance in time domain}\label{sec:6.2}

Here, we evaluate the prediction performance of our method by comparing it with
the traditional imputation techniques used in Subsection 4.1. With respect to the
comparison criteria, we consider (i) the square root of the mean squared
prediction error (MSPE), which, following the vectorization convention introduced
in Section \ref{bay_pred} and for fixed regions $s_1,\ldots,s_K$, $K\leq n$, is computed as,
$$
\sqrt{\mathrm{MSPE}}
=
\left\{
\frac{1}{Kh}
\left(
\widehat{\tilde{\mathbf{y}}}_{K}
-
\tilde{\mathbf{y}}_{K}
\right)^\top
\left(
\widehat{\tilde{\mathbf{y}}}_{K}
-
\tilde{\mathbf{y}}_{K}
\right)
\right\}^{1/2},
$$
where $\tilde{\mathbf{y}}_{K}$ denotes the vector of observed future responses
over the selected regions and forecasting times, and
$\widehat{\tilde{\mathbf{y}}}_{K}$ denotes the corresponding vector of point
predictions. (ii) mean lengths of the $95\%$ posterior predictive credible intervals for all predicted values. For each predicted value, this quantity is computed by obtaining the credible interval limits as the $2.5\%$ and $97.5\%$ percentiles of its predictive posterior distribution sample, and then calculating the interval length as the difference between the upper and lower credible interval limits for each simulation. We then computed the mean of the interval lengths across all predictions. Finally, we compute (iii) the mean coverage probability for all predictions. This measure was computed by obtaining the proportion of simulations in which each real value $Y(s_i,\tilde{t}_j)$ lies inside its credible interval limits, and then computing the mean of these proportions over all the predicted values.

\begin{figure}
\centering
\subfigure[]{\includegraphics[width=0.35\textwidth]{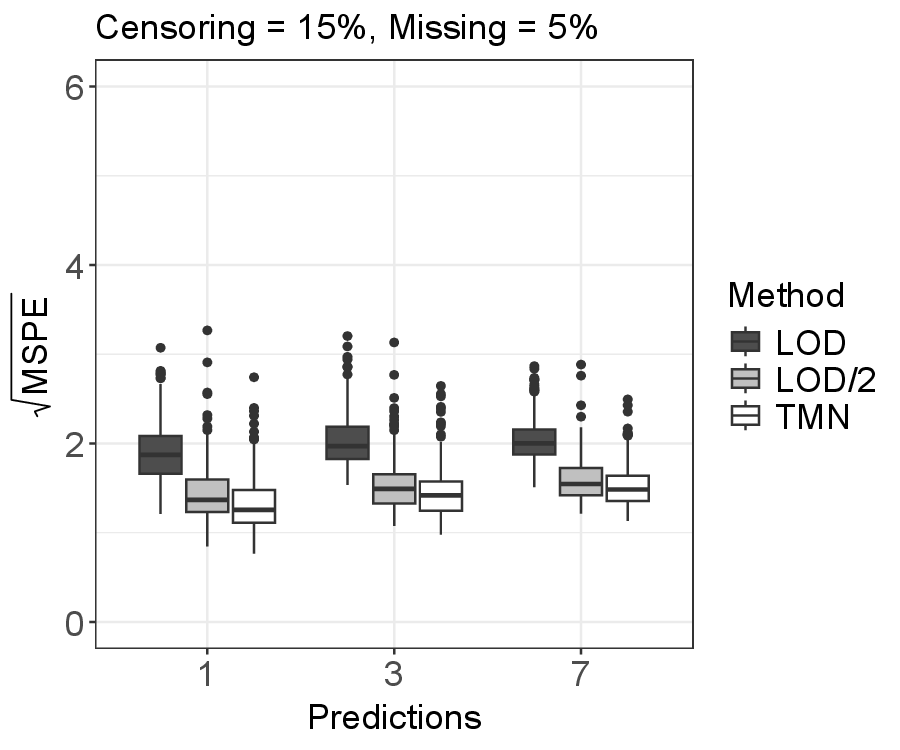}}, 
\subfigure[]{\includegraphics[width=0.35\textwidth]{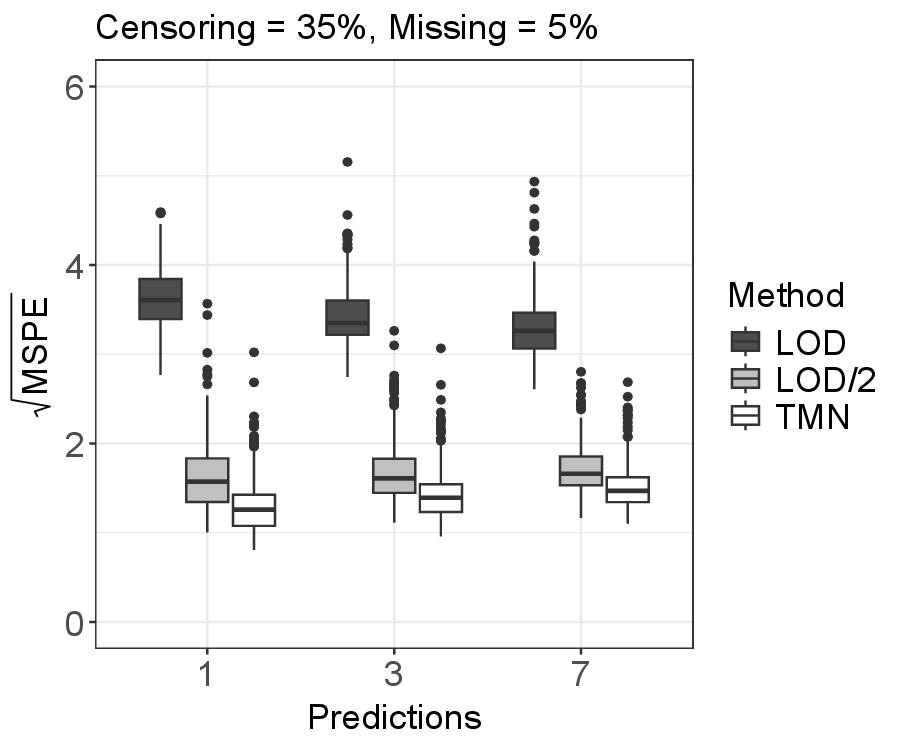}}, 
\subfigure[]{\includegraphics[width=0.35\textwidth]{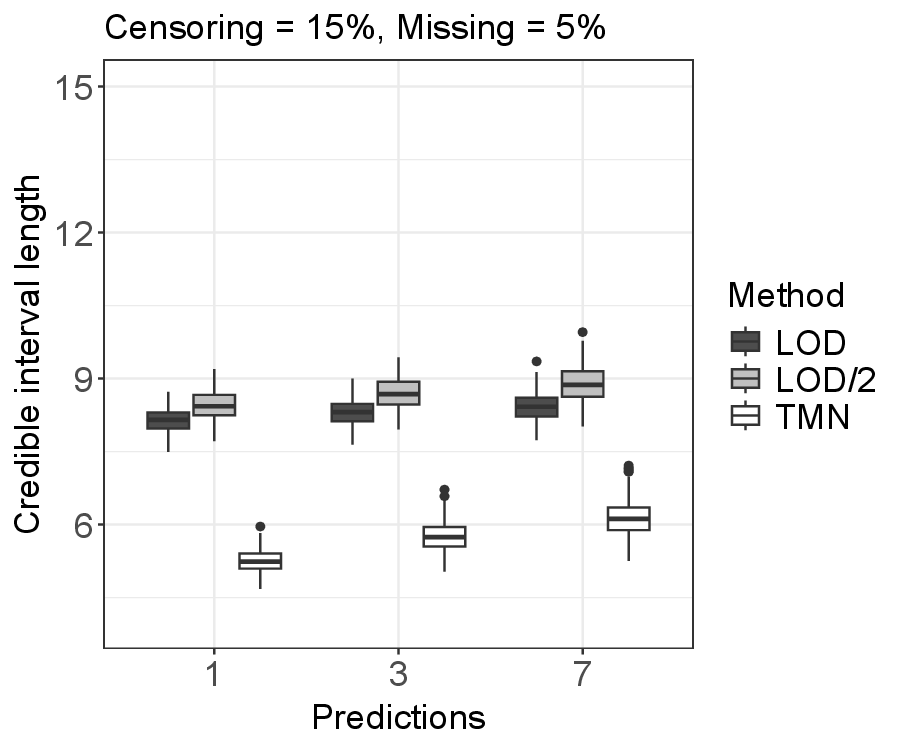}}
\subfigure[]{\includegraphics[width=0.35\textwidth]{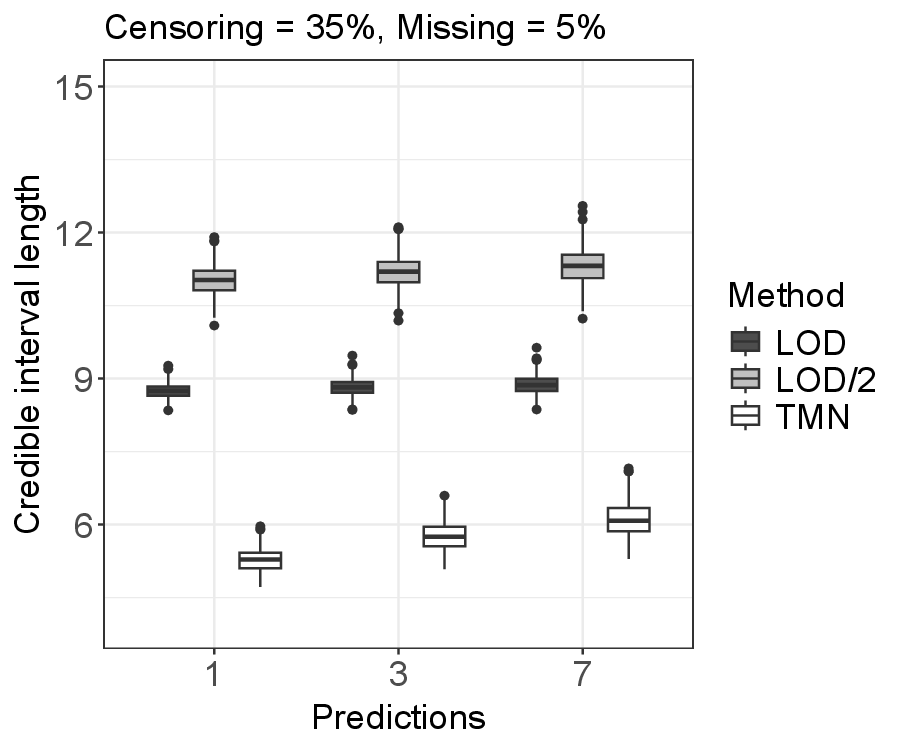}}
\caption{{\bf Simulation study II}. Comparison of our proposal (NST-CLG) with
methods that impute the censored observations by using the limit of detection. For this
scenario, we consider $N$= 1296.
}
\label{figsim2500}
\end{figure}

For this simulation study, we created three scenarios by computing all the comparison criteria for one-, three-, and seven-step-ahead predictions. Regarding the DAGAR model, Figure \ref{figsim2500} and Table \ref{tabsim2500} show the results for $N = 1296$ and both censoring levels. The results for $N = 625$ and $N = 2401$ are presented in Tables S.3.8 and S.3.9 and Figures S.3.8 and S.3.9. On the other hand, Figures S.3.10 - S.3.12 and Tables S.3.10 - S.3.12 show the simulation results for the SAR correlation.

With respect to the DAGAR model, the $\sqrt{\mbox{MSPE}}$ for the LOD method was the worst across all sample sizes, censoring levels, and prediction settings. The  NST-CLG model and LOD/2 are broadly similar, with the former being slightly better. The gap is larger for the prediction-interval length: the NST-CLG model produces the narrowest intervals, while LOD yields the widest. Moreover, LOD/2 returns coverage near 1.0 in every setting, indicating overly wide intervals. The LOD improved as the sample size approached the 95\% target at N = 750, after under- or overshooting at smaller sample sizes. By contrast, our model stays close to $95\%$ across all sample sizes, censoring levels, and prediction scenarios. Overall, the NST-CLG model delivers the best balance, with the smallest $\sqrt{\mbox{MSPE}}$ and shortest intervals, while maintaining coverage near the nominal 95\% level. Regarding the SAR model, all three methods yield comparable $\sqrt{\mbox{MSPE}}$ values at the 15\% and 5\% levels of censoring and missingness, across all sample sizes. However, when the proportion of censored data increases to 35\% (keeping a 5\% proportion of missingness), the performance of the LOD approach deteriorates noticeably, whereas LOD/2 and our approach exhibit similar behavior, with the NST-CLG model performing slightly better overall.
Regarding credible interval lengths and coverage probabilities, the LOD/2 method produces wider credible intervals but achieves coverage probabilities close to the nominal 95\% level across all sample sizes. In contrast, the LOD method provides intervals of similar width to those of the NST-CLG model, but its coverage probabilities remain noticeably lower than the 95\% nominal level. On the other hand, our proposed model maintains shorter intervals while still achieving coverage rates near the nominal level, demonstrating a clear advantage over the two ad hoc methods.

\begin{table}[!ht]
  \centering
  \caption{{\bf Simulation study II}. Average coverage probabilities of credible intervals as a function of the number of predicted observations, with $N = 1296$.}
    \begin{tabular}{ccccc}
\cmidrule{3-5}    \multicolumn{1}{r}{} &       & \multicolumn{3}{c}{Predicted time observations } \\
    \midrule
    Censoring-missignness & Method & One   & Three  & Seven  \\
    \midrule
    \multirow{3}[6]{*}{\raisebox{1\normalbaselineskip}[0pt][0pt]{15\%-5\%}} & NST-CLG   & 0.943 & 0.947 & 0.953 \\
         & LOD   &0.970 & 0.977 & 0.980 \\
         & LOD/2 & 0.997 & 0.997 & 0.997 \\ 
    \midrule
    \multirow{3}[6]{*}{\raisebox{1\normalbaselineskip}[0pt][0pt]{35\%-5\%}} & NST-CLG   & 0.950 & 0.950 & 0.950 \\ 
        & LOD   & 0.933 & 0.950 & 0.957 \\
         & LOD/2 & 1.000 & 1.000 & 1.000 \\ 
    \bottomrule
    \end{tabular}%
  \label{tabsim2500}%
\end{table}%

\section{Application: CO concentrations data}\label{section7}

In this section, we apply our proposal to the air quality
dataset previously analyzed in studies such as \citet{zhang2017} and
\citet{wardana2022}. We focus on CO concentrations ($\mu\text{g/m}^3$)
recorded twice daily, at 09:00 and 19:00, corresponding to periods of high
traffic activity. Although the original dataset includes twelve monitoring
stations distributed across the eight districts considered in the analysis, some
districts contain more than one station. Since our model is defined at the
district level, we selected one representative station per district to obtain a
balanced areal dataset.

\begin{figure}
    \centering
\includegraphics[scale = 0.75]{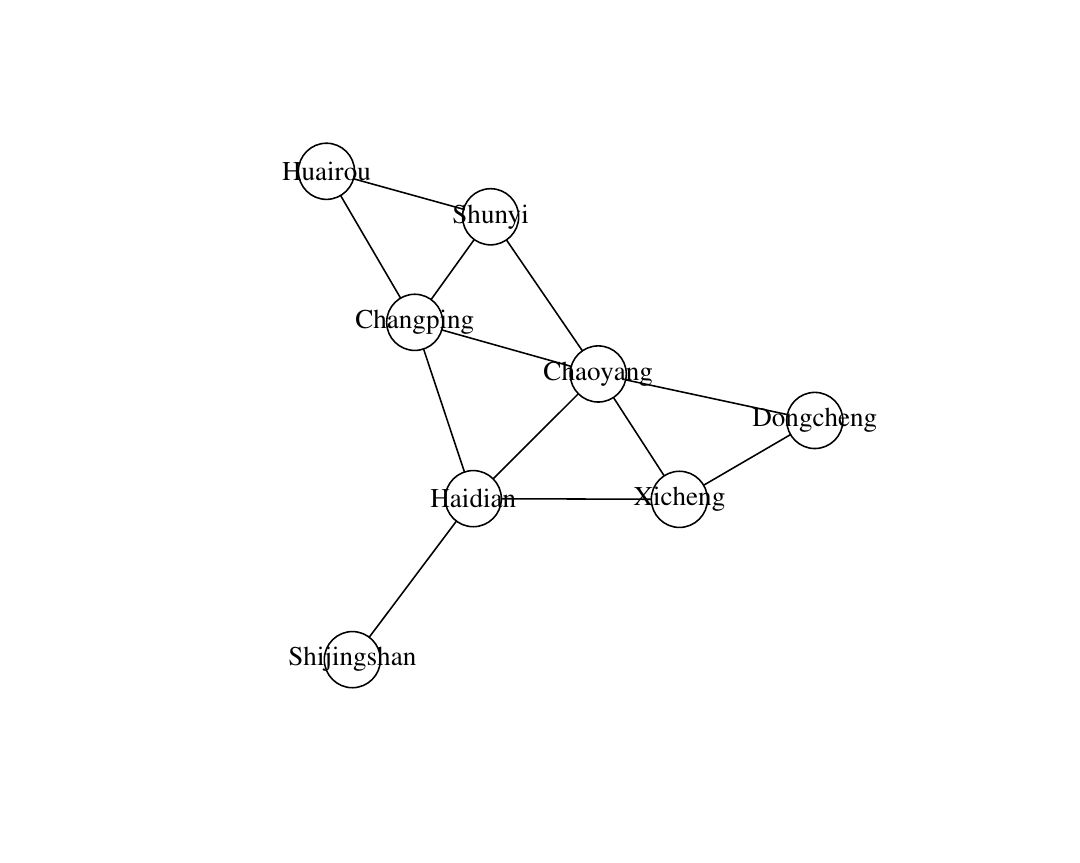}
    \caption{{\bf Air quality dataset}. Spatial network of the eight air-quality monitoring sites. The nodes correspond to monitoring stations used in the pollutant analysis, and the edges represent spatial neighborhood relationships.}
    \label{network}
\end{figure}

For districts with multiple stations, we chose the station located in the most
urbanized area. This criterion improves comparability across districts, as urban
stations are more directly affected by traffic and residential emissions, which
are important contributors to CO concentrations. Mixing urban, suburban, and
background stations could introduce heterogeneity due to station type rather than
district-level spatial dependence. Thus, the final dataset contains eight
monitoring locations, one per district. The spatial neighborhood structure is
then defined by considering districts as neighbors if they share a common
boundary, as illustrated in Figure \ref{network}.

\subsection{Descriptive analysis}

The analysis covered the period from February 10, 2016, to February 28, 2017, and included observations from eight districts. CO levels were recorded at 768 time points, yielding a total of 6,144 observations ($8 \times 768$). 
The dataset contains 144 missing values. The final 18 days (11/02/2017–28/02/2017) were reserved as a test set for all stations in order to assess the predictive performance of the proposed model. Accordingly, model training was conducted using $8 \times 732 = 5856$ observations, whereas testing involved $8 \times 36 = 288$ observations. 

\begin{figure}
    \centering
\includegraphics[scale=0.7]{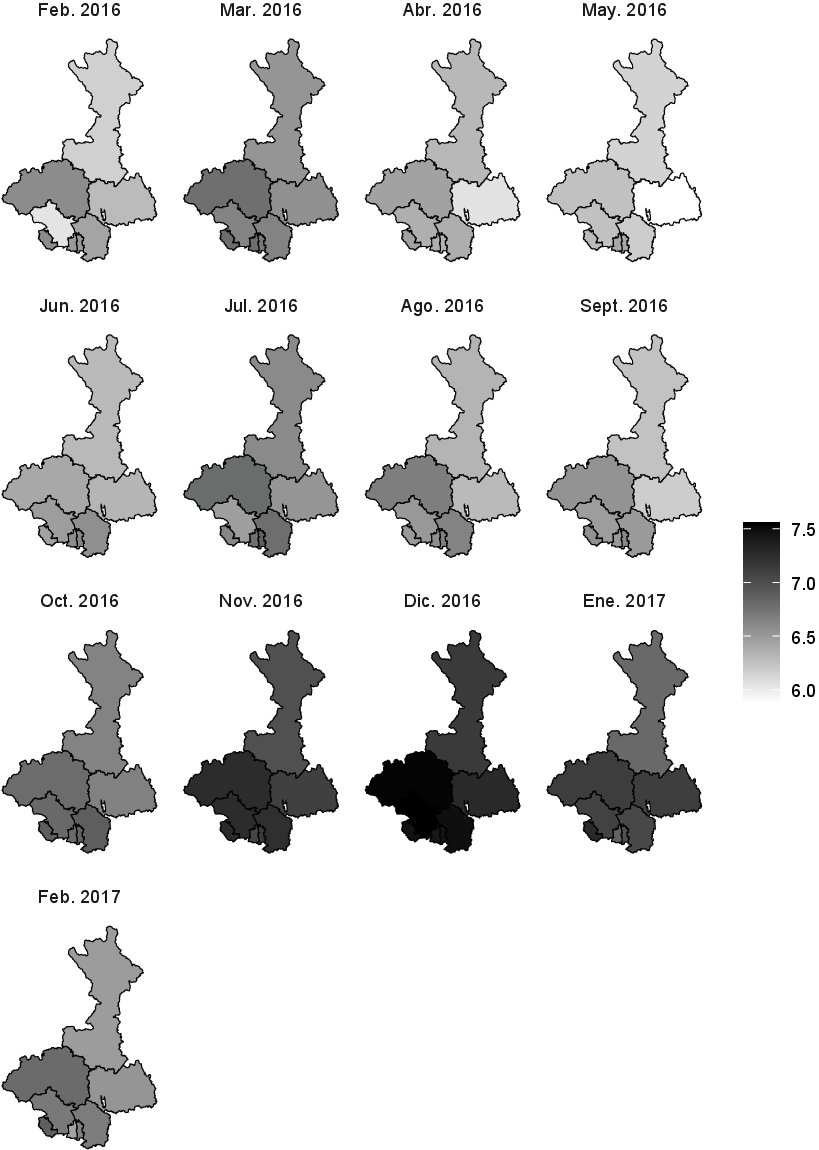}
    \caption{{\bf Air quality dataset}. Monthly mean CO concentrations (on the log scale) across the districts.}\label{concentrations_map}
\end{figure}

To mitigate the influence of extreme values and reduce issues related to variance heterogeneity, analysis was conducted using CO concentrations expressed on a logarithmic scale. Figures \ref{concentrations_map} and \ref{concentrations2}, show, respectively, the heat map and the raw scatter plot and  of the log-transformed CO concentrations across the eight districts. Together, both figures reveal clear short-term fluctuations and a seasonal pattern, with higher concentrations during winter, especially around December and January. They also show spatial differences in average levels and variability across districts, while similar patterns among neighboring areas suggest positive spatial correlation. Overall, these visual summaries support the presence of joint spatiotemporal dependence in the data.

Finally, Table \ref{descstat} presents summary statistics for the variables included in this dataset, namely $\log(\mathrm{CO})$, temperature (TEMP, °C), wind speed (WSP, m/s), and atmospheric pressure (PRES, Pa). These variables serve as suitable covariates for predicting CO concentrations. 

\begin{table}[!htbp]
\caption{{\bf Air quality dataset}. Summary statistics of the variables of interest}
\centering
\begin{tabular}{cccccc}\label{descstat}
 
Variable & Mean & Median & s.d & Min & Max \\ 
  \hline
log(CO) (response) &  6.687 & 6.685 & 0.853 &  4.605  & 9.200 \\ 
  TEMP & 13.086 & 13.700 &  11.674 & -12.520 & 35.800  \\ 
  PRES & 10.105 & 10.100 & 0.1051 &   9.855 & 10.367  \\ 
  WSP & 1.699 & 1.400 & 1.180 &   0.000 & 0.900 \\ 
   \hline
\end{tabular}
\end{table}

To gain insights into the temporal and spatial correlations, we imputed missing data using the mean value per region. We computed the autocorrelation (ACF) and partial autocorrelation (PACF) for each of the twelve sites  (see Figures S.4.14 and 
S.4.15). The ACF and PACF panels reveal temporal dependence across sites; the gradual ACF decay coupled with a dominant lag-1 spike in the PACF is consistent with the AR(1) specification for the temporal component.

\subsection{Model fitting and comparison}

After the descriptive analysis, we specified the mean structure as a particular
case of the tensor regression formulation in equation~\eqref{tensor_model}. For
this application, the covariate tensor
$\mathcal{X}\in\mathbb{R}^{n\times T\times 12}$ is constructed by stacking,
over all monitoring stations and time points, the covariate vectors associated
with each pair $(s_i,t_j)$. Thus, the mode-3 fiber of $\mathcal{X}$ at position
$(i,j)$ is given by $\mathcal{X}_{ij\cdot}=\mathbf{x}_{ij}$, where
\[
\mathbf{x}_{ij}
=
\left(
1,\,
\mathrm{TEMP}^*(s_i,t_j),\,
\mathrm{WSP}^*(s_i,t_j),\,
\mathrm{PRES}^*(s_i,t_j),\,
I_{\mathrm{winter}}(t_j),\,
I_{s_1}(s_i),\ldots,I_{s_7}(s_i)
\right)^\top .
\]

Here, $\mathrm{TEMP}^*$, $\mathrm{WSP}^*$, and $\mathrm{PRES}^*$ denote the
centered meteorological covariates defined in Table~\ref{descstat}. {The set
$
S^*=\{s_1,\ldots,s_7\}
=
\{\text{Dongsi},\text{Guanyuan},\text{Gucheng},\text{Huairou},
\text{Nongzhanguan},\text{Shunyi},\text{Wanliu}\}
$}
contains all monitoring stations. For each station $s_r\in S^*$, $I_{s_r}(s_i)$ is a binary
indicator equal to one when the observation corresponds to station $s_r$, and
zero otherwise. These station indicators account for baseline differences
across monitoring locations. In addition, we include the winter indicator
$I_{\mathrm{winter}}(t_j)$, defined as
\[
I_{\text{winter}}(t_j) =
\begin{cases}
1, & \text{if } t_j \in \{\text{Nov 15 -- Mar 15}\}, \\
0, & \text{otherwise},
\end{cases}
\]
which captures the systematic increase in CO concentrations observed during the winter period. The corresponding coefficient tensor $\mathcal{B}$ is specified so that its
entries are constant over space and time for each fixed effect. That is,
$
\mathcal{B}_{iq0}=\beta_0,
\mathcal{B}_{iq1}=\beta_{TEMP^{*}},
\mathcal{B}_{iq2}=\beta_{WSP^{*}},
\mathcal{B}_{iq3}=\beta_{PRES^{*}},
\mathcal{B}_{iq4}=\beta_{winter},
$
and, for the station-specific indicators,
$
\mathcal{B}_{iq,4+r}=\beta_{s_r},\, r=1,\ldots,7.
$

Thus, in this application, the general tensor regression term
$\left\langle \mathcal{B},\mathcal{X}\right\rangle_3$ corresponds to a
fixed-effects mean structure including meteorological covariates, a winter
effect, and station-specific intercept shifts. The station Changping is used as
the reference location, so the set $S^*$ contains the remaining seven monitoring
stations.

Regarding the covariance structure, we considered four models: DAGAR-AR(1), DAGAR-AR(2), SAR-AR(1), and SAR-AR(2). These models were estimated using three chains of 25,000 iterations each, with a burn-in period of 5000. For model comparison, we employed the expected Akaike, deviance and Bayesian information criteria (EAIC,  DIC and EBIC) to assess goodness of fit. We also compute the expected log predictive density (ELPD) to evaluate predictive performance.

Table \ref{criteriasel} presents the comparison of the different spatiotemporal specifications. The results show a consistent advantage of the DAGAR-based models over their SAR counterparts across all criteria. Both EAIC and EBIC are substantially lower under the DAGAR structure, indicating a better balance between model fit and complexity. A similar conclusion can be drawn from the DIC values, where the differences are large enough to reflect a noticeable improvement in overall fit. Within the DAGAR framework, the AR(1) and AR(2) temporal specifications perform quite similarly. Although the AR(2) version achieves a slightly lower DIC, the AR(1) model yields lower EAIC and EBIC values and also displays stronger predictive performance, as indicated by the highest (least negative) ELPD. Taken together, these findings suggest that the DAGAR–AR(1) specification provides the most balanced and robust performance among the models considered.

\begin{table}[htbp]
  \centering
  \caption{{\bf Air quality dataset}. Model comparison criteria under different correlation structures}\label{criteriasel}
    \begin{tabular}{lrrrr}
    \toprule
          & \multicolumn{1}{l}{EAIC} & \multicolumn{1}{l}{EBIC} & \multicolumn{1}{l}{DIC} & \multicolumn{1}{l}{ELPD} \\
    \midrule
    DAGAR -- AR(1) & \textbf{7292.796} & \textbf{21158.7} & \textbf{5215.576} & \textbf{-375.285} \\
    DAGAR -- AR(2) & 7296.01 & 21321.49 & 5194.885 & -384.140 \\
    SAR -- AR(1) & 7872.318 & 23416.91 & 5543.617 & -522.778 \\
    SAR -- AR(2) & 7891.614 & 23654.99 & 5530.138 & -528.172 \\
    \bottomrule
    \end{tabular}%
  \label{tab:addlabel}%
\end{table}%

Table 5 presents the posterior estimates for the selected DAGAR--AR(1) model. The meteorological variables display clear and meaningful effects on $\log(\mathrm{CO})$ concentrations. In particular, higher temperatures and stronger winds are associated with lower CO levels, which is consistent with improved atmospheric dispersion. Atmospheric pressure also shows a negative association once the other meteorological variables are taken into account. This effect should be interpreted with some caution, as pressure may be capturing broader atmospheric conditions beyond temperature and wind alone.

The station-specific coefficients indicate differences across monitoring locations. Some stations exhibit  higher baseline concentrations relative to Changping, while others show lower levels. These differences likely reflect variations in local emission sources and district-specific characteristics. 

The spatial parameter $\rho$ (0.852) indicates strong similarity between neighboring districts, while the temporal parameter $\gamma$ (0.695) reveals substantial persistence over time. Moreover, the combined spatiotemporal effect $\gamma\rho$ is estimated at 0.592 (95\% credible interval: 0.557--0.627). Numerically, this means that a considerable portion of the past spatial influence carries forward into the present. In practical terms, current CO concentrations are shaped not only by their own history but also by the previous behavior of nearby districts, highlighting the tightly interconnected spatiotemporal dynamics of CO.



\begin{table}[htbp]
  \centering
  \caption{{\bf Air quality dataset}. Estimated DAGAR -- AR(1) model}\label{finalmodel}
    \begin{tabular}{cccc}
    \toprule
    Parameter  & Estimate & \multicolumn{1}{c}{Parameter} & \multicolumn{1}{c}{Estimate} \\
    \midrule
    $\beta_0$ & 6.435 (6.289,6.571) & \multicolumn{1}{c}{$\beta_{s_5}$} & \multicolumn{1}{c}{0.273 (0.179,0.368)} \\
    $\beta_1$ & -0.034 (-0.039,-0.028) & \multicolumn{1}{c}{$\beta_{s_6}$} & \multicolumn{1}{l}{0.085 (0.010,0.162)} \\
    $\beta_2$ & -0.065 (-0.080,-0.049) & \multicolumn{1}{c}{$\beta_{s_7}$} & \multicolumn{1}{l}{0.084 (0.004,0.164)} \\
    $\beta_3$ & -5.868 (-6806,-4964) & \multicolumn{1}{c}{$\sigma^2$} & \multicolumn{1}{c}{0.397 (0.358,0.441)} \\
    $\beta_4$ & 0.449 (0.232,0.673) & \multicolumn{1}{c}{$\rho$} & \multicolumn{1}{c}{0.852 (0.835,0.868)} \\
    $\beta_{s_1}$ & 0.301 (0.220,0.383) & \multicolumn{1}{c}{$\gamma$} & \multicolumn{1}{c}{0.695 (0.659,0.730)} \\
    $\beta_{s_2}$ & 0.242 (0.146,0.343) & \multicolumn{1}{c}{$\gamma\rho$} & \multicolumn{1}{c}{0.592 (0.557,0.627)} \\
    $\beta_{s_3}$ & 0.064 (0.002,0.126) & \multicolumn{1}{c}{$\tau^2$} & \multicolumn{1}{c}{0.103 (0.096,0.109)} \\
    $\beta_{s_4}$ & -0.300 (-0.369,-0.233) & ---     & ---  \\
    \bottomrule
    \end{tabular}%
  \label{tab:addlabel}%
\end{table}%

\begin{figure}
\centering
\subfigure[]{\includegraphics[width=0.4\textwidth]{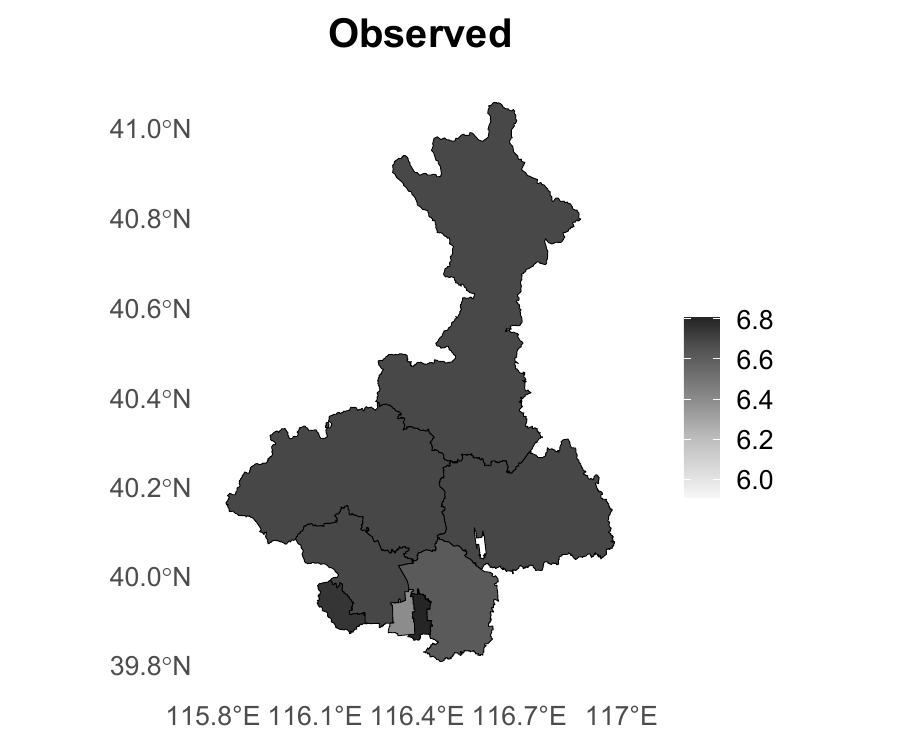}}
\subfigure[]{\includegraphics[width=0.4\textwidth]{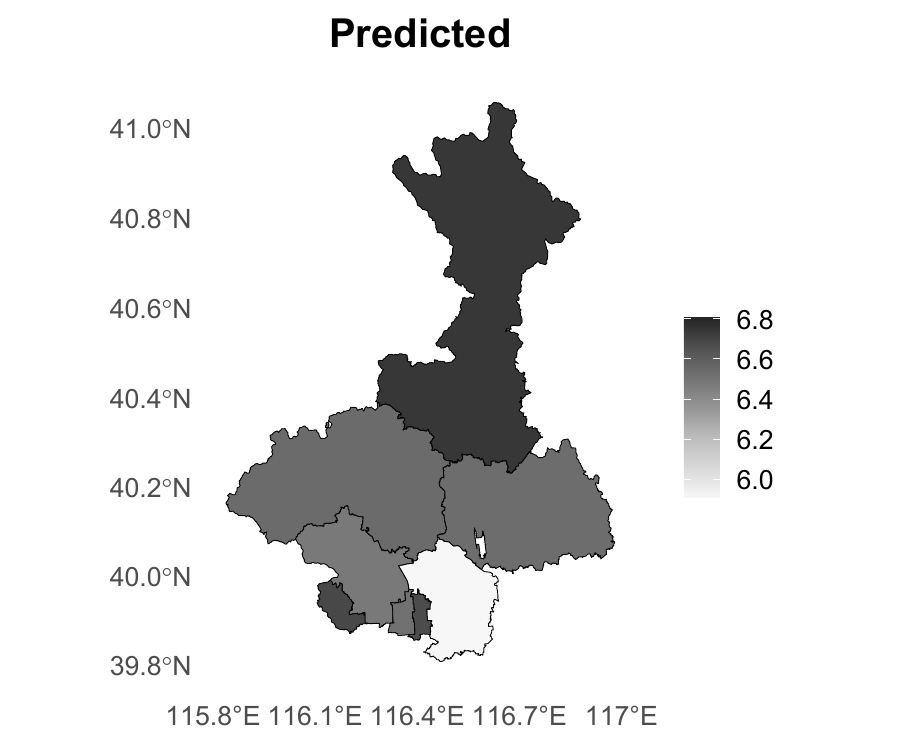}}
\subfigure[]{\includegraphics[width=0.4\textwidth]{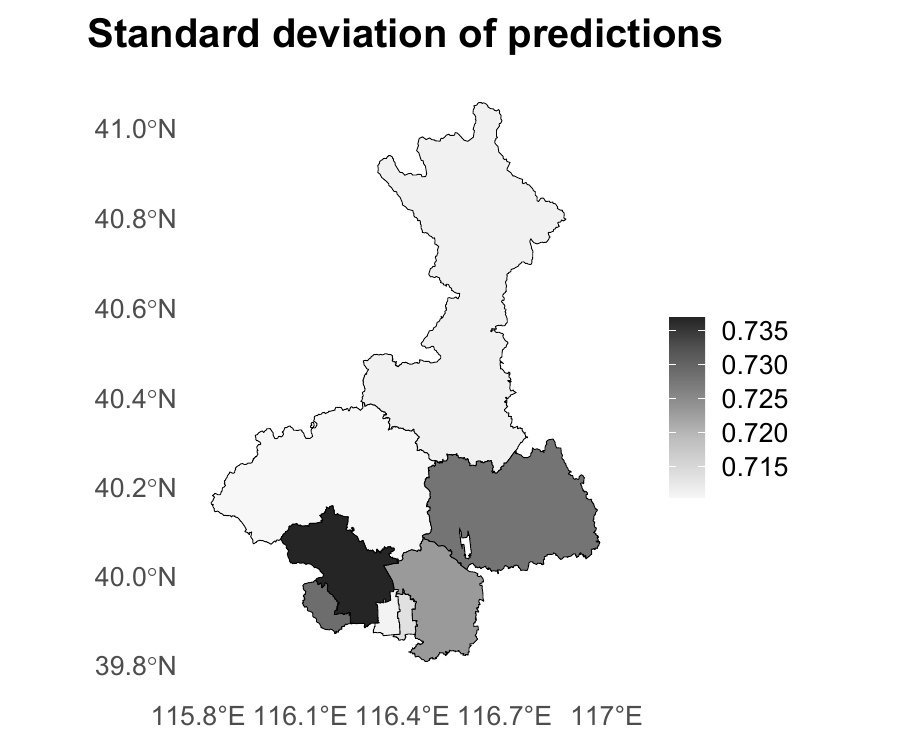}}
\caption{{\bf Air quality dataset}. Prediction for the analyzed districts}\label{predmap}
\end{figure}

\begin{figure}
    \centering
\includegraphics[scale = 0.5]{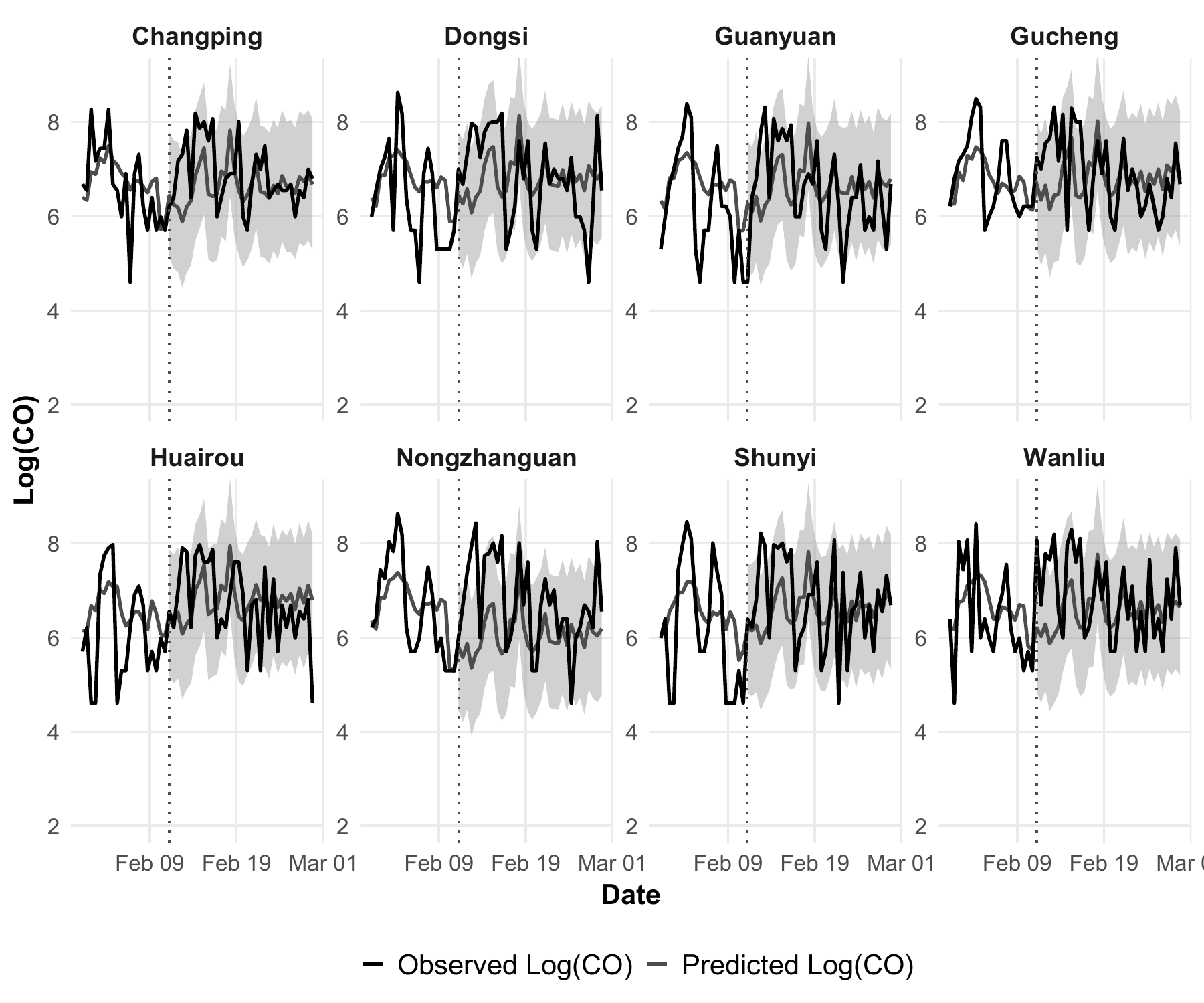}
    \caption{{\bf Air quality dataset}. Observed and predicted $\log(\mathrm{CO})$ concentrations for the DAGAR -- AR(1) model across twelve monitoring stations. The shaded areas represent 95\% predictive intervals.}
    \label{predictiondagarar1}
\end{figure}

Figure \ref{predmap} presents the spatial distribution of the predictions. The predicted averages closely match the observed spatial pattern, preserving the main differences across districts. The uncertainty map indicates moderate variability, with slightly higher uncertainty in some areas. On the other hand, Figure \ref{predictiondagarar1} shows the observed and predicted log(CO) concentrations across stations. The model follows the main temporal patterns closely, capturing the most relevant increases and decreases in CO levels. As expected, uncertainty grows slightly in the forecasting period, but most observations remain within the 95\% predictive intervals. This suggests that the model provides reasonable and well-calibrated predictions. Overall, the model performs consistently across both time and space, generating coherent predictions with realistic uncertainty levels.


\section{Discussion}\label{section8}

In this paper, we introduced a new methodology for analyzing spatiotemporal areal data with censored and missing responses. We propose a novel spatiotemporal random effect that jointly captures temporal autoregressive and spatial DAGAR dependencies, leading to an interpretable model structure in which the parameters can be directly understood as temporal, spatial, and spatiotemporal associations. Although our framework was developed under the multivariate normal assumption, the proposed random effect can readily be extended to non-Gaussian settings, such as count data.

Simulation studies demonstrated that our approach outperforms common ad hoc strategies for handling censored or missing data, such as replacing censored observations with LOD or LOD/2 and imputing missing values with the sample mean. Moreover, the proposed method achieved superior predictive accuracy for new observations compared to standard techniques that rely on partial information. All the codes used in this paper are available on GitHub at \url{https://github.com/joalor93/areal_paper_stan_miss_cens_results}, and can be used to fully replicate the results presented herein.

In the application to the $\log(\mathrm{CO})$ concentrations, the proposed model provided a better overall fit, while also offering a clearer interpretation of the underlying spatiotemporal dynamics. The estimated spatial parameter points to strong similarity across neighboring districts, and the temporal parameter reveals noticeable persistence over time. In addition, the combined spatiotemporal effect shows that current CO levels are shaped not only by their own past values but also by the previous behavior of nearby areas. This highlights the importance of accounting for space and time simultaneously. Overall, the DAGAR specification achieves a meaningful balance between interpretability and predictive performance.

Future research directions encompass extending the model to accommodate multiple variables, each characterized by its distinct spatiotemporal correlation structure (MDAGAR), as well as developing analogous methodologies for non-Gaussian data that maintain the proposed spatiotemporal dependence. Additionally, we aim to investigate non-separable spatiotemporal structures defined over graphs, which will be the primary focus of our forthcoming paper. However, in these instances, computational complexity must be addressed, necessitating the consideration of more efficient approaches as part of ongoing research.

\section*{Acknowledgments}
Jose A. Ordoñez acknowledges the financial support from the Agencia Nacional de Investigaci\'on y Desarrollo ANID, FONDECYT Postdoctorado 3240170. The work of Tsung-I Lin was partially supported by the National Science and Technology
Council of Taiwan under Grant Number NSTC 112-2118-M-005-004-MY3. Luis M. Castro acknowledges the financial support from the Agencia Nacional de Investigaci\'on y Desarrollo ANID, FONDECYT Regular 1260978, and Centros de Investigación y Desarrollo de Excelencia de Interés Nacional [Grant CIN250062].

\section{Appendix}

\subsection{Proof of Proposition 3.1}

Given the correlation matrix in equation (\textcolor{blue}{5}) of the main manuscript, the Durbin Levinson recursion, 
and a fixed position $s_i$, $\boldsymbol{\omega}_{s_i}$ can be written as,

\begin{eqnarray}\label{systemw}
 \epsilon(s_i,t_1) & =& \omega(s_i,t_1),\nonumber\\
\epsilon(s_i,t_2) & = & \omega(s_i,t_2) - \gamma^{(1)}_{1}\,\omega(s_i,t_1), \nonumber\\
\epsilon(s_i,t_3) &= & \omega(s_i,t_3) -\gamma^{(2)}_{1}\,\omega(s_i,t_2) - \gamma^{(2)}_{2}\,\omega(s_i,t_1)  ,\nonumber\\
&\ \vdots & \nonumber \\
\epsilon(s_i,t_p) &=& \omega(s_i,t_p) - \gamma^{(p-1)}_{1}\,\omega(s_i,t_{p-1})
               - \gamma^{(p-1)}_{2}\,\omega(s_i,t_{p-2})
               - \ldots \nonumber \\
               &&- \gamma^{(p-1)}_{p-1}\,\omega(s_i,t_{1})
                 , \nonumber\\
\epsilon(s_i,t_j) &= & \omega(s_i,t_j) - \gamma_{1}\,\omega(s_i,t_{j-1})
               - \gamma_{2}\,\omega(s_i,t_{j-2})
               - \cdots \nonumber \\
              && - \gamma_{p}\,\omega(s_i,t_{j-p}),\,\, j>p. \nonumber
\end{eqnarray}

with  $\epsilon({s_i},t_j) \sim N(0,\nu_t)$, $t>p$, $k = 1, \ldots,p$ and $\nu_t$ as defined in the statement of the proposition. Denote the backward shift matrices as $\mathbf{J}^1, \ldots,\mathbf{J}^p$, with $[\mathbf{J}^{k}]_{(t,t-k)} = 1,\,\, \text{for}\,\,t>k$ and $0$ elsewhere. Then, we can write the process in its vectorial form, 

\begin{equation}\label{autorreg}
\boldsymbol{\epsilon_{s_i}} = (\mathbf{I} - \mathbf{D}_1\mathbf{J}^1 - \ldots \mathbf{D}_p \mathbf{J}^p)\boldsymbol{\omega}_{s_i},     
\end{equation}

where $\mathbf{D}_k$ is a $T \times T$ diagonal matrix with components, 
\[
(\mathbf{D}_k)_{jj} =
\begin{cases}
0, & j \le k, \\[6pt]
\gamma^{(j-1)}_{k}, & k < j \le p, \\[6pt]
\gamma_{k}, & j > p,
\end{cases}
\]

$\boldsymbol{\epsilon}_{s_{i}}  = (\epsilon(s_i,t_1), \ldots, \epsilon(s_i, {t}_T))\sim N(0, \mathbf{D})$ and $\mathbf{D}= \text{diag}(\nu_1, \ldots, \nu_T)$. 
If we set $\mathbf{B}_{T} = \sum_{k = 1}^p \mathbf{D}_k \mathbf{J}^k$, and $\mathbf{F}_T = \text{diag}(\nicefrac{1}{\nu_1},\ldots,\nicefrac{1}{\nu_T})$, then, we can deduce from (\ref{autorreg}), that $\boldsymbol{\omega}_{s_i}$ can be written as a DAG and that 
\[
\begin{aligned}
\boldsymbol{\omega}_{s_i} \sim N_{T}(\mathbf{0}, \left[(\text{\bf I}_{T}- \text{\bf B}_{T})^\top \text{\bf F}_{T}(\text{\bf I}_{T}- \text{\bf B}_{T})\right]^{-1}).
\end{aligned}
\]
Therefore, $\boldsymbol{\omega}_{s_i}$ follows a T-variate normal distribution as settled in Proposition 
3.1 of the main manuscript.  $\qed$

\subsection{Proof of Theorem 
3.1}

From Proposition 3.1 of the main manuscript, 
we can use the representation 

\[\sigma^2
\Phi = \left [(\mathbf{I}_{T} - \mathbf{B}_{T})^{\top}\mathbf{F}_{T}(\mathbf{I}_{T} - \mathbf{B}_{T}) \right]^{-1}.\] 

Also,  given the SAR/DAGAR structure, we can represent the spatial covariance as $\Sigma = \left[(\mathbf{I}_{n} - \mathbf{B}_{n})^{\top}\mathbf{F}_{n}(\mathbf{I}_{n} - \mathbf{B}_{n})\right]$ with $\mathbf{B}_n$ and $\mathbf{F}_n$ defined for both cases in Subsection 
3.1 of the main manuscript. Therefore, by the Kronecker product properties,  

\begin{eqnarray*}
    \Sigma \otimes\sigma^2\Phi &=&   \left[(\mathbf{I}_{n} - \mathbf{B}_{n})^{\top}\mathbf{F}_{n}(\mathbf{I}_{n} - \mathbf{B}_{n})\right]^{-1} \otimes \left [(\mathbf{I}_{T} - \mathbf{B}_{T})^{\top}\mathbf{F}_{T}(\mathbf{I}_{T} - \mathbf{B}_{T}) \right]^{-1}\\
     & = &  \left[ (\mathbf{I}_{n} - \mathbf{B}_{n})^{-1} \otimes (\mathbf{I}_{T} - \mathbf{B}_{T})^{-1} \right] \mathbf{F}^{-1}_{n} \otimes \mathbf{F}^{-1}_{T} \left[ (\mathbf{I}_{n} - \mathbf{B}_{n})^{-1} \otimes (\mathbf{I}_{T} - \mathbf{B}_{T})^{-1} \right]^{\top}\\
     & = & [(\mathbf{I}_{n} - \mathbf{B}_{n}) \otimes (\mathbf{I}_{T} - \mathbf{B}_T)]^{-1} (\mathbf{F}_{n} \otimes \mathbf{F}_{T} )^{-1}  \lbrace [(\mathbf{I}_{n} - \mathbf{B}_{n}) \otimes (\mathbf{I}_{T} - \mathbf{B}_T)]^{-1}\rbrace^{\top}\\
     & = &   \lbrace [(\mathbf{I}_{n} - \mathbf{B}_{n}) \otimes (\mathbf{I}_{T} - \mathbf{B}_T)]^{\top} (\mathbf{F}_{n} \otimes \mathbf{F}_{T} ) [(\mathbf{I}_{n} - \mathbf{B}_{n}) \otimes (\mathbf{I}_{T} - \mathbf{B}_T)]\rbrace^{-1}
\end{eqnarray*}

Now, 

\begin{eqnarray*}
(\mathbf{I}_{n} - \mathbf{B}_{n}) \otimes (\mathbf{I}_{T} - \mathbf{B}_T) &=& (\mathbf{I}_{n} \otimes \mathbf{I}_{T} - \mathbf{I}_{n}\otimes\mathbf{B}_{T} - \mathbf{B}_{n} \otimes \mathbf{I}_T + \mathbf{B}_{n}\otimes \mathbf{B}_{T})\\
& = & \mathbf{I}_{nT} - \mathbf{B}_{nT},
\end{eqnarray*}

where $\mathbf{I}_{nT}$ is the identity matrix of order $nT$ and $\mathbf{B}_{nT} = \mathbf{I}_{n}\otimes\mathbf{B}_{T} + \mathbf{B}_{n} \otimes \mathbf{I}_T - \mathbf{B}_{n}\otimes \mathbf{B}_{T}$. Then, 

\begin{eqnarray*}
    \Sigma \otimes \sigma^2\Phi &=& [(\mathbf{I}_{nT} - \mathbf{B}_{nT})^{\top}\mathbf{F}_{nT}(\mathbf{I}_{nT} - \mathbf{B}_{nT})]^{-1}
\end{eqnarray*}

with 
$\mathbf{F}_{nT} = \mathbf{F}_{n} \otimes \mathbf{F}_{T}$. Then  $\boldsymbol{\omega} \sim N_{n\times T} (\mathbf{0}, [(\mathbf{I}_{nT} - \mathbf{B}_{nT})^{\top}\mathbf{F}_{nT}(\mathbf{I}_{nT} - \mathbf{B}_{nT})]^{-1} )$ and therefore, it can be expressed as a SAR/DAGAR process depending on the definition of $\mathbf{B}_n$, $\mathbf{F}_n$. Note that we can express equations (\textcolor{blue}{10}) and (\textcolor{blue}{11}) of the main manuscript 
 in its matrix form, i.e, 

$$\boldsymbol{\omega} = (\mathbf{I}_{n}\otimes\mathbf{B}_{T})\boldsymbol{\omega} + (\mathbf{B}_{n} \otimes \mathbf{I}_T)\boldsymbol{\omega} - (\mathbf{B}_{n}\otimes \mathbf{B}_{T})\boldsymbol{\omega} + \boldsymbol{\epsilon}\,\,\,\,\,\,\,\,\,\,\,\,\,\,\,\,\,\,\,\,\qed$$

\subsection{Additional Simulation results}

\begin{table}[!htbp]
  \centering
  \caption{{\bf Simulation study I}. DAGAR model. Coverage probabilities of the mean structure parameters considering a level of censoring of $15\%$} \label{supp:tabest15meanpar}%
    \begin{tabular}{ccccc}
    \toprule
    \multicolumn{1}{l}{N} & Parameter & NST-CLG   & LOD   & LOD/2 \\
    \midrule
    \multirow{3}[2]{*}{625} & $\beta_0$ & 0.960  & 0.003  & 0.850 \\
          & $\beta_1$ & 0.930  & 0.000  & 0.220 \\
          & $\beta_2$ & 0.933  & 0.000  & 0.263 \\
    \midrule
    \multirow{3}[2]{*}{1296} & $\beta_0$ & 0.927  & 0.000  & 0.597 \\
          & $\beta_1$ & 0.967  & 0.000  & 0.033 \\
          & $\beta_2$ & 0.963  & 0.000  & 0.000 \\
    \midrule
    \multirow{3}[2]{*}{2401} & $\beta_0$ & 0.927  & 0.000  & 0.613 \\
          & $\beta_1$ & 0.947  & 0.000  & 0.003 \\
          & $\beta_2$ & 0.940  & 0.000  & 0.000 \\
    \bottomrule
    \end{tabular}%
\end{table}%

\begin{table}[!htbp]
  \centering
  \caption{{\bf Simulation study I}. DAGAR model. Coverage probabilities of the mean structure parameters considering a level of censoring of $35\%$}
    \begin{tabular}{ccccc}
    \toprule
    \multicolumn{1}{l}{N} & Parameter & \multicolumn{1}{l}{NST-CLG} & \multicolumn{1}{l}{LOD} & \multicolumn{1}{l}{LOD/2} \\
    \midrule
    \multirow{3}[2]{*}{625} & $\beta_0$ & 0.950 & 0.000 & 1.000 \\
          & $\beta_1$ & 0.940 & 0.000 & 0.330 \\
          & $\beta_2$ & 0.970 & 0.000 & 0.740 \\ 
    \midrule
    \multirow{3}[2]{*}{1296} & $\beta_0$ & 0.920 & 0.000 & 0.880 \\
          & $\beta_1$ & 0.940 & 0.000 & 0.000 \\ 
          & $\beta_2$ & 0.940 & 0.000 & 0.000 \\ 
    \midrule
    \multirow{3}[2]{*}{2401} & $\beta_0$ & 0.960 & 0.000 & 1.000 \\ 
          & $\beta_1$ &  0.940 & 0.000 & 0.260 \\ 
          & $\beta_2$ & 0.937 & 0.000 & 0.050 \\ 
    \bottomrule
    \end{tabular}%
  \label{supp:tabest35meanpar}%
\end{table}%

\begin{table}[!htbp]
  \centering
  \caption{{\bf Simulation study I}. DAGAR model. Coverage probabilities of the covariance structure parameters considering a level of censoring of $35\%$}
    \begin{tabular}{ccccc}
    \toprule
    \multicolumn{1}{l}{N} & Parameter & NST-CLG   & LOD   & LOD/2 \\
    \midrule
    \multirow{4}[2]{*}{625} & $\sigma^2$ & 0.953 & 0.337 & 0.973 \\
          & $\rho$   & 0.960 & 0.877 & 0.927 \\ 
          & $\tau^2$  & 0.970 & 0.000 & 0.000 \\ 
          & $\phi$   & 0.943 & 0.970 & 0.970 \\ 
    \midrule
    \multirow{4}[2]{*}{1296} & $\sigma^2$ & 0.953 & 0.070 & 0.950 \\ 
          & $\rho$   & 0.950 & 0.957 & 0.967 \\ 
          & $\tau^2$  & 0.963 & 0.000 & 0.000 \\ 
          & $\phi$   & 0.940 & 0.883 & 0.937 \\ 
    \midrule
    \multirow{4}[2]{*}{2401} & $\sigma^2$ & 0.960 & 0.013 & 0.950 \\
          & $\rho$   & 0.977 & 0.940 & 0.943 \\ 
          & $\tau^2$  & 0.963 & 0.000 & 0.000 \\ 
          & $\phi$   & 0.937 & 0.953 & 0.943 \\ 
    \bottomrule
    \end{tabular}%
  \label{supp:tabest35covpar}%
\end{table}%

\begin{table}[!htbp]
  \centering
  \caption{{\bf Simulation study}. SAR model. Coverage probabilities of the mean structure parameters considering a level of censoring of $15\%$}
    \begin{tabular}{ccccc}
    \toprule
    \multicolumn{1}{l}{N} & Parameter & \multicolumn{1}{l}{NST-CLG} & \multicolumn{1}{l}{LOD} & \multicolumn{1}{l}{LOD/2} \\
    \midrule
    \multirow{3}[2]{*}{625} & $\beta_0$ &0.937 & 0.000 & 0.693 \\ 
          & $\beta_1$ & 0.963 & 0.000 & 0.110 \\ 
          & $\beta_2$  & 0.940 & 0.000 & 0.040 \\ 
    \midrule
    \multirow{3}[2]{*}{1296} & $\beta_0$ &  0.933 & 0.000 & 0.503 \\
          & $\beta_1$ & 0.943 & 0.000 & 0.017 \\ 
          & $\beta_2$ & 0.947 & 0.000 & 0.007 \\ 
    \midrule
    \multirow{3}[2]{*}{2401} & $\beta_0$ & 0.917 & 0.000 & 0.287 \\ 
          & $\beta_1$ & 0.940 & 0.000 & 0.000 \\ 
          & $\beta_2$ & 0.950 & 0.000 & 0.000 \\
    \bottomrule
    \end{tabular}%
  \label{supp:tabest15meanparsar}%
\end{table}%

\begin{table}[!htbp]
  \centering
  \caption{{\bf Simulation study I}. SAR model. Coverage probabilities of the mean structure parameters considering a level of censoring of $35\%$}
    \begin{tabular}{ccccc}
    \toprule
    \multicolumn{1}{l}{N} & Parameter & \multicolumn{1}{l}{NST-CLG} & \multicolumn{1}{l}{LOD} & \multicolumn{1}{l}{LOD/2} \\
    \midrule
    \multirow{3}[2]{*}{625} & $\beta_0$ & 0.927 & 0.000 & 0.997 \\ 
          & $\beta_1$ &  0.957 & 0.000 & 0.613 \\ 
          & $\beta_2$  & 0.940 & 0.000 & 0.373 \\ 
    \midrule
    \multirow{3}[2]{*}{1296} & $\beta_0$ &0.940 & 0.000 & 0.937 \\ 
          & $\beta_1$ &0.957 & 0.000 & 0.097 \\
          & $\beta_2$ & 0.930 & 0.000 & 0.013 \\ 
    \midrule
    \multirow{3}[2]{*}{2401} & $\beta_0$ & 0.920 & 0.000 & 0.270 \\ 
          & $\beta_1$ & 0.953 & 0.000 & 0.000 \\ 
          & $\beta_2$ & 0.923 & 0.000 & 0.000 \\ 
    \bottomrule
    \end{tabular}%
  \label{supp:tabest35meanparsar}%
\end{table}%

\begin{table}[!htbp]
  \centering
  \caption{{\bf Simulation study I}. SAR model. Coverage probabilities of the covariance structure parameters considering a level of censoring of $15\%$}
    \begin{tabular}{ccccc}
    \toprule
    \multicolumn{1}{l}{N} & Parameter & NST-CLG   & LOD   & LOD/2 \\
    \midrule
    \multirow{4}[2]{*}{625} & $\sigma^2$ & 0.963 & 0.760 & 0.957 \\ 
          & $\rho$   & 0.967 & 0.917 & 0.910 \\ 
          & $\tau^2$  & 0.947 & 0.000 & 0.000 \\ 
          & $\phi$   & 0.970 & 0.933 & 0.917 \\
    \midrule
    \multirow{4}[2]{*}{1296} & $\sigma^2$ & 0.947 & 0.467 & 0.927 \\ 
          & $\rho$   & 0.957 & 0.933 & 0.937 \\ 
          & $\tau^2$  &0.943 & 0.000 & 0.000 \\ 
          & $\phi$   & 0.960 & 0.900 & 0.930 \\ 
    \midrule
    \multirow{4}[2]{*}{2401} & $\sigma^2$ & 0.963 & 0.217 & 0.913 \\ 
          & $\rho$   & 0.950 & 0.913 & 0.910 \\ 
          & $\tau^2$  & 0.960 & 0.000 & 0.000 \\ 
          & $\phi$   & 0.963 & 0.933 & 0.933 \\ 
    \bottomrule
    \end{tabular}%
  \label{supp:tabest15covparsar}%
\end{table}%

\begin{table}[!htbp]
  \centering
  \caption{{\bf Simulation study I}. SAR model. Coverage probabilities of the covariance structure parameters considering a level of censoring of $35\%$}
    \begin{tabular}{ccccc}
    \toprule
    \multicolumn{1}{l}{N} & Parameter & NST-CLG   & LOD   & LOD/2 \\
    \midrule
    \multirow{4}[2]{*}{625} & $\sigma^2$ &  0.940 & 0.713 & 0.963 \\ 
          & $\rho$  & 0.963 & 0.793 & 0.873 \\
          & $\tau^2$  & 0.907 & 0.000 & 0.000 \\
          & $\phi$   & 0.970 & 0.843 & 0.863 \\ 
    \midrule
    \multirow{4}[2]{*}{1296} & $\sigma^2$ & 0.943 & 0.137 & 0.930 \\ 
          & $\rho$   & 0.957 & 0.900 & 0.920 \\
          & $\tau^2$  & 0.930 & 0.000 & 0.000 \\
          & $\phi$   & 0.973 & 0.917 & 0.913 \\ 
    \midrule
    \multirow{4}[2]{*}{2401} & $\sigma^2$  & 0.937 & 0.010 & 0.870 \\ 
          & $\rho$   & 0.923 & 0.940 & 0.923 \\  
          & $\tau^2$  & 0.967 & 0.000 & 0.000 \\
          & $\phi$   & 0.943 & 0.953 & 0.913 \\ 
    \bottomrule
    \end{tabular}%
  \label{supp:tabest35covparsar}%
\end{table}%

\begin{table}[!htbp]
  \centering
  \caption{{\bf Simulation study II}. DAGAR model. Average coverage probabilities of credible intervals varying the number of predicted observations, considering $N = 625$.}
    \begin{tabular}{ccccc}
\cmidrule{3-5}    \multicolumn{1}{r}{} &       & \multicolumn{3}{c}{Predicted time observations } \\
    \midrule
    Censoring/missignness & Method & One   & Three  & Seven  \\
    \midrule
    \multirow{3}[6]{*}{\raisebox{1\normalbaselineskip}[0pt][0pt]{20\%/5\%}} & MTN   & 0.953 & 0.953 & 0.947 \\ 
         & LOD   &  0.987 & 0.977 & 0.973 \\ 
         & LOD/2 &  0.997 & 0.993 & 0.990 \\
    \midrule
    \multirow{3}[6]{*}{\raisebox{1\normalbaselineskip}[0pt][0pt]{40\%/5\%}} & MTN   & 0.953 & 0.950 & 0.950 \\ 
        & LOD   & 0.950 & 0.930 & 0.913 \\ 
         & LOD/2 & 1.000 & 1.000 & 1.000 \\ 
    \bottomrule
    \end{tabular}%
  \label{supp:tabsim2250}%
\end{table}%

\begin{table}[!htbp]
  \centering
  \caption{{\bf Simulation study II}. DAGAR model. Average coverage probabilities of credible intervals varying the number of predicted observations, considering $N = 2401$.}
    \begin{tabular}{ccccc}
\cmidrule{3-5}    \multicolumn{1}{r}{} &       & \multicolumn{3}{c}{Predicted time observations } \\
    \midrule
    Censoring/missignness & Method & One   & Three  & Seven  \\
    \midrule
    \multirow{3}[6]{*}{\raisebox{1\normalbaselineskip}[0pt][0pt]{20\%/5\%}} & MTN   & 0.950 & 0.943 & 0.943 \\ 
         & LOD   & 0.973 & 0.967 & 0.967 \\ 
         & LOD/2 & 0.997 & 0.993 & 0.990 \\ 
    \midrule
    \multirow{3}[6]{*}{\raisebox{1\normalbaselineskip}[0pt][0pt]{40\%/5\%}} & MTN   & 0.966 & 0.973  & 0.973 \\
        & LOD   & 0.936 & 0.923  & 0.923 \\
         & LOD/2 & 1.000 & 1.000 & 1.000 \\
    \bottomrule
    \end{tabular}%
  \label{supp:tabsim2750}%
\end{table}%

\begin{table}[!htbp]
  \centering
  \caption{{\bf Simulation study II}. SAR model. Average coverage probabilities of credible intervals varying the number of predicted observations, considering $N = 625$.}
    \begin{tabular}{ccccc}
\cmidrule{3-5}    \multicolumn{1}{r}{} &       & \multicolumn{3}{c}{Predicted time observations } \\
    \midrule
    Censoring/missignness & Method & One   & Three  & Seven  \\
    \midrule
    \multirow{3}[6]{*}{\raisebox{1\normalbaselineskip}[0pt][0pt]{20\%/5\%}} & MTN   & 0.950 & 0.953 & 0.953 \\ 
         & LOD   & 0.967 & 0.960 & 0.950 \\ 
         & LOD/2 & 0.987 & 0.980 & 0.973 \\ 
    \midrule
    \multirow{3}[6]{*}{\raisebox{1\normalbaselineskip}[0pt][0pt]{40\%/5\%}} & MTN   &  0.950 & 0.963 & 0.963 \\ 
        & LOD   & 0.930 & 0.903 & 0.897 \\ 
         & LOD/2 & 0.997 & 0.997 & 0.993 \\ 
    \bottomrule
    \end{tabular}%
  \label{supp:tabsim2250sar}%
\end{table}%

\begin{table}[!htbp]
  \centering
  \caption{{\bf Simulation study II}. SAR model. Average coverage probabilities of credible intervals varying the number of predicted observations, considering $N = 1296$.}
    \begin{tabular}{ccccc}
\cmidrule{3-5}    \multicolumn{1}{r}{} &       & \multicolumn{3}{c}{Predicted time observations } \\
    \midrule
    Censoring/missignness & Method & One   & Three  & Seven  \\
    \midrule
    \multirow{3}[6]{*}{\raisebox{1\normalbaselineskip}[0pt][0pt]{20\%/5\%}} & MTN   & 0.967 & 0.967 & 0.967 \\ 
         & LOD   &  0.957 & 0.947 & 0.943 \\ 
         & LOD/2 & 0.990 & 0.983 & 0.983 \\
    \midrule
    \multirow{3}[6]{*}{\raisebox{1\normalbaselineskip}[0pt][0pt]{40\%/5\%}} & MTN   & 0.963 & 0.963 & 0.963 \\ 
        & LOD   & 0.807 & 0.833 & 0.857 \\ 
         & LOD/2 & 0.997 & 0.993 & 0.993 \\ 
    \bottomrule
    \end{tabular}%
  \label{supp:tabsim2500sar}%
\end{table}%

\begin{table}[!htbp]
  \centering
  \caption{{\bf Simulation study II}. SAR model. Average coverage probabilities of credible intervals varying the number of predicted observations, considering $N = 2401$.}
    \begin{tabular}{ccccc}
\cmidrule{3-5}    \multicolumn{1}{r}{} &       & \multicolumn{3}{c}{Predicted time observations } \\
    \midrule
    Censoring/missignness & Method & One   & Three  & Seven  \\
    \midrule
    \multirow{3}[6]{*}{\raisebox{1\normalbaselineskip}[0pt][0pt]{20\%/5\%}} & MTN   & 0.967 & 0.967 & 0.967 \\ 
         & LOD   &  0.957 & 0.947 & 0.943 \\
         & LOD/2 & 0.990 & 0.983 & 0.983 \\ 
    \midrule
    \multirow{3}[6]{*}{\raisebox{1\normalbaselineskip}[0pt][0pt]{40\%/5\%}} & MTN   & 0.957 & 0.953 & 0.957 \\
        & LOD  & 0.840 & 0.900 & 0.913 \\ 
         & LOD/2 & 1.000 & 1.000 & 1.000 \\
    \bottomrule
    \end{tabular}%
  \label{supp:tabsim2750sar}%
\end{table}%

\begin{figure}[!htbp]
\centering
\subfigure[]{\includegraphics[width=0.45\textwidth]{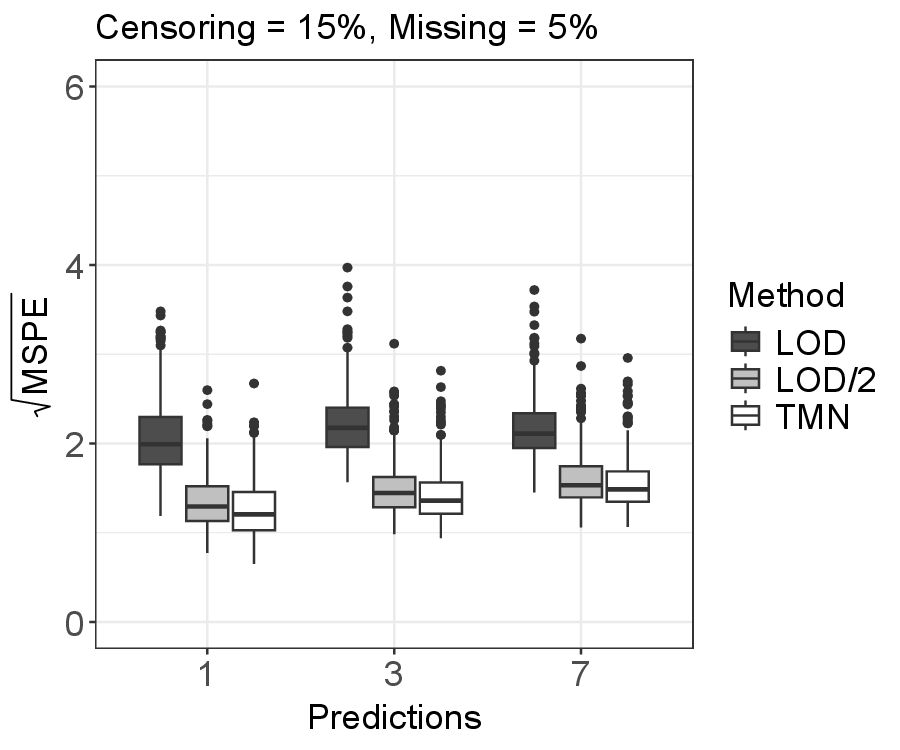}}, 
\subfigure[]{\includegraphics[width=0.45\textwidth]{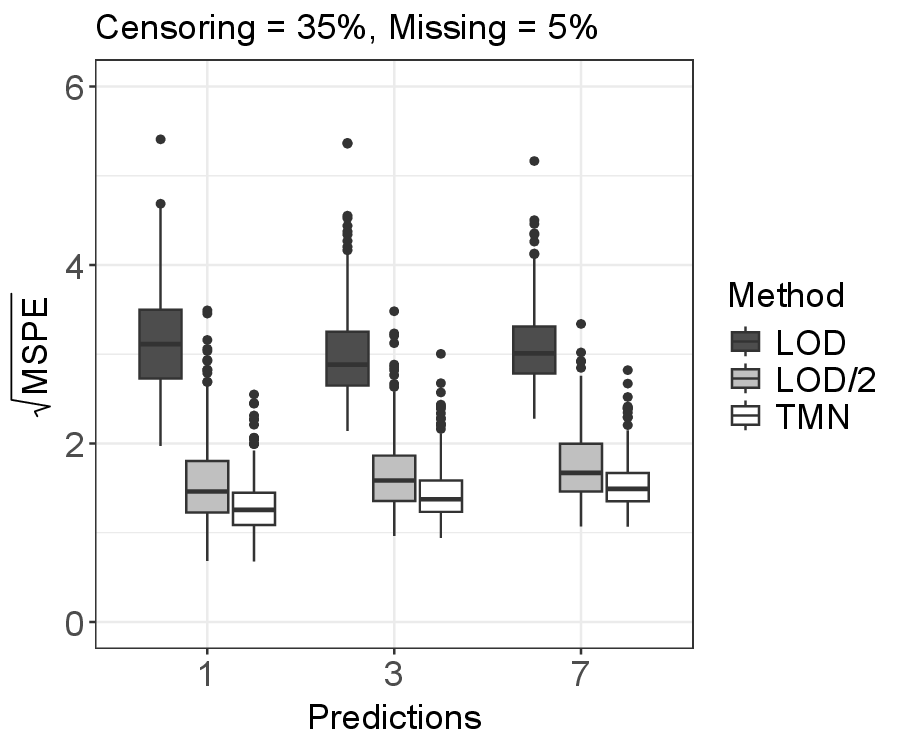}}, 
\subfigure[]{\includegraphics[width=0.45\textwidth]{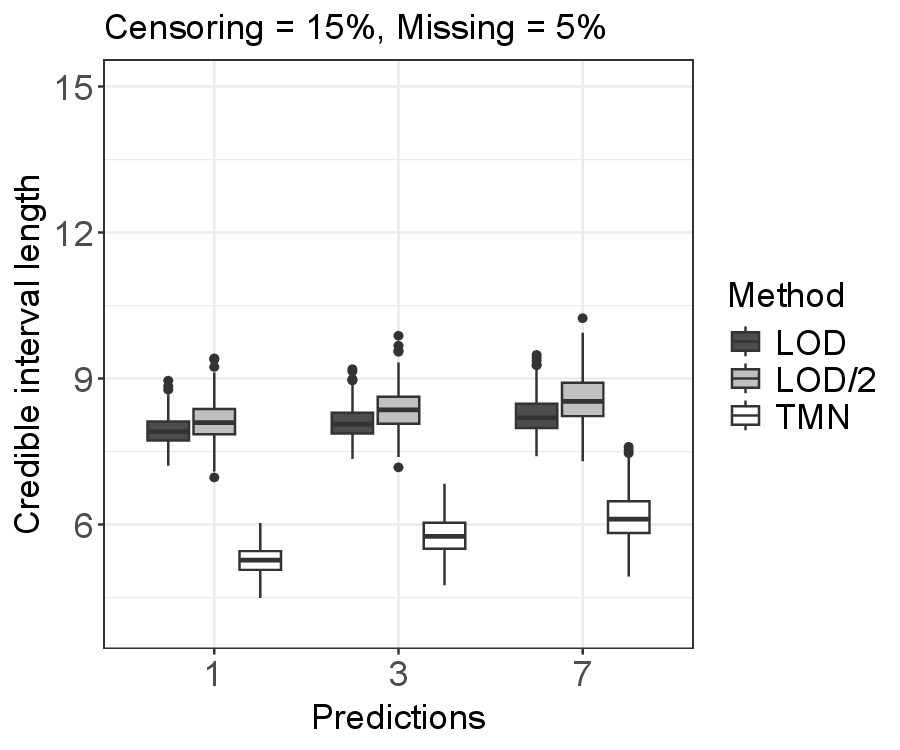}}
\subfigure[]{\includegraphics[width=0.45\textwidth]{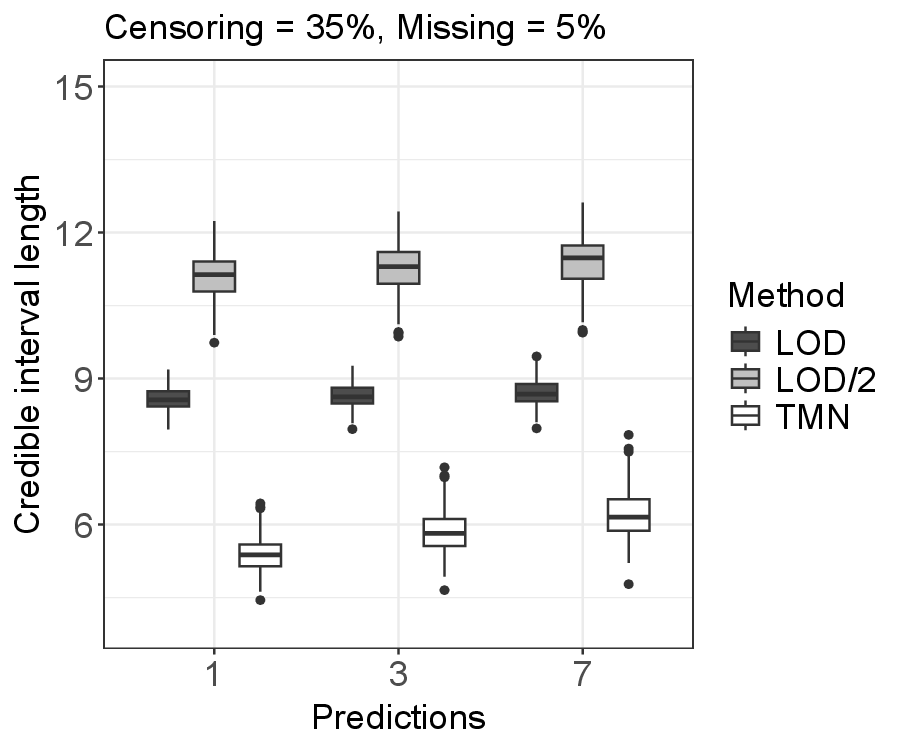}}
\caption{{\bf Simulation study II}. DAGAR model. Comparison of our proposal (NST-CLG) with methods that impute the censored observations by using the limit of detection. For this scenario, we consider $N =625$.}
\label{supp:figsim2250}
\end{figure}

\begin{figure}[!htbp]
\centering
\subfigure[]{\includegraphics[width=0.45\textwidth]{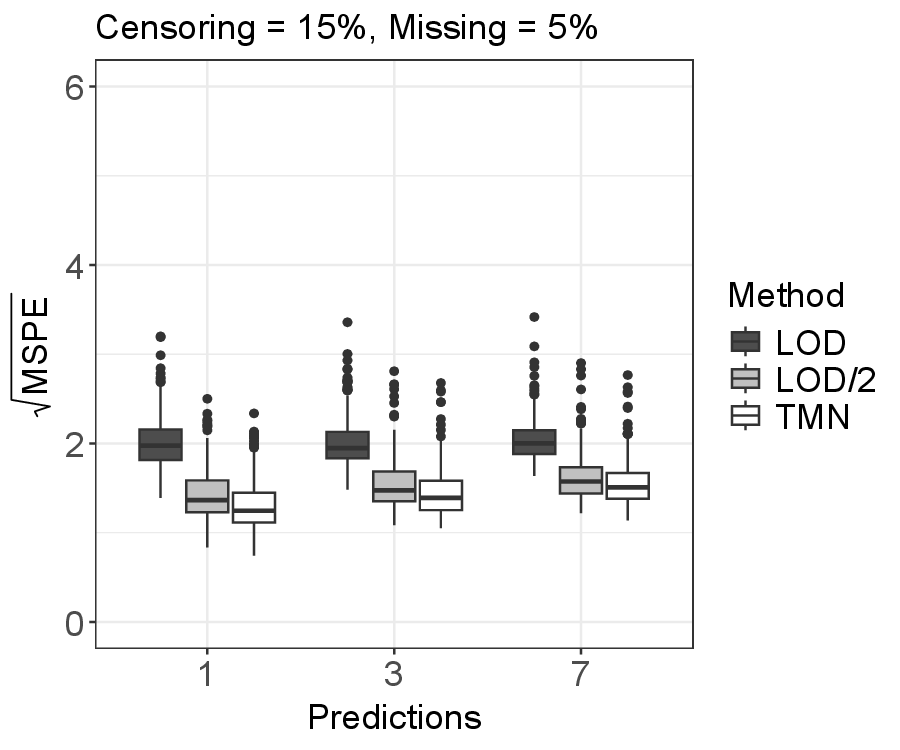}}, 
\subfigure[]{\includegraphics[width=0.45\textwidth]{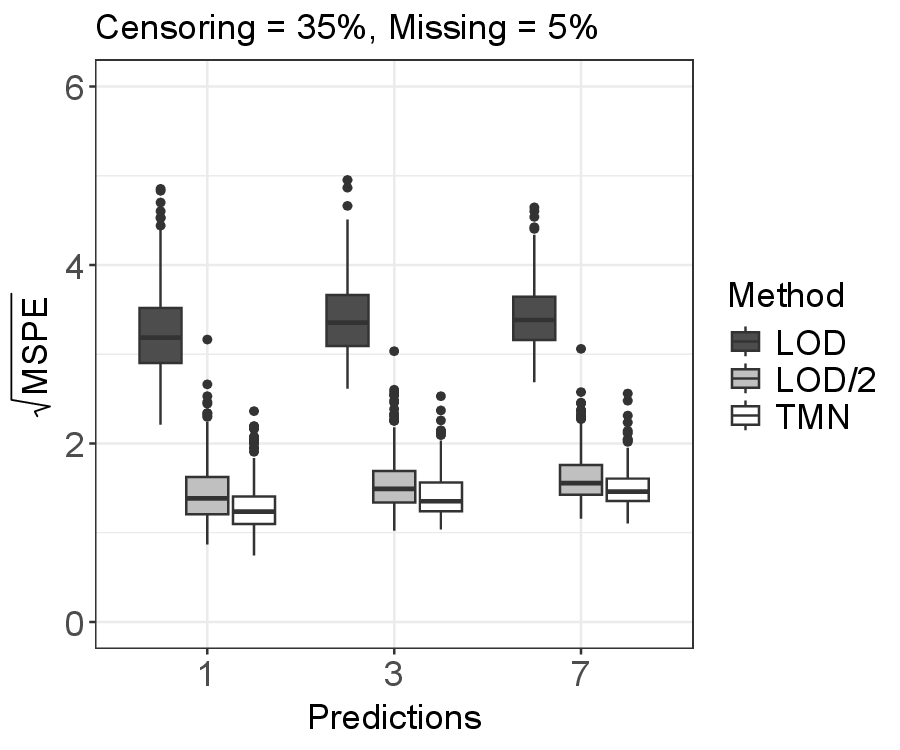}}, 
\subfigure[]{\includegraphics[width=0.45\textwidth]{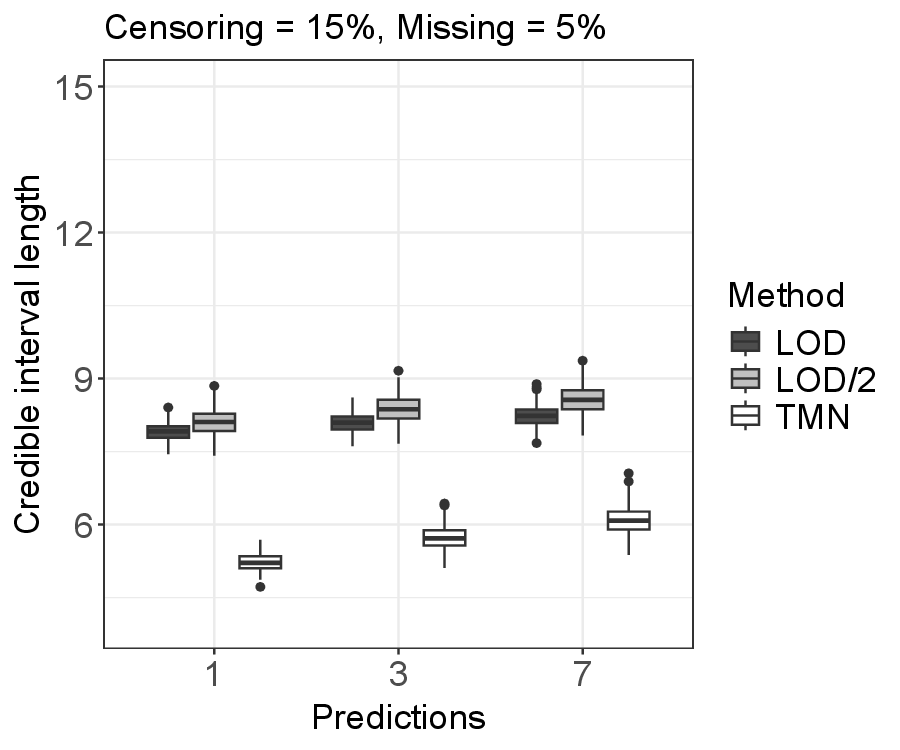}}
\subfigure[]{\includegraphics[width=0.45\textwidth]{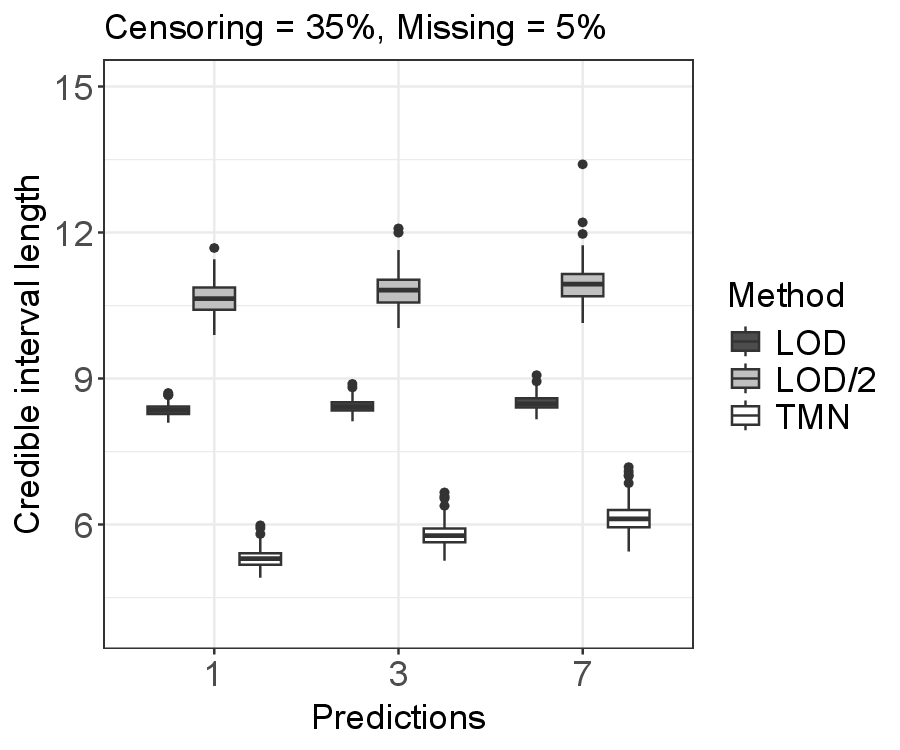}}
\caption{{\bf Simulation study II}. DAGAR model. Comparison of our proposal (NST-CLG) with methods that impute the censored observations by using the limit of detection. For this scenario, we consider $N =2401$.}
\label{supp:figsim2750}
\end{figure}

\begin{figure}[!htbp]
\centering
\subfigure[]{\includegraphics[width=0.45\textwidth]{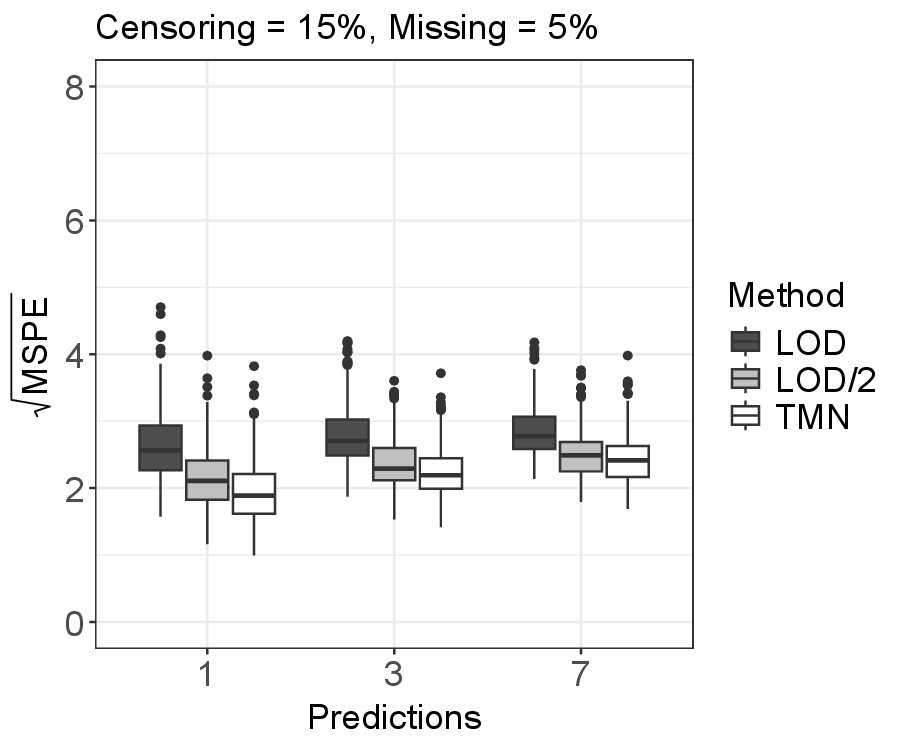}}, 
\subfigure[]{\includegraphics[width=0.45\textwidth]{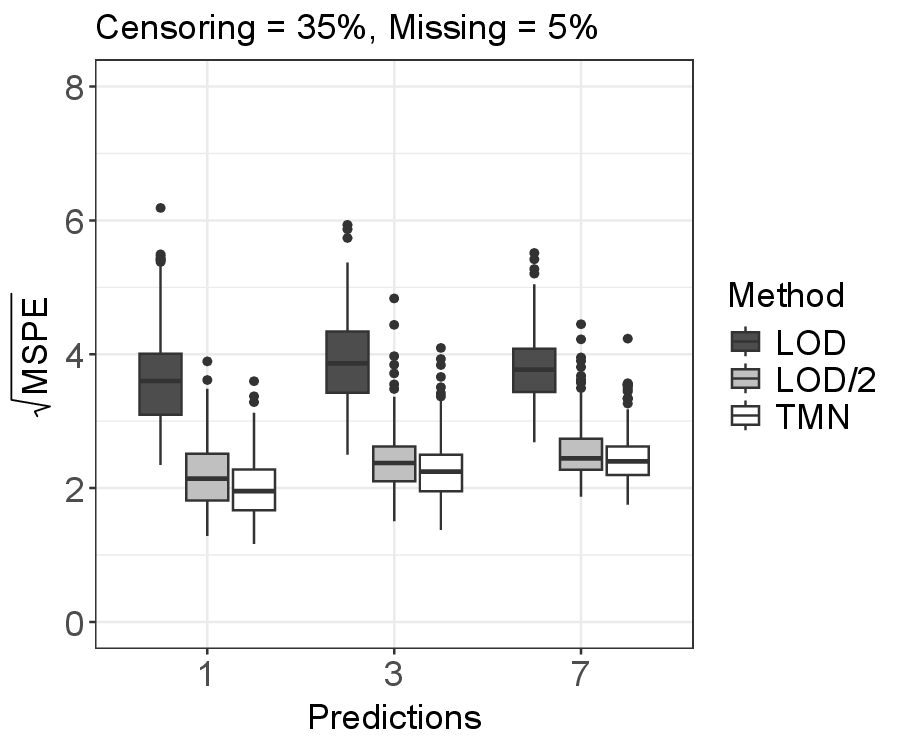}}, 
\subfigure[]{\includegraphics[width=0.45\textwidth]{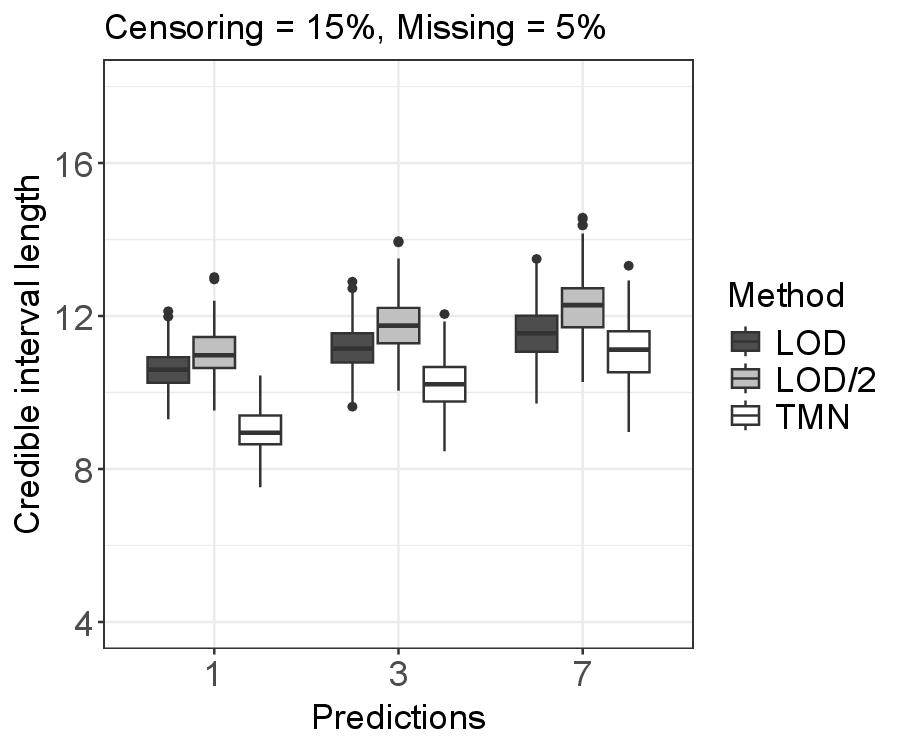}}
\subfigure[]{\includegraphics[width=0.45\textwidth]{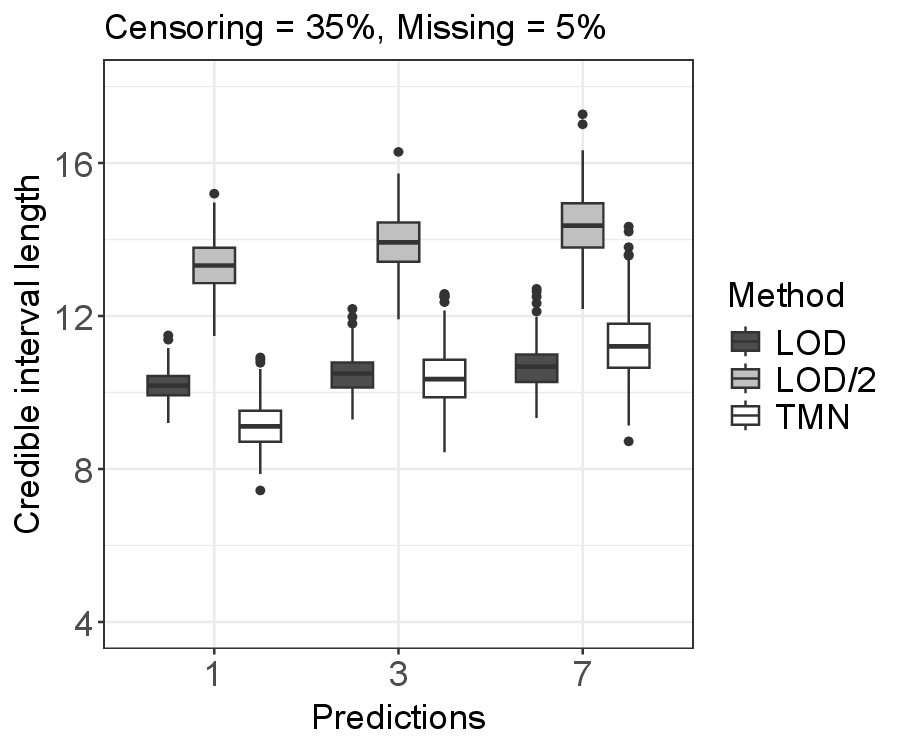}}
\caption{{\bf Simulation study II}. SAR model. Comparison of our proposal (NST-CLG) with methods that impute the censored observations by using the limit of detection. For this scenario, we consider $N =625$.}
\label{supp:figsim2250sar}
\end{figure}

\begin{figure}[!htbp]
\centering
\subfigure[]{\includegraphics[width=0.45\textwidth]{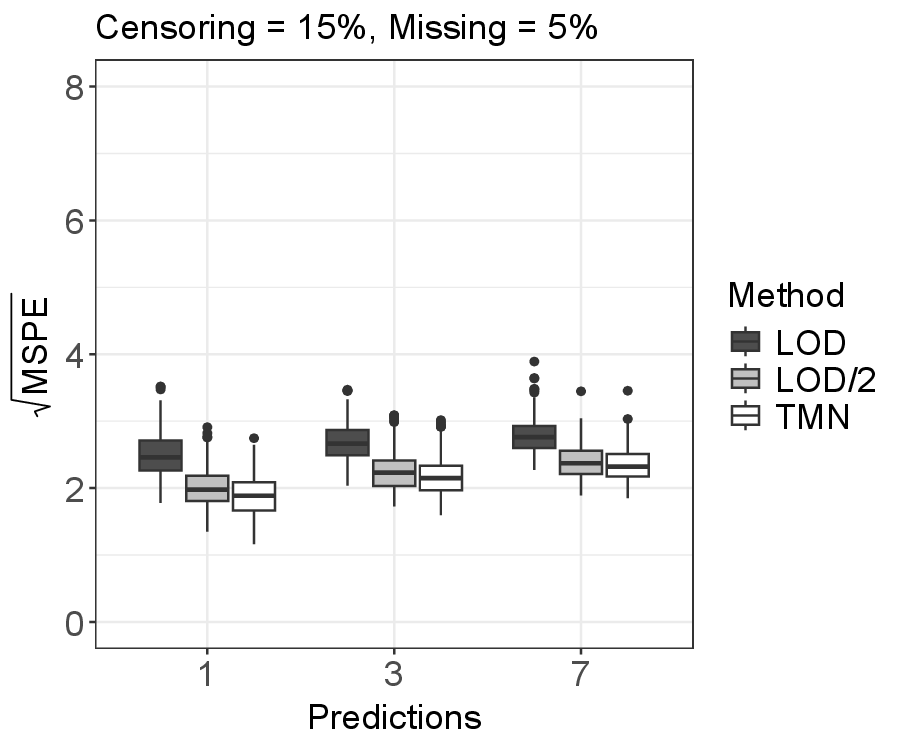}}, 
\subfigure[]{\includegraphics[width=0.45\textwidth]{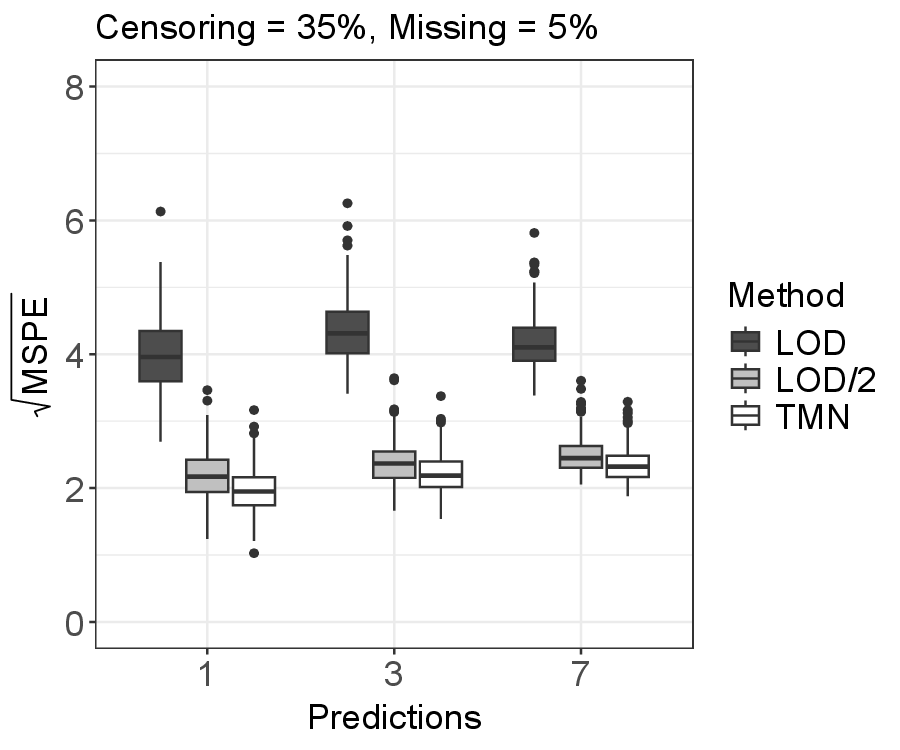}}, 
\subfigure[]{\includegraphics[width=0.45\textwidth]{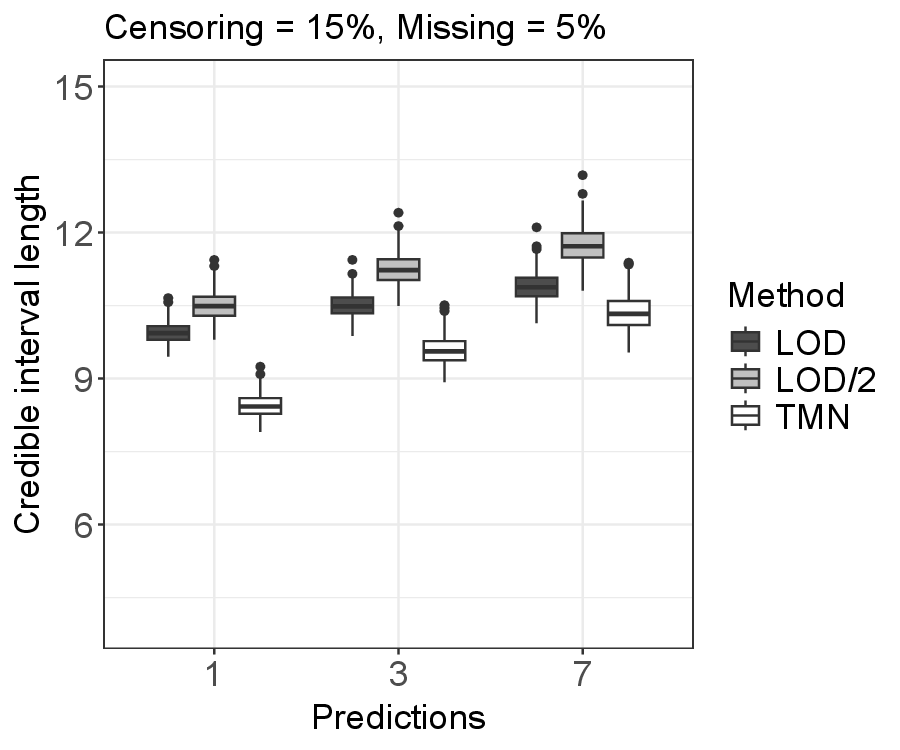}}
\subfigure[]{\includegraphics[width=0.45\textwidth]{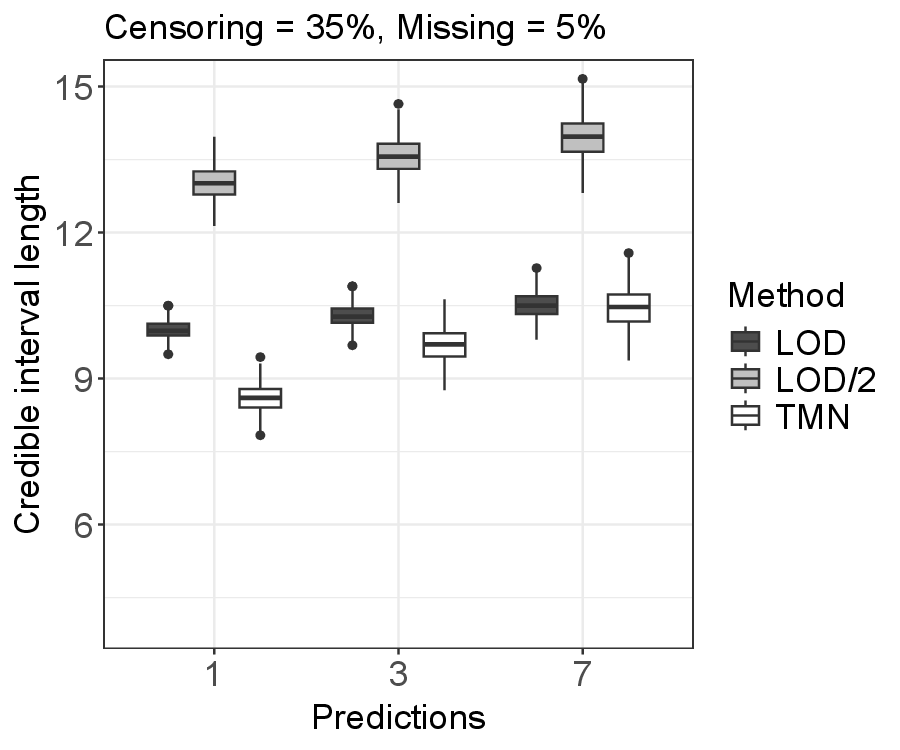}}
\caption{{\bf Simulation study II}. SAR model. Comparison of our proposal (NST-CLG) with methods that impute the censored observations by using the limit of detection. For this scenario, we consider $N =1296$.}
\label{supp:figsim2500sar}
\end{figure}

\begin{figure}[!htbp]
\centering
\subfigure[]{\includegraphics[width=0.45\textwidth]{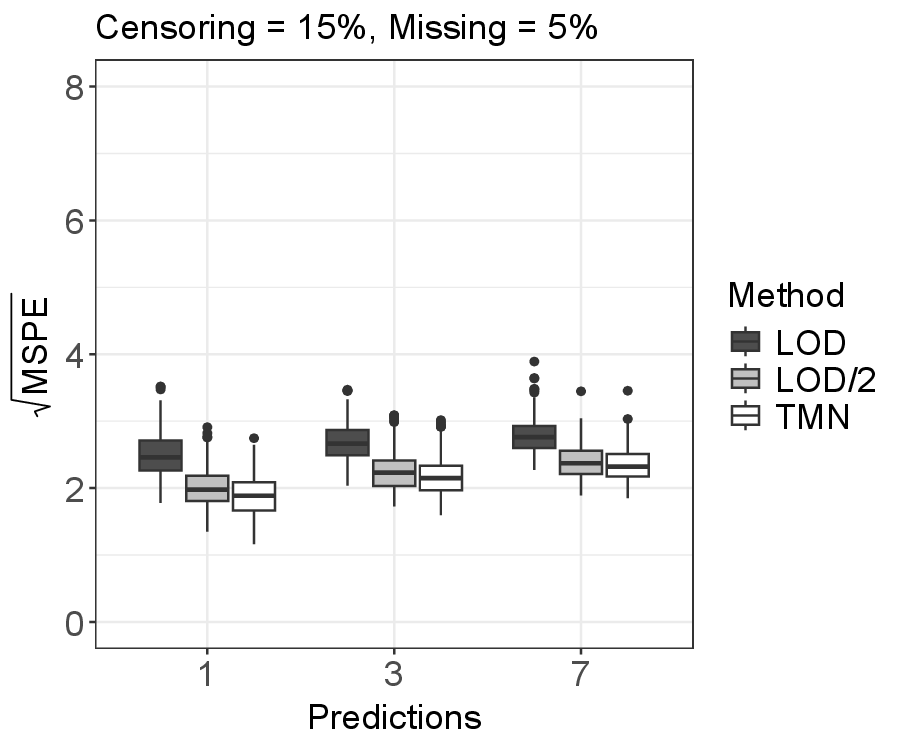}}, 
\subfigure[]{\includegraphics[width=0.45\textwidth]{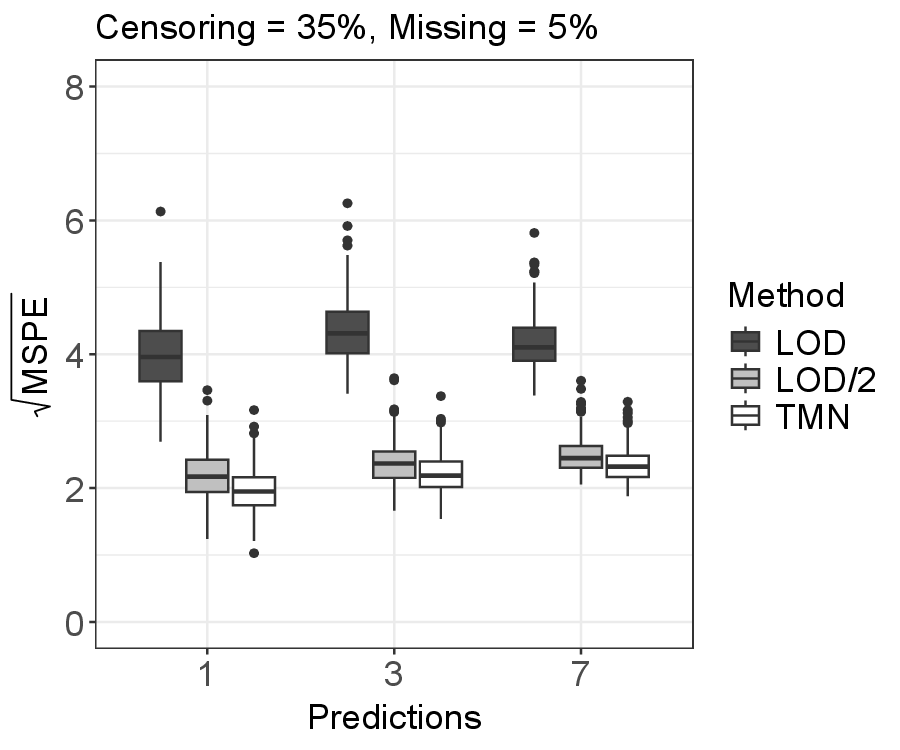}}, 
\subfigure[]{\includegraphics[width=0.45\textwidth]{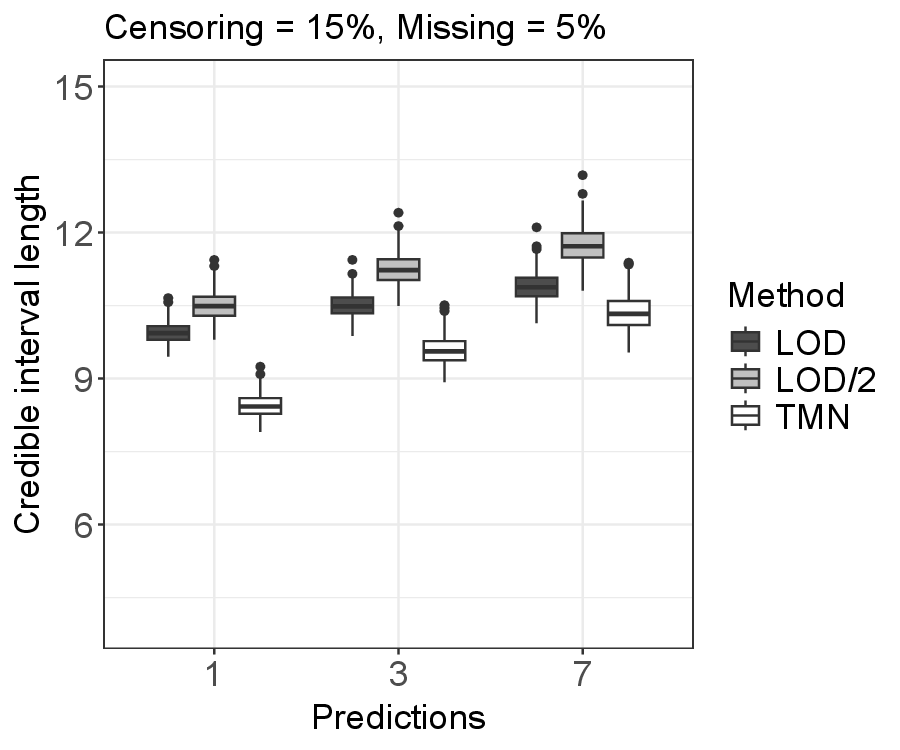}}
\subfigure[]{\includegraphics[width=0.45\textwidth]{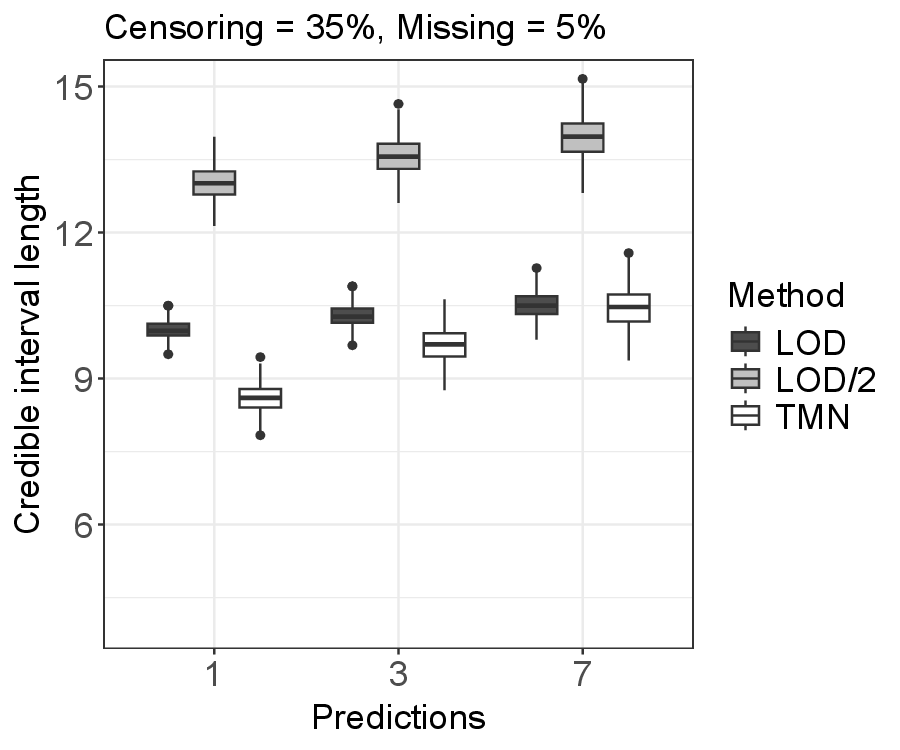}}
\caption{{\bf Simulation study II}. SAR model. Comparison of our proposal (NST-CLG) with methods that impute the censored observations by using the limit of detection. For this scenario, we consider $N =2401$.}
\label{supp:figsim2750sar}
\end{figure}

\clearpage

\begin{samepage}
\subsection{Air quality dataset: Additional results}
\vspace{4cm}
\begin{center}
\includegraphics[width=0.7\linewidth]{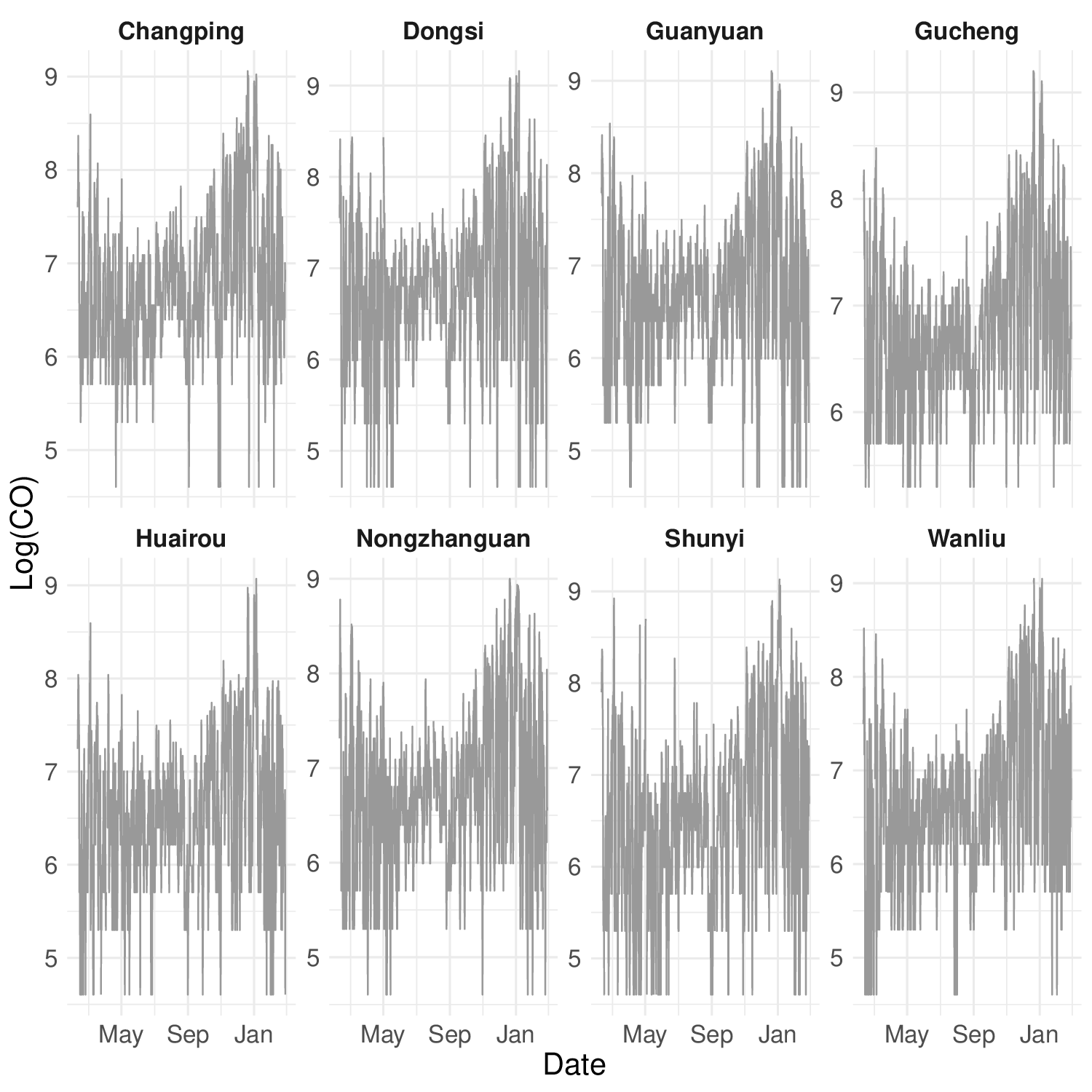}
\captionof{figure}{{\bf Air quality dataset} Time series of CO concentrations 
(on the log scale) at the monitoring stations from February 
2016 to February 2017.}
\label{concentrations2}
\end{center}
\end{samepage}

\begin{figure}[!htbp]
    \centering
\includegraphics[width=1\linewidth]{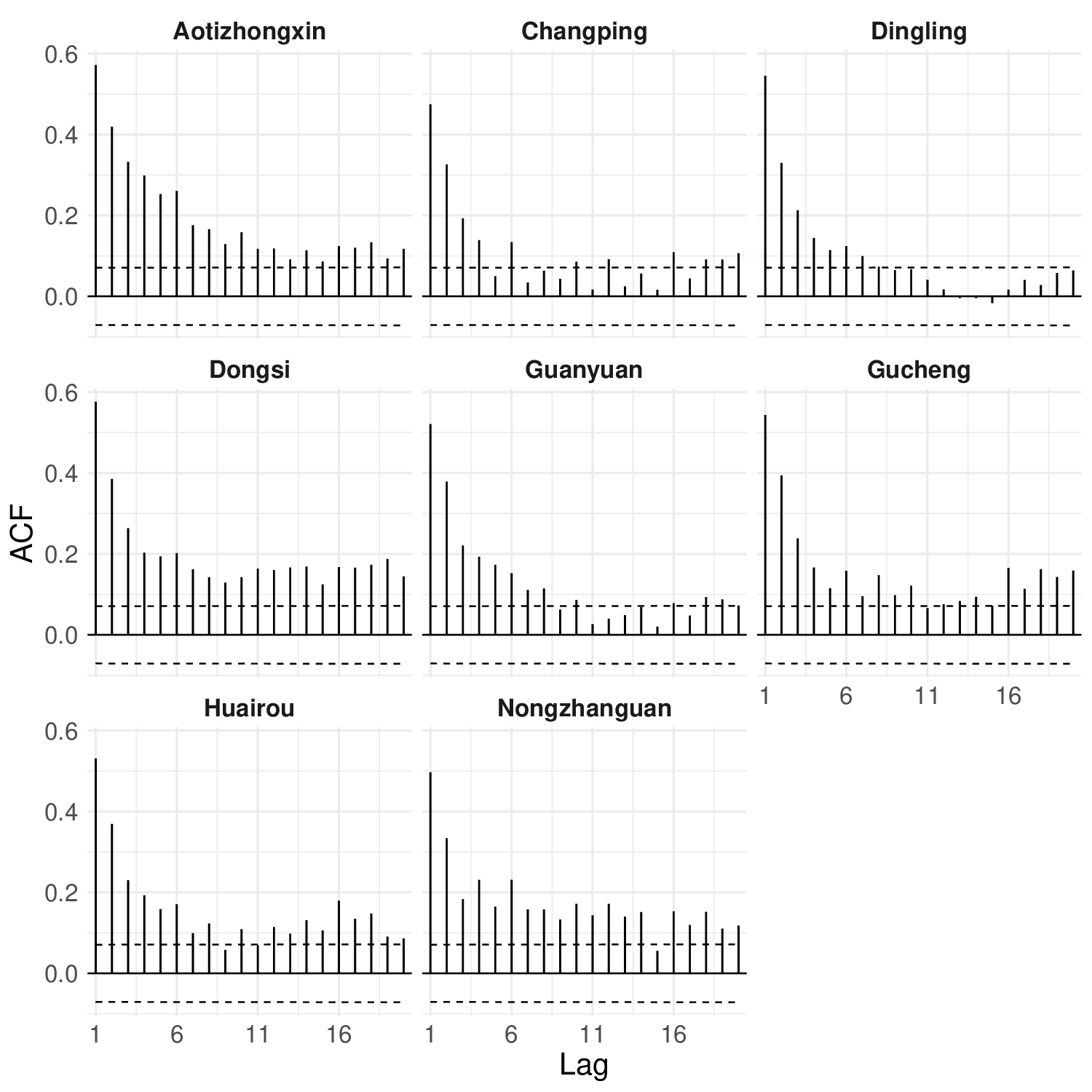}
    \caption{{\bf Air quality dataset}. Autocorrelation function of log-transformed CO concentrations, with missing values imputed using the station-wise mean.}
    \label{supp:acfco}
\end{figure}

\begin{figure}
    \centering
\includegraphics[width=1\linewidth]{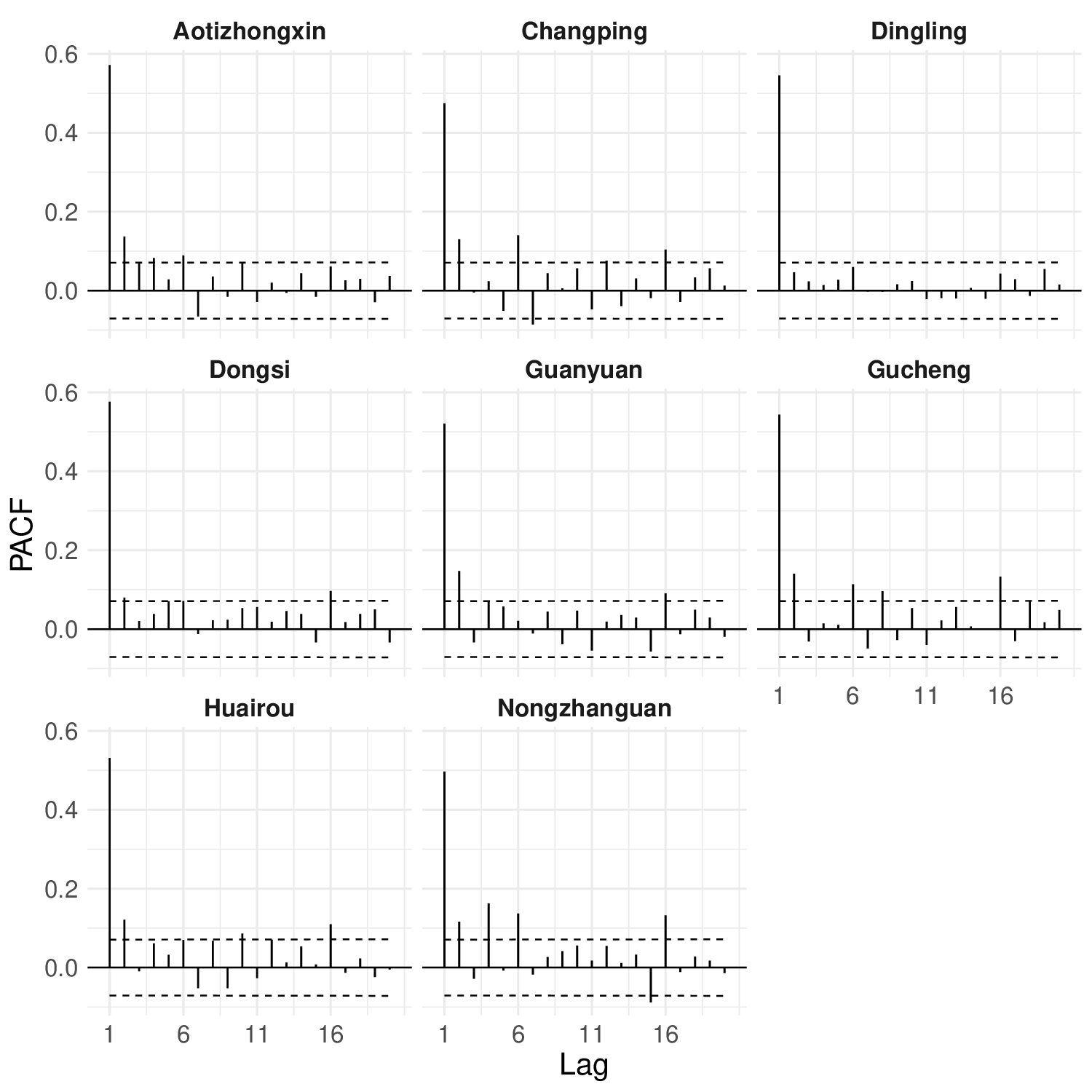}
    \caption{{\bf Air quality dataset}. Partial autocorrelation function of log-transformed CO concentrations, with missing values imputed using the station-wise mean.}
    \label{supp:pacfco}
\end{figure}





\bibliographystyle{chicago}
\bibliography{sn-bibliography}%

\end{document}